\documentclass[a4paper,fleqn]{article}
\usepackage[numbers,sort&compress]{natbib}
\usepackage{graphicx}
\usepackage{latexsym}
\usepackage{amsmath,amssymb}
\usepackage{dcolumn}
\usepackage{float}
\usepackage{caption}

\begin{document}
\captionsetup[figure]{labelfont={bf},labelformat={default},labelsep=space,name={Fig.}}

\title{{\bf Probing cosmic anisotropy from galaxy clusters via the dipole fitting method}
\author{\normalsize Jian-Ping Hu$^{1,4}$\thanks{Corresponding Author(J. P. Hu): Email: hjp1206@163.com; hjp2022@nju.edu.cn},
       Chao Geng$^{2}$, Xuan-Dong Jia$^{3}$, Zhao-Yu Zuo$^{1}$, Tao-Zhi Yang$^{1}$
        and Fa-Yin Wang$^{3.4}$ \\
   \normalsize \emph{$^{1}${Ministry of Education Key Laboratory for Nonequilibrium Synthesis and Modulation of Condensed Matter, }}\\ \normalsize \emph {School of Physics, Xi'an Jiaotong University, Xianning West Road, Xi'an, China}\\
   \normalsize \emph{$^{2}${Department of Astronomy, University of Science and Technology of China, JinZhai Road, Hefei, China}} \\ \normalsize 
   \normalsize \emph{$^{3}${School of Astronomy and Space Science, Nanjing University, Xianlin Road, Nanjing, China}}\\
   \normalsize \emph{$^{4}${Key Laboratory of Modern Astronomy and Astrophysics (Nanjing University), }} \\
   \normalsize \emph{Ministry of Education, Xianlin Road, Nanjing, China}
   }}

\date{}
\maketitle \baselineskip 0.3in

{\bf Abstract} \ {The cosmological principle, as the cornerstone of the standard cosmological model, requires that the universe be homogeneous and isotropic on large scales. As a fundamental assumption, it is constantly subjected to testing via various datasets and methods. In this work, we used the dipole fitting (DF) method to correct the logarithmic luminosity ($\log{L_{X}}$) of galaxy clusters, search for cosmic anisotropic signals, and establish a statistical isotropic analysis scheme. Compared to the type Ia supernovae (SNe Ia), the galaxy clusters offer a superior spatial distribution, which enhances the reliability of the identified anisotropic signals. Using a sample of 313 galaxy clusters (observed by Chandra and XMM-Newton), we identified the preferred direction (l, b) = (${257.82^{\circ}}_{-52.88}^{+58.01}$, $-31.30{^{\circ}}_{-39.46}^{+35.92}$) of the cosmic anisotropy. The corresponding magnitude of anisotropy is $A$ = $-5.4 \times 10^{-4}$. Subsample reanalyses categorized by instrumentation (Chandra and XMM-Newton) and redshift (low-redshift, $z \leq 0.10$; high-redshift, $z > 0.10$) revealed more significant anisotropic signals. The XMM-Newton dataset yields a statistical significance of $1.8\sigma$ (Bootstrap) and $1.9\sigma$ (Randomized), which are considerably higher than those from the Chandra or total datasets. Meanwhile, the reanalyses also reveal that the choice of equipment and the sample redshift influence the preferred direction, anisotropic magnitude, and statistical significance obtained from galaxy clusters. Overall, the DF method can be well integrated with galaxy clusters and applied to the detection of cosmic anisotropy. Our results suggest the presence of anisotropic signals in galaxy clusters; however, they do not reach a statistical confidence level sufficient to challenge the cosmological principle. Further tests are still needed to better understand these signals.
}


\section{Introduction}\label{sec:intro}
The $\rm \Lambda$ cold dark matter (CDM) model is widely recognized as the standard cosmological model, as it aligns well with the majority of astronomical observations \cite{2018ApJ...859..101S,2019PhRvL.122q1301A,2020MNRAS.492.4456K,2022ApJ...938..110B,2022MNRAS.513.5686C,2022MNRAS.514.1828D,2022MNRAS.516.2575J,2022ApJ...931...50L,2022ApJ...941...84L,2022PhRvD.106j3530P,2022ApJ...924...97W,2023PhRvD.107f3522D,2023MNRAS.522.1247K,2023MNRAS.521.4406L,2023arXiv231112098R,2025JCAP...02..021A}. Despite its remarkable success in modeling the evolution of the Universe, this model is encountering increasing challenges as the volume and precision of observations continue to grow. In addition to the recent hot issues such as Hubble constant ($H_{0}$) tension \cite{2019NatAs...3..891V,2020NatRP...2...10R,2021CQGra..38o3001D,2021A&ARv..29....9S,2022NewAR..9501659P,2023Univ....9...94H,2023CQGra..40i4001K,2023Univ....9..393V,2024IAUS..376...15R}, $S_{8}$ tension \cite{2021APh...13102604D,2021MNRAS.505.5427N,2024MNRAS.528L..20A,2024NatAs...8..405H,2024PhRvD.110f3501S}, dynamic dark energy \cite{2012PhRvL.109q1301Z,2017NatAs...1..627Z,2023arXiv231116862T,2024arXiv241212905C,2024PhRvD.110l3519J,2024JCAP...06..047V,2025JCAP...02..021A,2025JCAP...03..026L,2025arXiv250308658O,2025arXiv251014390W,2025arXiv250420664S}, evolutionary behavior of $H_{0}$ \cite{2020MNRAS.498.1420W,2020PhRvD.102j3525K,2021ApJ...912..150D,2022MNRAS.517..576H,2023A&A...674A..45J,2024PDU....4401464O,2025ApJ...979L..34J,2025MNRAS.536.3232M}, there are also the growth tension \cite{2017PhRvD..96f3517B,2018PhRvD..98d3526A,2018MNRAS.474.4894J}, age of the universe \cite{2013PDU.....2..166V,2019JCAP...03..043J,2022JHEAp..36...27V}, Baryon Acoustic Oscillation (BAO) curiosities \cite{2017JCAP...04..024E,2018ApJ...853..119A,2019JCAP...10..044C}, small-scale curiosities \cite{2017ARA&A..55..343B,2019A&ARv..27....2S} and cosmic anisotropy \cite{2003ApJS..148....1B,2005A&A...441..915H,2006PhRvL..96s1302B,2008JCAP...06..018K,2011MNRAS.414..264C,2011PhRvL.107s1101W,2014A&A...565A.111R,2020PhRvD.102l4059A,2023JCAP...07..020K}. Research related to the above topics indicates that the $\rm \Lambda$CDM model might have limitations in describing the structure and evolution of the Universe. Thus, its status as the standard cosmological model remains contingent upon validation through more precise astronomical observations \cite{2022JHEAp..34...49A,2025arXiv250401669D}.

The cosmological principle, as a basic assumption of the $\rm \Lambda$CDM model, requires that the universe is statistically isotropic and on sufficiently large scales. Given the significance of this hypothesis, it is necessary to rigorously examine and verify it using the latest data samples and different detection methods \cite{2020A&A...636A..15M}. Numerous studies suggest that the Universe might be inhomogeneous and anisotropic; such as the cosmic microwave background (CMB)
\cite{2003PhRvD..68l3523T,2004MNRAS.355.1283B,2011ApJS..192...17B,2011MNRAS.411.1445G,2020MNRAS.492.3994G,2020A&A...641A...7P}, fine-structure constant \cite{2012MNRAS.422.3370K,2017ChPhC..41f5102L,2023MmSAI..94b.270M}, direct measurement of the Hubble parameter \cite{2021PhRvL.126w1101K}, anisotropic dark energy \cite{2012PhRvD..86h3517M,2020JCAP...10..019B,2021PDU....3200806M}, anisotropic matter density \cite{2020PhRvD.102b3520K,2024A&A...681A..88H}, anisotropic Hubble constant \cite{2007NewA...12..533M,2020A&A...636A..15M,2022PhRvD.105f3514K,2022PhRvD.105j3510L,2023PhRvD.108l3533M,2024A&A...681A..88H,2024A&A...689A.215H}, large dipole of radio source counts \cite{2017MNRAS.471.1045C,2019PhRvD.100f3501S,2023MNRAS.524.3636S}, and so on. These studies are based on different types of observations, including galaxies \cite{2017MNRAS.468.1953P,2018JCAP...04..031B,2018MNRAS.477.1772R,2024MNRAS.531.4545O,2025MNRAS.537....1O}, galaxy spin direction \cite{2024MNRAS.534.1553P}, galaxy cluster \cite{2018A&A...611A..50M,2020A&A...636A..15M,2021A&A...649A.151M,2024A&A...691A.355P,2025RSPTA.38340030M,2025arXiv250401745H,2026arXiv260206007Y}, dwarf galaxies \cite{2024MNRAS.532.2490S}, quasar \cite{2016A&A...590A..53P,2019A&A...622A.113T,2021ApJ...908L..51S,2021EPJC...81..948Z,2021EPJC...81..694Z,2023MNRAS.525..231D,2023ApJ...953..144G,2024MNRAS.527.8497M,2024MNRAS.tmp.2660O,2024JCAP...06..019P}, CosmicFlows-4 \cite{2024arXiv241212637K,2025arXiv251002510K}, gamma-ray burst \cite[GRB;][]{2014MNRAS.443.1680W,2022MNRAS.511.5661Z,2025arXiv251020705S,2026arXiv260528662Z}, type Ia supernovae \cite[SNe Ia;][]{2013PhRvD..87l3522C,2014MNRAS.437.1840Y,2015ApJ...810...47J,2016RMxAA..52..133M,2018EPJC...78..755D,2018MNRAS.478.5153S,2018MNRAS.474.3516W,2019A&A...631L..13C} and their joint datasets \cite{2020A&A...643A..93H,2023PDU....3901162A,2023PhRvD.107b3507K,2024JCAP...09..077D,2024PDU....4601575O,2026arXiv260404408B}. Recently, some researches have shown that the local cosmic anisotropy might significantly affect the local constraints on $H_{0}$ \cite{2023PDU....4201365Y,2024PDU....4601626Y,2024ApJ...975L..36H}. The evolution behavior of $H_{0}$ and the Hubble tension might be closely related to the violation of cosmological principle in the local universe \cite{2024A&A...681A..88H,2024MNRAS.tmp.2636M,2026SSPMA..5640021H}. Therefore, precise testing of the cosmological principle can not only help us better understand the current state of the Universe, but also aid in exploring the physical origin of the $H_{0}$ evolutionary behavior and the Hubble tension.

Currently, the SNe Ia are the most commonly used late-time cosmological probe to test the cosmological principle \cite{Soltis:2019ryf,2020A&A...643A..93H,2022PhRvD.105f3514K,2022MNRAS.514..139R,2022A&A...668A..34H,2022PhRvD.106j3527Z,2023ChPhC..47l5101T,2024PhRvD.109l3533B,2024ApJ...971...19C,2024A&A...681A..88H,2024A&A...689A.215H,2024ApJ...967...47L,2025EPJC...85..339Y,2025JCAP...03..066B,2025arXiv250400903M,2025EPJC...85..596S,2026arXiv260108505R,2026arXiv260622302Z}. Its combination with the hemispheric comparison (HC) method \cite{2007A&A...474..717S,2023PhRvD.108f3509P,2024A&A...689A.215H}, dipole fitting (DF) method \cite{2012MNRAS.422.3370K,2018MNRAS.478.3633C,2024ApJ...975L..36H}, and regional fitting method \cite{2024A&A...681A..88H} can well reflect the anisotropy level of the local universe. The SNe Ia reveal departures from isotropy at confidence levels from 1.4 to 4.7$\sigma$ \cite{2019A&A...631L..13C,2020A&A...643A..93H,2023JCAP...11..054S,2023ChPhC..47l5101T,2024A&A...681A..88H,2024A&A...689A.215H,2024ApJ...967...47L}. It is worth noting that SNe Ia samples (Pantheon or Pantheon+) are still not uniformly distributed in the sky; some SNe are very concentrated, forming a belt-like structure that is the SDSS sample \cite{2011ApJS..194...45S}. A vast body of research has explored the impact of inhomogeneous distribution of SNe on the results of cosmic anisotropy, and found that the unevenness of the spatial distribution of samples will affect the preferred direction of anisotropy \cite{2018ApJ...865..119A,2019MNRAS.486.5679Z} and contribute a certain level of anisotropy \cite{2024A&A...689A.215H,2024A&A...681A..88H}. Consequently, the spatial inhomogeneity of SNe Ia may lead to an overestimation of the statistical significance of simulation results. If the simulated data are generated merely by reshuffling existing data on the celestial sphere rather than creating new data, the resulting randomness may be insufficient, thereby inflating the statistical significance \cite{2024A&A...681A..88H,2024ApJ...967...47L}. To circumvent this problem, on the one hand, a subsample with uniform spatial distribution can be selected from the existing observations. On the other hand, the influence of such spatial inhomogeneity can be weakened by incorporating new measurements; for instance, combining Cepheids with SNe either with Pantheon or Pantheon+ does not lead to a $H_{0}$ variation much beyond 2.0$\sigma$ \cite{2022PhRvD.106j3527Z,2023PhRvD.108l3533M}. Additionally, the most straightforward approach is to utilize other types of uniformly distributed observations, such as quasars \cite{2016ApJ...819..154L,2019NatAs...3..272R,2020A&A...642A.150L} and galaxy clusters \cite{2011A&A...534A.109P,2020A&A...636A..15M}. Therefore, the development of other new uniformly distributed probes that can be used to test the cosmological principle should be encouraged. Not long ago, Migkas et al. \cite{2020A&A...636A..15M} compiled a new sample of galaxy clusters consisting of 313 homogeneously selected sources from the Meta-Catalogue of X-ray detected Clusters of galaxies. Using the sky scanning method, they found a strong anisotropy toward Galactic coordinates (280$^{\circ}$, -20$^{\circ}$) with $\geq$4.0$\sigma$ confidence level. To date, no study has considered utilizing the dipole fitting (DF) method to probe the cosmological principle from the galaxy cluster observations. Furthermore, there is currently no statistical isotropic analysis scheme applicable to the DF method.

In this paper, we combined the galaxy cluster to an anisotropic model constructed using the DF method and the $\rm \Lambda$CDM model for the first time, to test the cosmological principle. Our goal is to probe anisotropic signals from the galaxy clusters, assess the statistical significance of anisotropy, and analyze the potential impact of factors such as data spatial distribution, observation equipment, and redshift on the probed results. The paper is organized as follows: In Section \ref{data} we briefly introduce the basic information of the galaxy cluster sample used and give its redshift distribution and spatial location distribution. Section \ref{method} details how to combine the $L_{X}-T$  relation of galaxy clusters with the DF method to probe the preferred direction of the cosmic anisotropy. It also describes how to construct the statistical isotropic analyses to evaluate the significance of the results. In Section \ref{result}, we present the main results, including the constraints of $L_{X}-T$ correlation, the anisotropic signals derived from various galaxy cluster datasets, and the corresponding statistical results. Afterwards, the corresponding investigation and discussion are given in Section \ref{dis}. Finally, a brief summary is given in Section \ref{sum}.

\section{Data} \label{data}
Galaxy cluster sample used in this work was screened by Migkas et al. \cite{2020A&A...636A..15M} from the Meta-Catalogue as of July 2019 of X-ray detected Clusters of galaxies \cite[MCXC;][]{2011A&A...534A.109P}. The parent catalogs of the clusters consists of the ROSAT extended Brightest Cluster Sample \cite[eBCS;][]{2000MNRAS.318..333E}, the Northern ROSAT All-Sky (NORAS) Galaxy Cluster Survey \cite{2000ApJS..129..435B} and the ROSAT-ESO Flux-Limited X-Ray (REFLEX) Galaxy Cluster Survey Catalog \cite{2004A&A...425..367B}. They are all based on the ROSAT All-Sky Survey \cite[RASS;][]{1999A&A...349..389V}. The basic selection criteria are that these clusters have high-quality Chandra \cite{2000SPIE.4012....2W} or XMM-Newton \cite{2001A&A...365L...1J} public observations. More detailed screening conditions can be found from the section 2 in \cite{2020A&A...636A..15M}. 

\begin{figure*}[htp]
    \centering
    \includegraphics[width=0.37\textwidth]{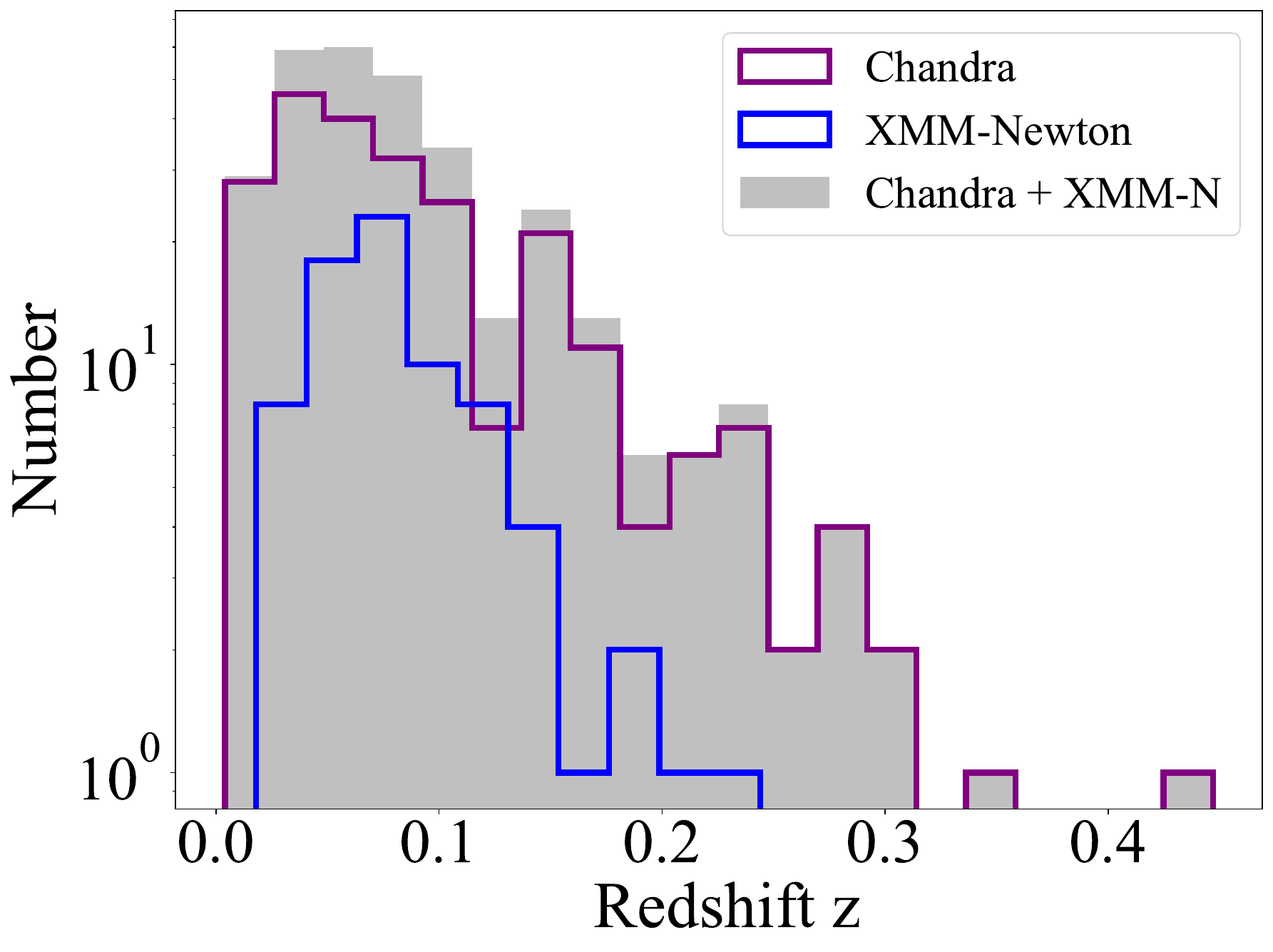} \ \
    \includegraphics[width=0.50\textwidth]{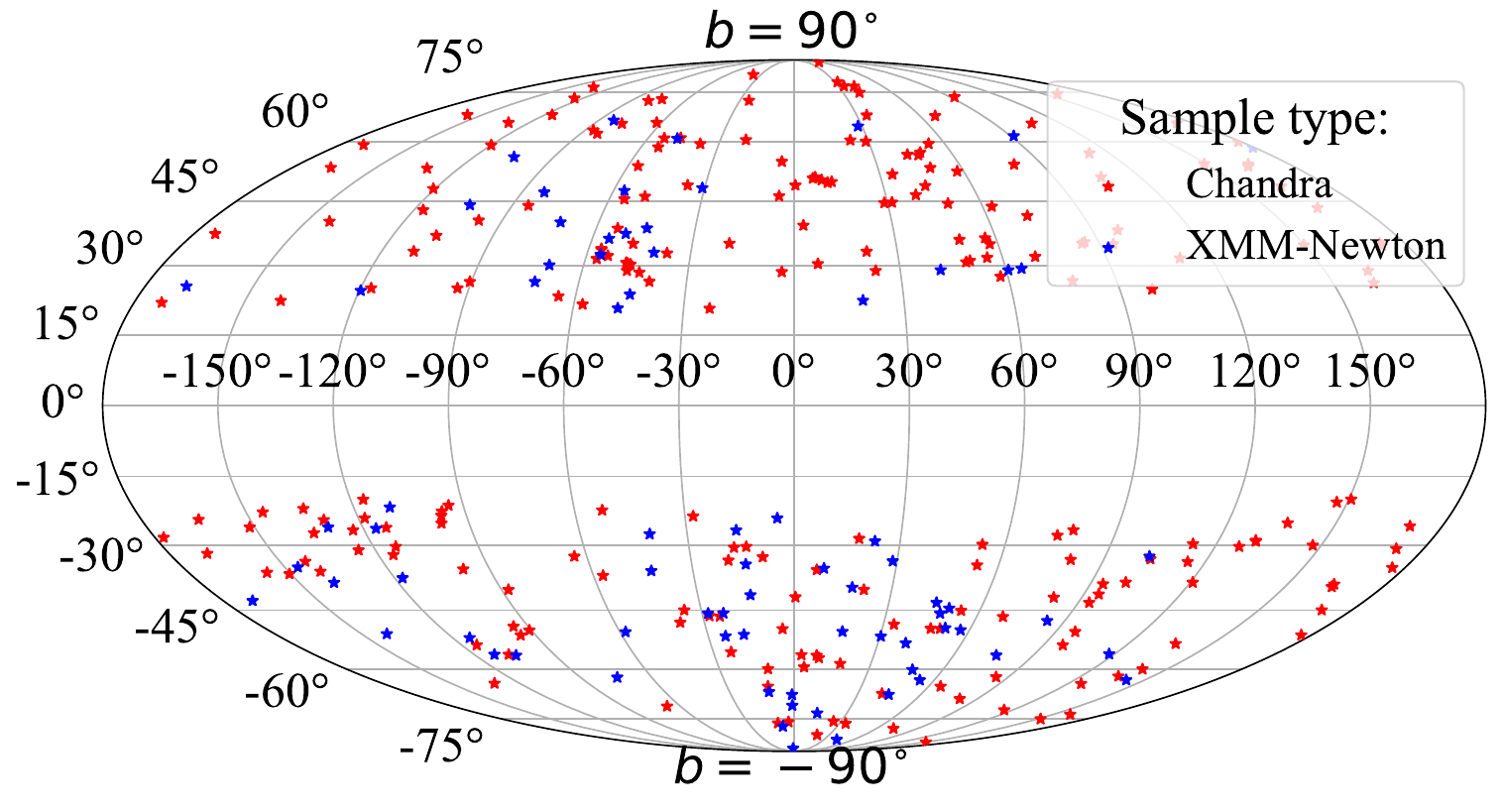}
    \caption{\label{Figzloc} Basic information of the galaxy cluster dataset. Redshift distribution (left panel) and location distribution (right panel) in the galactic coordinate system.}
\end{figure*}

The total sample, Chandra + XMM-Newton, used in this paper comprises 313 galaxy clusters, spanning a redshift range of 0.004 to 0.447. For brevity, Chandra + XMM-Newton is abbreviated as Chandra + XMM-N throughout this paper, particularly in figures and tables. This sample is derived from two datasets: Chandra and XMM-Newton. The former contains 237 clusters with a redshift range of (0.004, 0.447). The latter contains 76 clusters with a redshift range of (0.018, 0.244). By comparing the redshift range, it can be found that compared with the XMM-Newton dataset, the Chandra dataset cover a wider redshift range and have more high-redshift observations. We provide the redshift distributions of different types datasets in the left panel of Fig. \ref{Figzloc}. It can help us better understand the composition structure and redshift distribution of the datasets used. From the redshift distribution of the total sample, it is easy to find that the most of observed redshifts are less than 0.10. For convenience, we refer to the dataset with redshift below 0.10 as LR dataset. It consists of 219 clusters accounting for about 70\% of the total sample. It is mainly contributed by the Chandra observation (162 clusters) and a small part by the XMM-Newton observation (57 clusters). There are 94 clusters with redshift greater than 0.10, including 75 Chandra measurements and 19 XMM-Newton measurements. This sample is called the HR dataset. Meanwhile, we also mapped their location distributions in the galactic coordinate system, as shown in the right panel of Fig. \ref{Figzloc}. Compared to SNe Ia, the spatial distribution of galaxy cluster is more uniform. In the SNe Ia samples (for example Pantheon and Pantheon+), there is a data-rich band structure \cite{2022ApJ...938..110B}, which significantly affect the overall uniformity of the observations \cite{2024A&A...689A.215H,2024A&A...681A..88H}. From the comparison of Chandra and XMM-Newton datasets, we find that the Chandra observation is more uniform than the XMM-Newton observation in terms of spatial distribution. Overall, the spatial distribution of galaxy clusters is relatively uniform, making them suitable for testing the cosmological principle. 

\section{Methodology} \label{method}
\subsection{X-ray luminosity-temperature relation}
It is well-known that the galaxy clusters are the most massive gravitationally bound systems in the universe, strongly emitting X-ray photons due to the large amounts of hot gas they
contain ($\sim$10\% of their total mass) in their intra-cluster medium (ICM). Their physical quantities follow tight scaling relations, for which Kaiser \cite{1986MNRAS.222..323K} provided mathematical expressions. Specifically, the correlation between the X-ray luminosity ($L_{X}$) and the ICM gas temperature ($T$) of galaxy clusters is of particular interest since it can be used to test the cosmological principle. 

The general properties of the $L_{X}-T$ scaling relation have been extensively scrutinized in the past by several authors \cite{2002ApJ...578L.107V,2007MNRAS.382.1289P,2009A&A...498..361P,2011A&A...532A.133M,2011A&A...535A...4R,2012MNRAS.424.2086H,2012MNRAS.421.1583M,2015A&A...573A..75B,2015A&A...573A.118L,2016A&A...592A...3G,2016MNRAS.463..820Z,2018A&A...611A..50M,2020A&A...636A..15M,2021A&A...649A.151M}. A standard power-law form of the $L_{X}-T$ relation are as follows \cite{2011A&A...532A.133M}:
\begin{eqnarray}
\label{eq:lt}
    \frac{L_{X}}{10^{44} \rm{erg/s}} E(z)^{-1} = k \times (\frac{T}{4 \rm{keV}})^{s},
\end{eqnarray} 
where term $E(z)$ = $\sqrt{\Omega_{m}(1+z)^{3}+\Omega_{\Lambda}}$ scales $L_{X}$ accordingly to explain the redshift evolution of the $L_{X}-T$ relation. Parameters $k$ and $s$ are the free parameters to be fitted. Parameters $L_{X}$ and $T$ represent the X-ray luminosity and temperature, respectively. $L_{X}$ is strongly related to cosmology, and can be derived from the formula $L_X =  4\pi d_{L}^{2}F_0$. Here, $F_{0}$ represents the k-corrected flux, and $d_{L}$ represents the luminosity distance. Considering a spatially flat $\rm \Lambda$CDM model, the luminosity distance $d_{L}$ can be derived from the following formal
\begin{equation}
d_{L} = \frac{c(1+z)}{H_{0}} \int_{0}^{z} 
\frac{dz'}{\sqrt{\Omega_{m} (1+z')^{3} + \Omega_{\Lambda}}},
\label{dl}
\end{equation}
where $c$ is the speed of light, $H_{0}$ is the Hubble constant,
$\Omega_{m}$ and $\Omega_{\Lambda}$ are the matter parameter and the dark energy (DE) parameter respectively, satisfying $\Omega_{m}$ + $\Omega_{\Lambda}$ = 1. Equation (\ref{eq:lt}) can be written in logarithmic form as follows:
\begin{eqnarray}
\label{eq:ltlog}
    \log{\mathcal{L}_{X}} = \log{k} + s\log{\mathcal{T}},
\end{eqnarray} 
where 
\begin{eqnarray}
\label{eq:l-t-}
    \mathcal{L}_{X} = \frac{L_{X}}{10^{44} \rm{erg/s}} E(z)^{-1} \ \ \rm{and} \ \ \mathcal{T} = \frac{T}{4 \rm{keV}}.
\end{eqnarray}
Constraints on the free parameters of $L_{X}-T$ relation to be fitted can be derived by minimizing corresponding $\chi^{2}_{\rm Cluster}$,
\begin{equation}
        \chi^{2}_{\rm Cluster} = \sum_{i=1}^{N} \frac{\log{[\mathcal{L}_{X, obs}}] - \log{[\mathcal{L}_{X, th}(\mathcal{T}, P_{n})}]}{ \sigma_{\log{L_{i}}}^{2} + s^{2}\times\sigma_{\log{T_{i}}}^{2} + \sigma_{int}^{2}}, 
        \label{chi_G}
\end{equation}
where $N$ represents the number of clusters used in analyzing, $\mathcal{L}_{X, th}$ is the theoretical X-ray luminosity based on the measured temperature $T$, and $P_{n}$ represents the parameters ($H_{0}$, $\Omega_{m}$, $k$ $s$, and $\sigma_{int}$) to be fitted. $\mathcal{L}_{X, obs}$ is the observed X-ray luminosity which derived by the observed flux based on a fix cosmological model. Furthermore, $\sigma_{\log{L_{i}}}$ and $\sigma_{\log{T_{i}}}$ are the corresponding 1$\sigma$ errors. $\sigma_{int}$ is the intrinsic scatter of the $L_{X}-T$ correlation, which is usually used to describe the compactness of the correlation; the smaller the value, the more compact the relationship. More information on the $L_{X}-T$ relation and its cosmological applications can be found from the review article \cite{2025RSPTA.38340030M}.

It is worth noting that the cosmological parameters and the $L_{X}-T$ correlation parameters cannot be constrained simultaneously since they are degenerate. Therefore, one needs to fix one of the parameters to investigate the behavior of the other. In this paper, the observed X-ray luminosity $\mathcal{L}_{X, obs}$ is derived utilizing $L_{X}$ which provided by Migkas et al. \cite{2020A&A...636A..15M}. They calculated $L_{X}$ based on the flat $\rm \Lambda$CDM cosmology with $\Omega_{m}$ = 0.30, $\Omega_{\Lambda}$ = 0.70, and $H_{0}$ = 70.0 km/s/Mpc. In the process of searching for anisotropic signals, they used a fixed $H_{0}$ = 70.0 km/s/Mpc to investigate the behavior of normalization parameter ($s$), then fixed the normalization parameter $s$ to its best-fit value and study the directional behavior of $H_{0}$. In addition, they also discussed the potential impact of fixing the slope parameter.


\subsection{Dipole fitting method} \label{dfmethod}
Dipole fitting (DF) method was proposed to find the fine structure constant ($\Delta \alpha$/$\alpha$) dipole \cite{2012MNRAS.422.3370K}. This approach constructs a dipole model which constitutes the first two term of the spherical harmonic expansion. Instead of ($\Delta \alpha$/$\alpha$), which corresponds to fine structure constant deviations from its earth measured value, Mariano \& Perivolaropoulos \cite{2012PhRvD..86h3517M} first applied this method to probe the dark energy dipole based on the distance modulus $\mu$. The basic equation is as follows
\begin{equation}
(\frac{\Delta \mu(z)}{\bar{\mu}(z)}) \equiv \frac{\mu(z) - \bar{\mu}(z)} {\bar{\mu}(z)} = A\cos{\theta}+B,
\label{ddd}
\end{equation}
where $\bar{\mu}(z)$ is the best fit distance modulus in the context of the $\rm \Lambda$CDM model, $\cos{\theta}$ is the angle with the dipole axis, $A$ is the dipole magnitude, and $B$ is the monopole magnitude. Afterwards, it was widely used to probe the preferred direction of the cosmic anisotropy \cite{2018ChPhC..42k5103C,2018PhRvD..97l3515D,2024ApJ...975L..36H}. Taking SNe Ia as an example, the DF method corrects the distance modulus $\mu$ expected by the $\rm \Lambda$CDM model. In this paper, we applied this method for the galaxy clusters to test the cosmological principle. In order to make the DF method applicable to galaxy cluster, we replaced ($\Delta \log{L_{X}}$/$\log{L_{X}}$) with ($\Delta \alpha$/$\alpha$), and the modified form is as follows:
\begin{equation}
(\frac{\Delta \log{L_{X}}}{\log{\bar{L}_{X}(z)}}) \equiv \frac{\log{L_{X}(z)} - \log{\bar{L}_{X}}(z)} {\log{\bar{L}_{X}(z)}} = A\cos{\theta}+B,
\label{dfgc}
\end{equation} 
where $\log{\bar{L}_{X}(z)}$ represents the best fits in the context of the $\rm \Lambda$CDM model. According to Eq. (\ref{dfgc}), we can obtain the following form
\begin{equation}
\log{\tilde{L}_{X,th}} = \log{L_{X,th}}\times(1+A\cos{\theta}+ B),
\label{dflx}
\end{equation}
where $\log{\tilde{L}_{X,th}}$ represents the modified theoretical X-ray luminosity. The angle with the dipole axis, $\cos{\theta}$, is defined by 
\begin{equation}
\cos{\theta} = \hat{\textbf{n}}\cdot\hat{\textbf{p}},
\label{theta}
\end{equation}
here, $\hat{\textbf{n}}$ and $\hat{\textbf{p}}$ correspond to the
dipole direction and the unit vector pointing to the position of galaxy clusters, respectively. In the galactic coordinate, the form of $\hat{\textbf{n}}$ can be written as
\begin{equation}
\hat{\textbf{n}} = \cos{(b)}\cos{(l)}\hat{\textbf{i}}+\cos{(b)}\sin{(l)}\hat{\textbf{j}}+sin{(b)}\hat{\textbf{k}}.
\label{n}
\end{equation}
For any observational source whose location is $(l_{i},b_{i})$, $\hat{\textbf{p}}$ is given by
\begin{equation}
\hat{\textbf{p}_{i}} = \cos{(b_{i})}\cos{(l_{i})}\hat{\textbf{i}}+\cos{(b_{i})}\sin{(l_{i})}\hat{\textbf{j}}+sin{(b_{i})}\hat{\textbf{k}}.
\label{p}
\end{equation}
Substituting Eqs. (\ref{dflx}) into (\ref{eq:ltlog}), we can derive the modified form of $\log{\mathcal{L}_{X}}$; that is, $\log{\tilde{\mathcal{L}}_{X,th}}$. Substitute $\log{\tilde{\mathcal{L}}_{X,th}}$ into Eq. (\ref{chi_G}), we can obtain a new chi-square function ($\chi^{2}_{\rm Cluster, DF}$), which takes into account the dipole-monopole correction and has the following form:
\begin{equation}
        \chi^{2}_{\rm Cluster, DF} = \sum_{i=1}^{N} \frac{\log{[\mathcal{L}_{X, obs}}] - \log{[\tilde{\mathcal{L}}_{X, th}(\mathcal{T}, P_{n})}]}{ \sigma_{\log{L_{i}}}^{2} + s^{2}\times\sigma_{\log{T_{i}}}^{2} + \sigma_{int}^{2}}.
        \label{chi_GDF}
\end{equation}
There are total 7 parameters to be fitted, including the correlation parameters ($k$, $s$ and $\sigma_{int}$) and the dipole parameters ($l$, $b$, $A$ and $B$). The coordinate ($l$, $b$) represents the preferred direction (longitude, latitude) in the galactic coordinate system. The parameters $A$ and $B$ are used to describe the level of anisotropy.

In summary, the dipole correction parameters are introduced as phenomenological corrections to the $L_{X}-T$ relation. If the galaxy cluster observations are more favorable to an isotropic universe, the values of $A$ and $B$ will be closer to zero. The non-zero $A$ or $B$ may indicate that the observations favor an anisotropic universe. In general, the DF method yields two sets of constraints corresponding to two distinct directions: one set characterized by brighter luminance and the other by darker luminance. It is generally believed that the direction with darker luminance is the preferred direction of anisotropy in the universe. In directions corresponding to negative $A$ values, the observed luminosity is lower than that estimated by the $L_X-T$ relation at a given temperature $T$. Similarly, for parameter $B$, a negative value indicates that the clusters are underluminous relative to the expected values calculated from the $L_X-T$ relation.

\subsection{Statistical isotropic analyses}
To assess whether the dipole magnitude $A$ derived from the galaxy cluster sample is consistent with statistical isotropy, we developed an isotropic analysis scheme specifically for the DF method. First, we fixed the parameters $l$ and $b$ based on the preferred direction constrained by the real data. In this analysis, we chose the direction $(l, b)$ associated with the negative $A$ value, representing luminance darker than the surrounding regions. Subsequently, we utilized simulated datasets to constrain parameters $A$ and $B$. The simulation results ($A_{B}$ and $A_{R}$) can be used to assess the statistical significance ($D$) of the real results. The values of $D$ can be obtained from the following formal
\begin{equation}
        D(\sigma) = \frac{A_{real} - A_{mock}}{\sqrt{\sigma_{A_{real}}^{2} + \sigma_{A_{mock}}^{2}}},
        \label{D}
\end{equation}
where $A_{real}$ represents the real results and $A_{mock}$ represents the simulation results ($A_{B}$ or $A_{R}$).

In our statistical analyses, we implemented two schemes. The first scheme preserves the spatial coordinates of the real data but randomly reshuffles the observed values among the galaxy cluster positions; for convenience, we refer to this scheme as "Bootstrap". From this, we obtain the constraints on the correction parameters ($A$ and $B$) for the simulated datasets. By comparing the $A$ values, we can estimate the contribution of cosmic structure to anisotropy. As a comparison, we considered a second scheme, referred to as "Randomized", which involves removing the original location information and distributing the data points uniformly across the celestial sphere. By investigating the $A$ values obtained from both schemes, we can estimate the degree to which the spatial distribution influences our results. 

Furthermore, we can use real results ($A_{real}$) and simulation results ($A_{B}$ and $A_{R}$) to obtain a $D$-distribution, which can quantify the statistical significance of the deviation from isotropy. Considering the limitations of computation time, multiple simulations were conducted to generate 1000 sets of simulated data. This yielded acceptable statistics. 

In this work, the minimization was performed employing a Bayesian Markov chain Monte Carlo \citep[MCMC;][]{2013PASP..125..306F} method with the $emcee$ package\footnote{https://emcee.readthedocs.io/en/stable/}. All the fittings in this paper were obtained adopting this python package. For the MCMC analysis, we employ 300 walkers run for 5000 steps each. The first 50 steps of each walker are discarded as burn-in to eliminate initial state dependencies. Convergence is confirmed by ensuring that the total chain length per walker well exceeds 50 times the integrated autocorrelation time ($\tau$) for all free parameters.

\section{Results} \label{result}
\subsection{Constraints of the $L_{X}-T$ correlation} \label{LTC}
\begin{figure}[htp]
    \centering
     \includegraphics[width=0.33\textwidth]{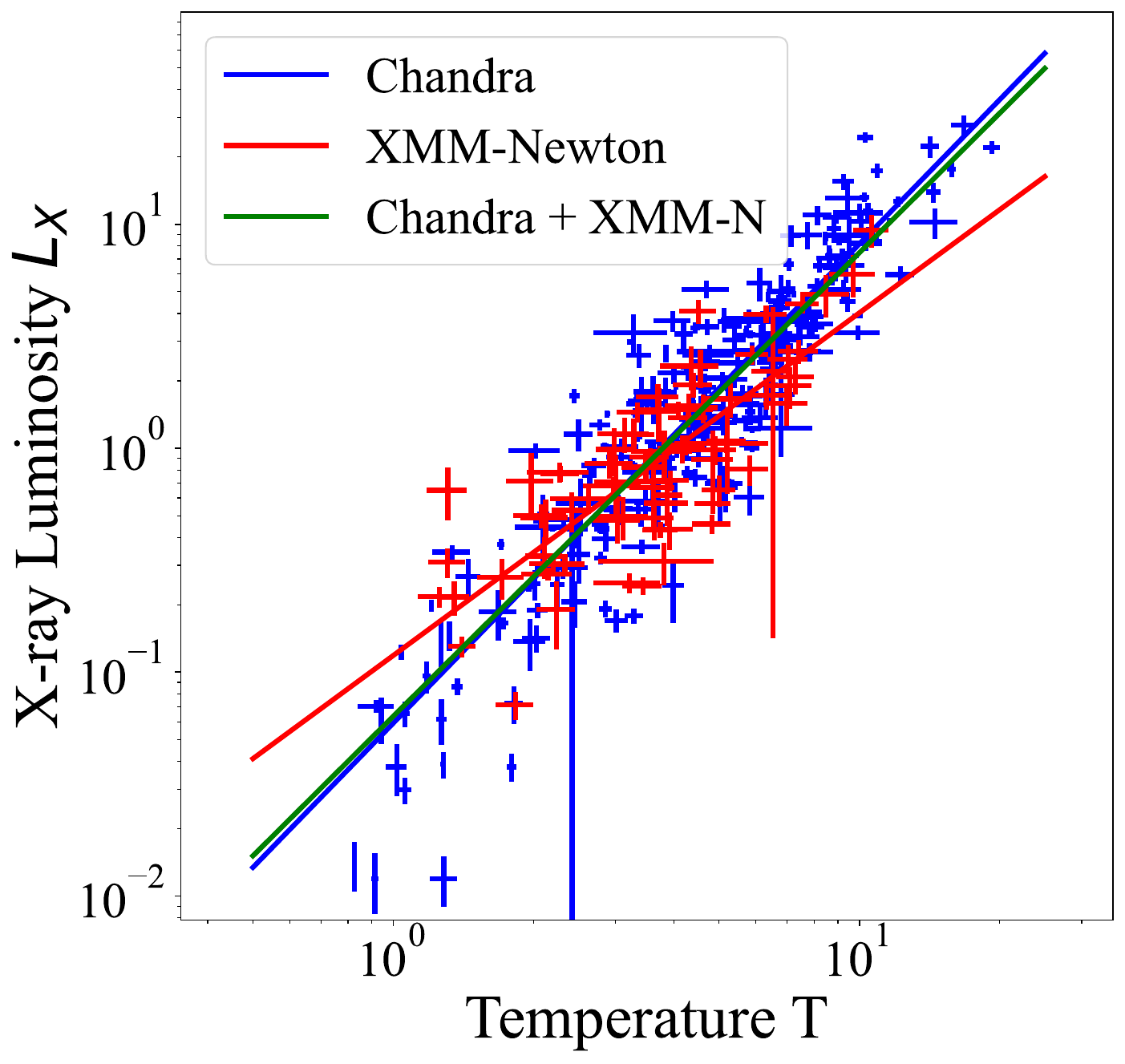}
     \includegraphics[width=0.33\textwidth]{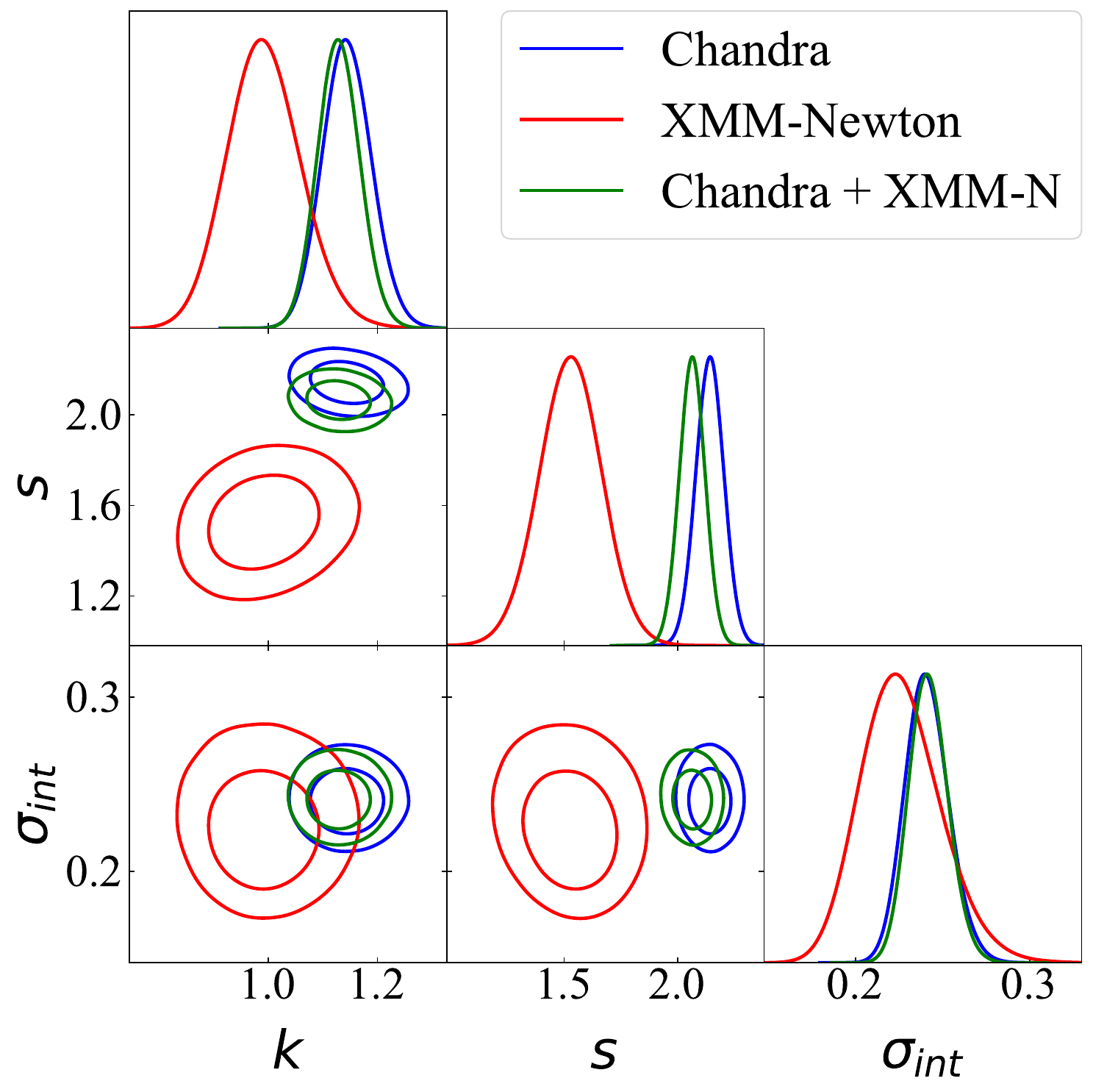}
    \caption{\label{FigLT} $L_{X}-T$ diagrams (left panel) and the corresponding parameter confidence contours (right panel) for different types datasets. Blue, red and green represent the results from Chandra, XMM-Newton, and Chandra + XMM-Newton datasets, respectively. }
\end{figure}

First, we gave the constraints of $L_{X}-T$ correlation using the different galaxy cluster datasets, including Chandra, XMM-Newton and Chandra + XMM-Newton. For the Chandra dataset, the best fitting results are $k$ = 1.15$\pm$0.05, $s$ = 2.14$\pm$0.06 and $\sigma_{int}$ = 0.24$\pm$0.01. For the XMM-Newton dataset, the best fitting results are $k$ = 0.99$_{-0.07}^{+0.06}$, $s$ = 1.53$\pm$0.14 and $\sigma_{int}$ = 0.23$_{-0.03}^{+0.02}$. For the total sample (Chandra + XMM-Newton), the corresponding results are $k$ = 1.12$\pm$0.04, $s$ = 2.07$\pm$0.06 and $\sigma_{int}$ = 0.24$\pm$0.01. The corresponding $L_{X}-T$ diagrams and confidence contours are displayed in Fig. \ref{FigLT}. 

\begin{figure}[htp]
    \centering
     \includegraphics[width=0.33\textwidth]{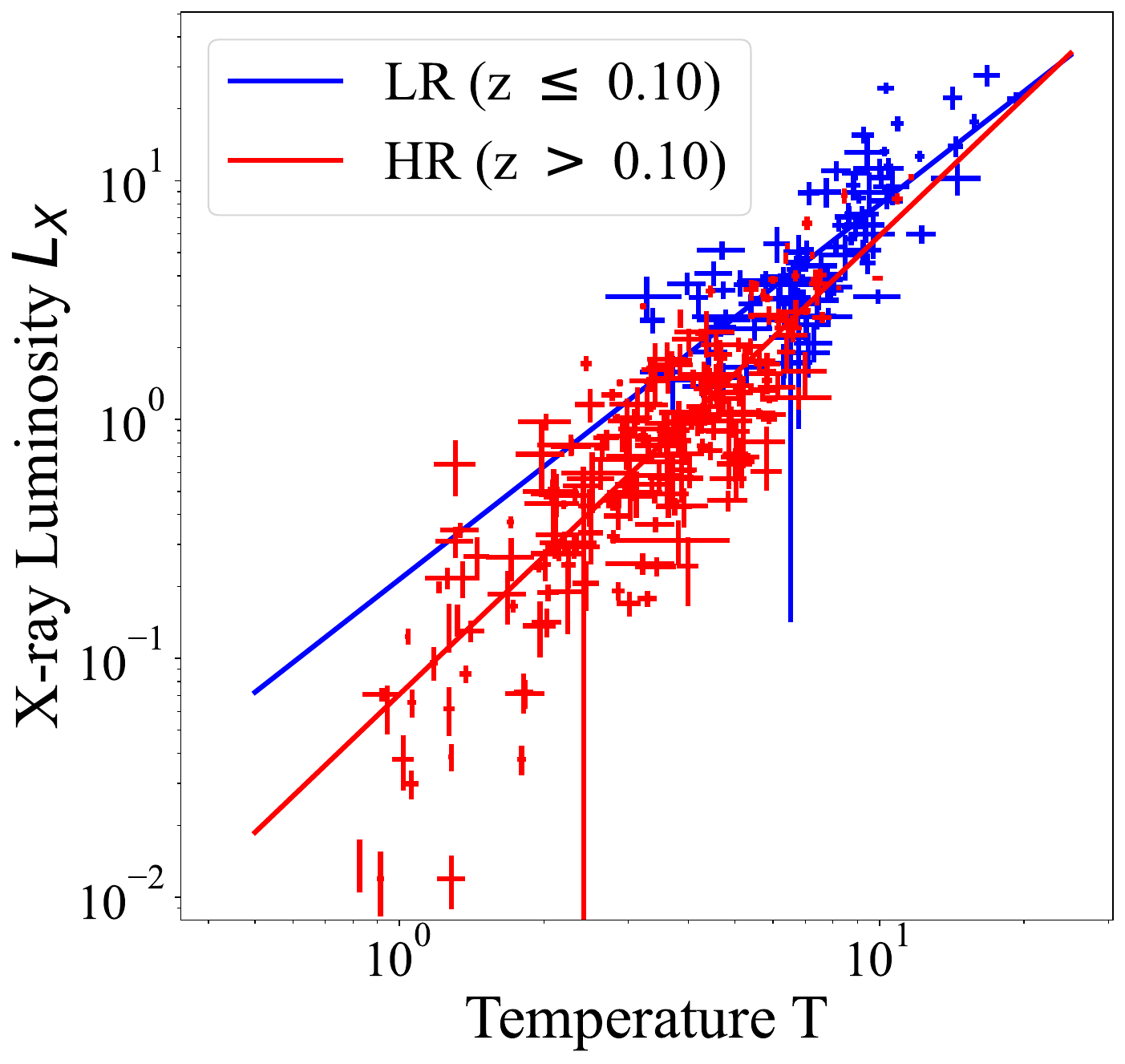}
     \includegraphics[width=0.33\textwidth]{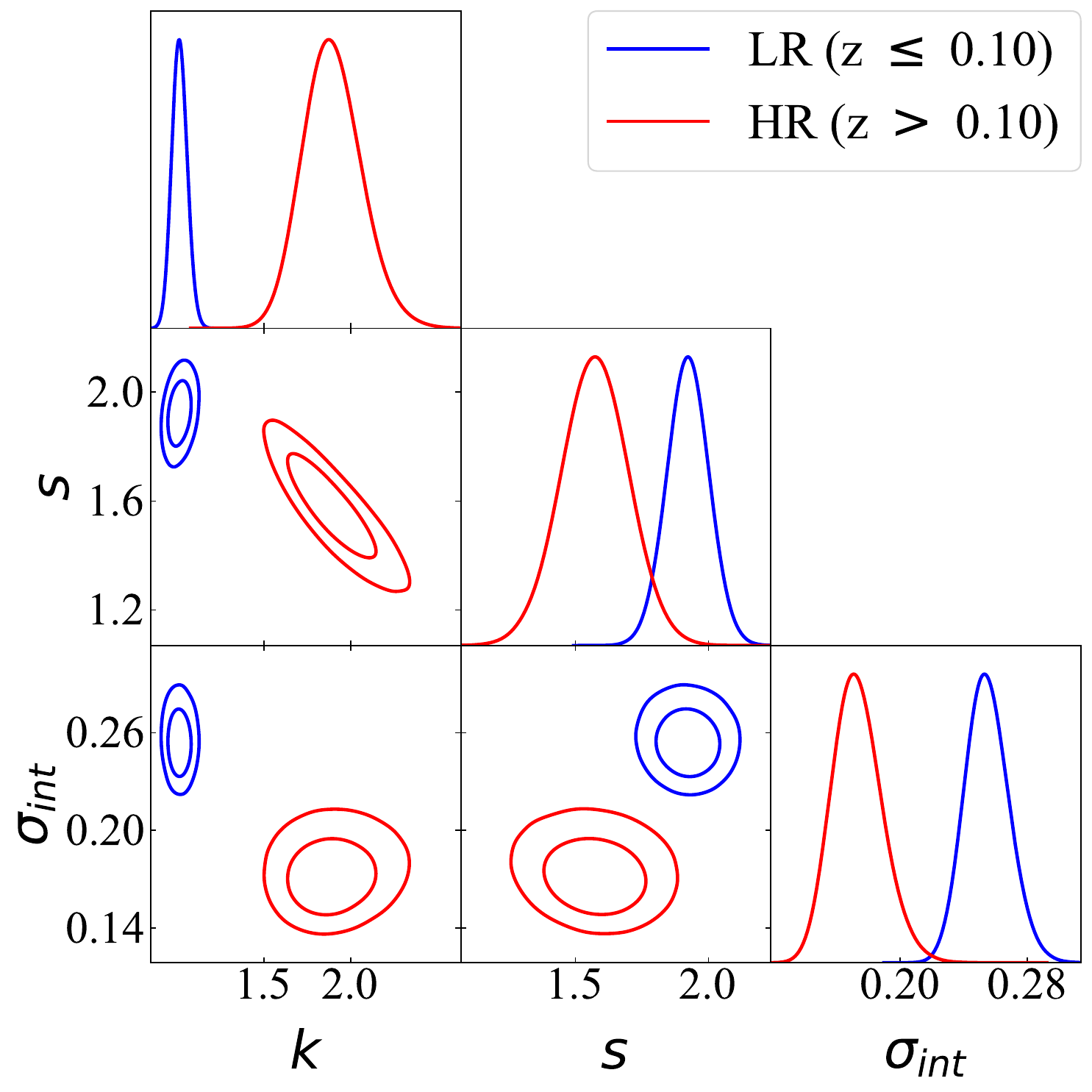}
    \caption{\label{FigLT1} $L_{X}-T$ diagrams (left panel) and the corresponding parameter confidence contours (right panel) for different redshift ranges. Blue and red represent the results from LR and HR datasets, respectively. }
\end{figure}

Afterwords, we also gave the constraints in different redshift intervals. The total sample is divided into two sub-samples according to redshift: low redshift subsample (LR; redshift $\leq$ 0.10) and high redshift subsample (HR; redshift $>$ 0.10). For the LR dataset, the best fits are $k$ = 1.01$\pm$0.04, $s$ = 1.92$\pm$0.08 and $\sigma_{int}$ = 0.26$_{-0.02}^{+0.01}$. For the HR dataset, the corresponding results are $k$ = 1.89$_{-0.18}^{+0.15}$, $s$ = 1.57$\pm$0.13 and $\sigma_{int}$ = 0.17$_{-0.02}^{+0.01}$. The corresponding $L_{X}-T$ diagrams and confidence contours are showed in Fig. \ref{FigLT1}. All numerical results are summary in Table \ref{T1}. According to the constraints of the $L_{X}-T$ relation, we can use galaxy clusters to test cosmological principle and search for the preferred direction of the cosmic anisotropy.

\begin{table}[htp]
\centering
	\caption{Constraints of the $L_{X}-T$ correlation ($k$, $s$, and $\sigma_{int}$) from different galaxy cluster datasets. \label{T1}}
    \renewcommand{\arraystretch}{1.5}
		\begin{tabular}{c|ccc}
			\hline\hline
		Sample & $k$ & $s$ & $\sigma_{int}$ \\ \hline
		Chandra   & 1.15$\pm$0.05 & 2.14$\pm$0.06 & 0.24$\pm$0.01 \\
        XMM-Newton & 0.99$_{-0.07}^{+0.06}$  & 1.53$\pm$0.14 &  0.23$_{-0.03}^{+0.02}$ \\
        Chandra + XMM-N  & 1.12$\pm$0.04  & 2.07$\pm$0.06 & 0.24$\pm$0.01 \\ 
        LR (z $\leq$ 0.10) & 1.01$\pm$0.04  & 1.92$\pm$0.08 & 0.26$_{-0.02}^{+0.01}$ \\ 
        HR (z $>$ 0.10) & 1.89$_{-0.18}^{+0.15}$ & 1.57$\pm$0.13 & 0.17$_{-0.02}^{+0.01}$ \\ 
			\hline\hline
		\end{tabular}
\end{table}

\subsection{Anisotropic signal from different galaxy cluster samples}
We applied the DF method to the galaxy cluster to search for the preferred direction of the cosmic anisotropy. From the description of the DF method in Sect. \ref{dfmethod}, we can see that there are a total of 7 free parameters: correlation parameters ($k$, $s$ and $\sigma_{int}$), and dipole parameters ($l$, $b$, $A$ and $B$). In order to obtain more accurate constraints on the preferred direction ($l$, $b$), it is better to fix the correlation parameters ($k$, $s$ and $\sigma_{int}$). Thus, we fixed the correlation parameters according to the constraints of $L_{X}-T$ relation in Section \ref{LTC} to search for cosmic anisotropic signals.
 
\begin{figure*}[ht]
    \centering
    \includegraphics[width=0.39\textwidth]{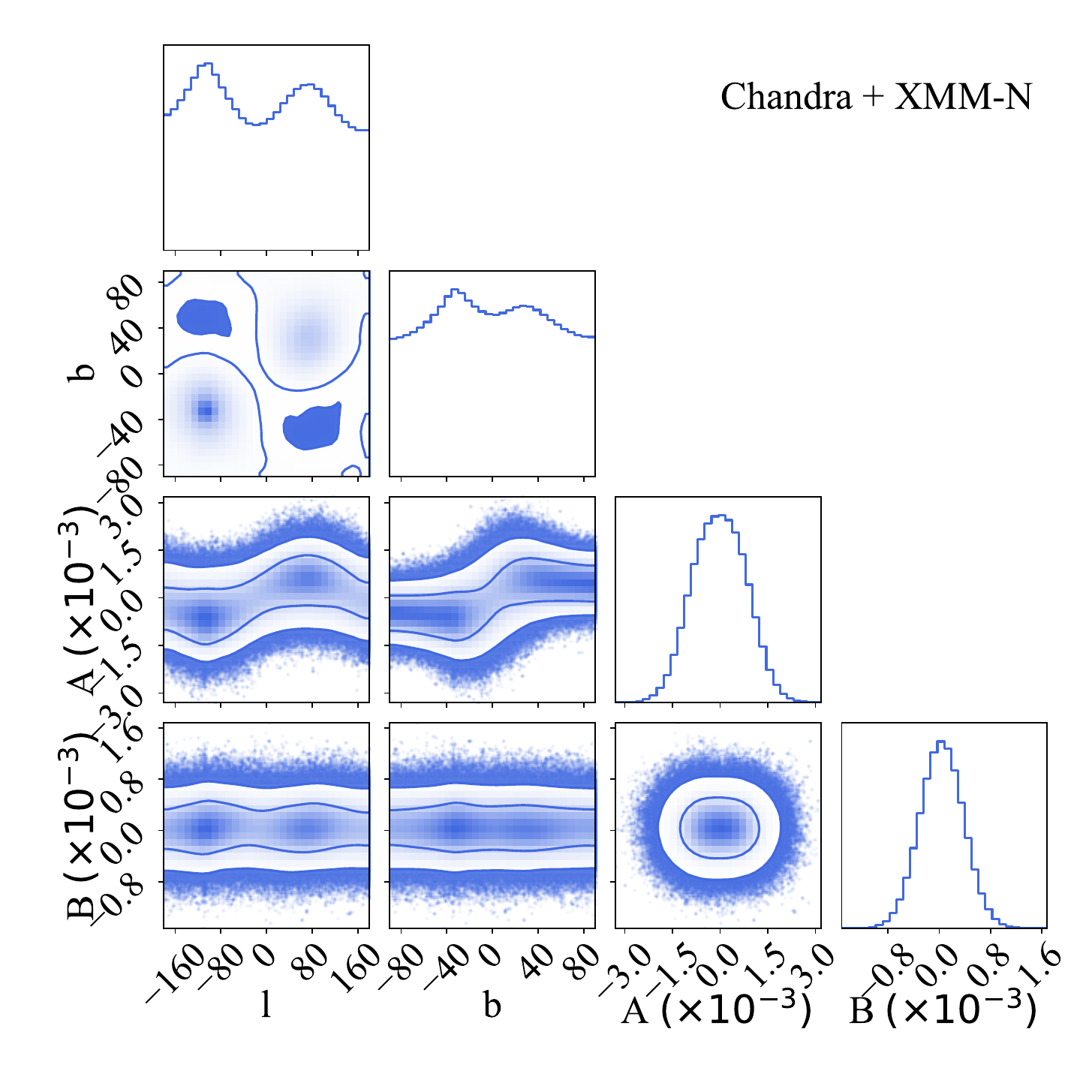}
    \includegraphics[width=0.39\textwidth]{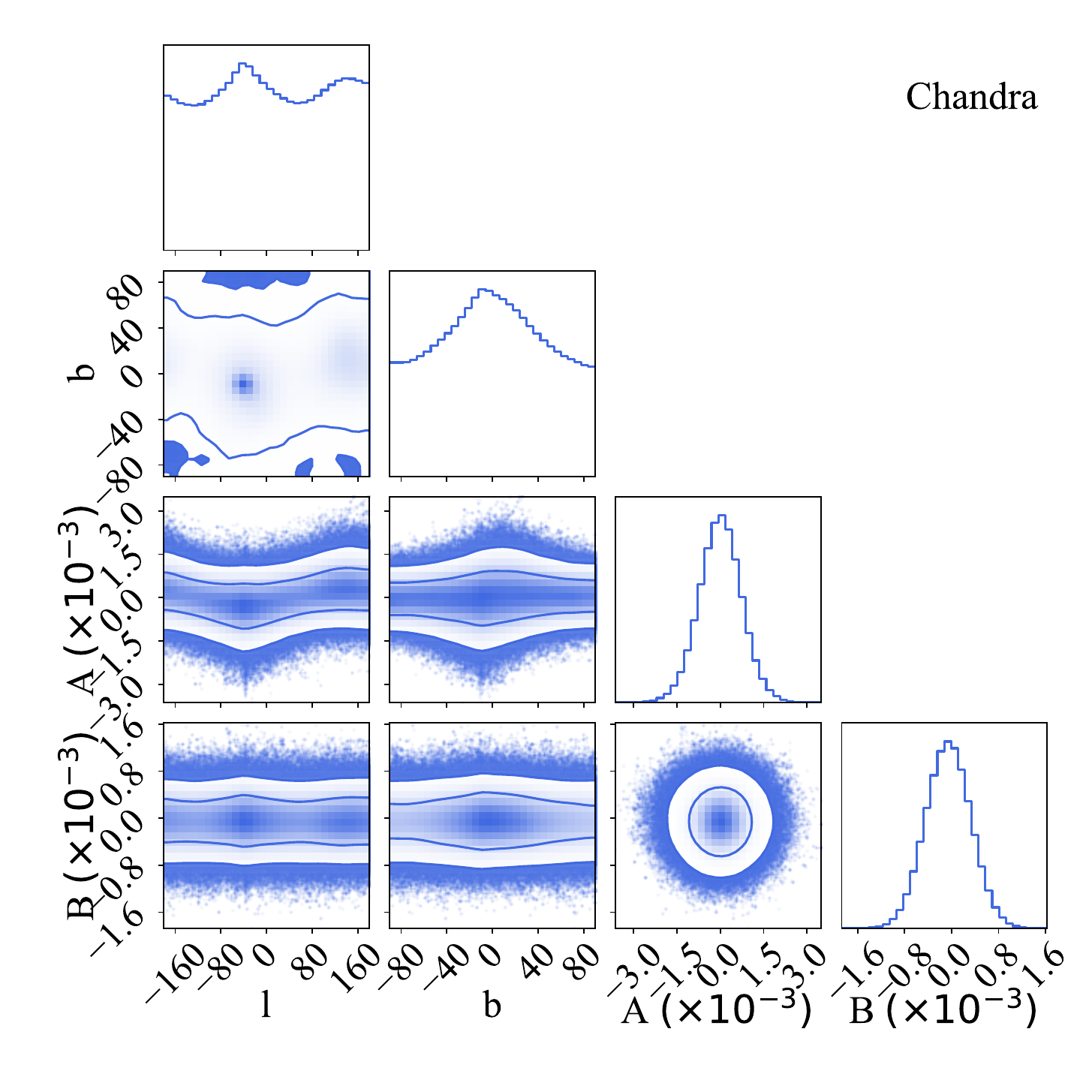}
    \includegraphics[width=0.39\textwidth]{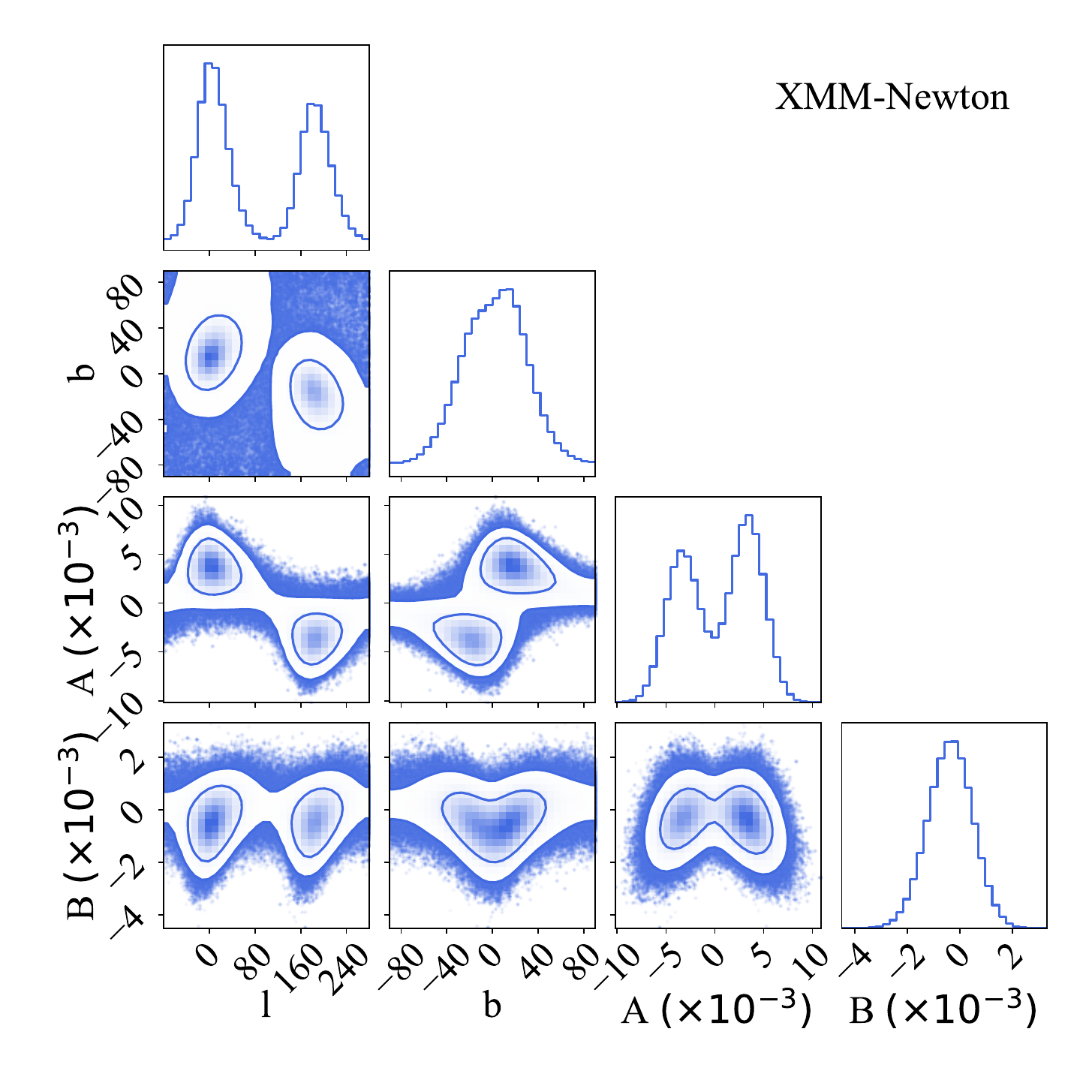} 
    \caption{\label{Figdy} Confidence contours (1$\sigma$ and 2$\sigma$) of the dipole parameters (l, b, A, and B) for different types datasets, including Chandra + XMM-Newton, Chandra and XMM-Newton. }
\end{figure*}

The constraints of dipole parameters obtained from different types datasets (Chandra + XMM-Newton, Chandra, and XMM-Newton) are displayed in Fig. \ref{Figdy}. As expected, two best fits (two peaks) for parameters $l$, $b$, and $A$ are obtained. These two directions ($l$, $b$) are almost symmetrical in the galactic coordinate system. The directions corresponding to the negative $A$ values are the preferred directions for the cosmic anisotropy. For Fig. \ref{Figdy}, we need to emphasize that the two best-fit values of $A$ are very close to zero, making the result appear to have only one best-fit value, which is zero. This phenomenon is particularly prominent in the constraints of parameter $A$ \cite{2024ApJ...975L..36H}. By narrowing the prior range of the parameter $l$, we search over half a sphere to obtain a single direction ($l$, $b$) and their corresponding correction parameters ($A$ and $B$). This means that we perform two separate constraints, resulting in two sets of constraints, as shown in Fig. \ref{fig4}.

In order to understand the potential impact of fixing the $L_{X}-T$ relation on the detection of anisotropic signals, we conducted a dedicated analysis using the Chandra + XMM-Newton dataset as an example. We simultaneously fitted the $L_{X}-T$ relation parameters ($k$, $s$ and $\sigma_{int}$) with dipole fitting parameters ($l$, $b$, $A$ and $B$), and the fitting results are shown in Fig. \ref{figfree}. Figure \ref{figfree} reveals a clear coupling between parameters $k$ and $B$, making simultaneous constraint impossible. The optimal fitting results for the $L_{X}-T$ relation parameters $s$ and $\sigma_{int}$ are 2.05$\pm$0.06 and 0.25$\pm$0.01, respectively. The two preferred directions are 
\begin{eqnarray*} 
(l, b) = ({256.93^{\circ}}_{-60.06}^{+52.10}, \ \  -31.71{^{\circ}}_{-39.65}^{+34.50}) 
\end{eqnarray*} 
and 
\begin{eqnarray*} 
(l, b) = (76.87{^{\circ}}_{-61.95}^{+49.27}, \ \ 31.71{^{\circ}}_{-34.77}^{+40.07}),
\end{eqnarray*}  
respectively. The corresponding constraints on the correction parameter ($A$) are -5.4$\pm$6.4$\times10^{-4}$ and -5.4$\pm$6.8$\times10^{-4}$.

\begin{figure*}[htp]
    \centering 
    \includegraphics[width=0.39\textwidth]{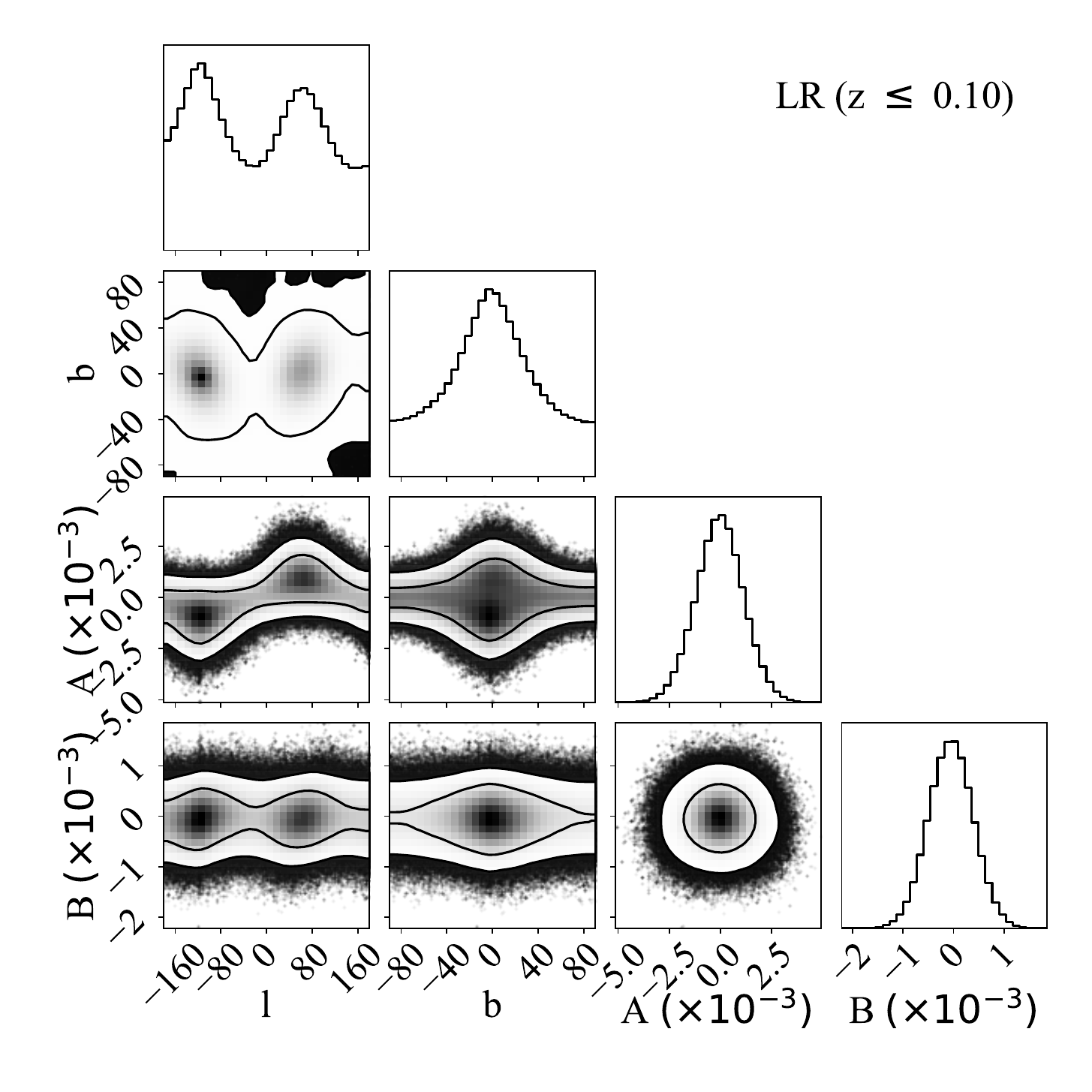} \ \
    \includegraphics[width=0.39\textwidth]{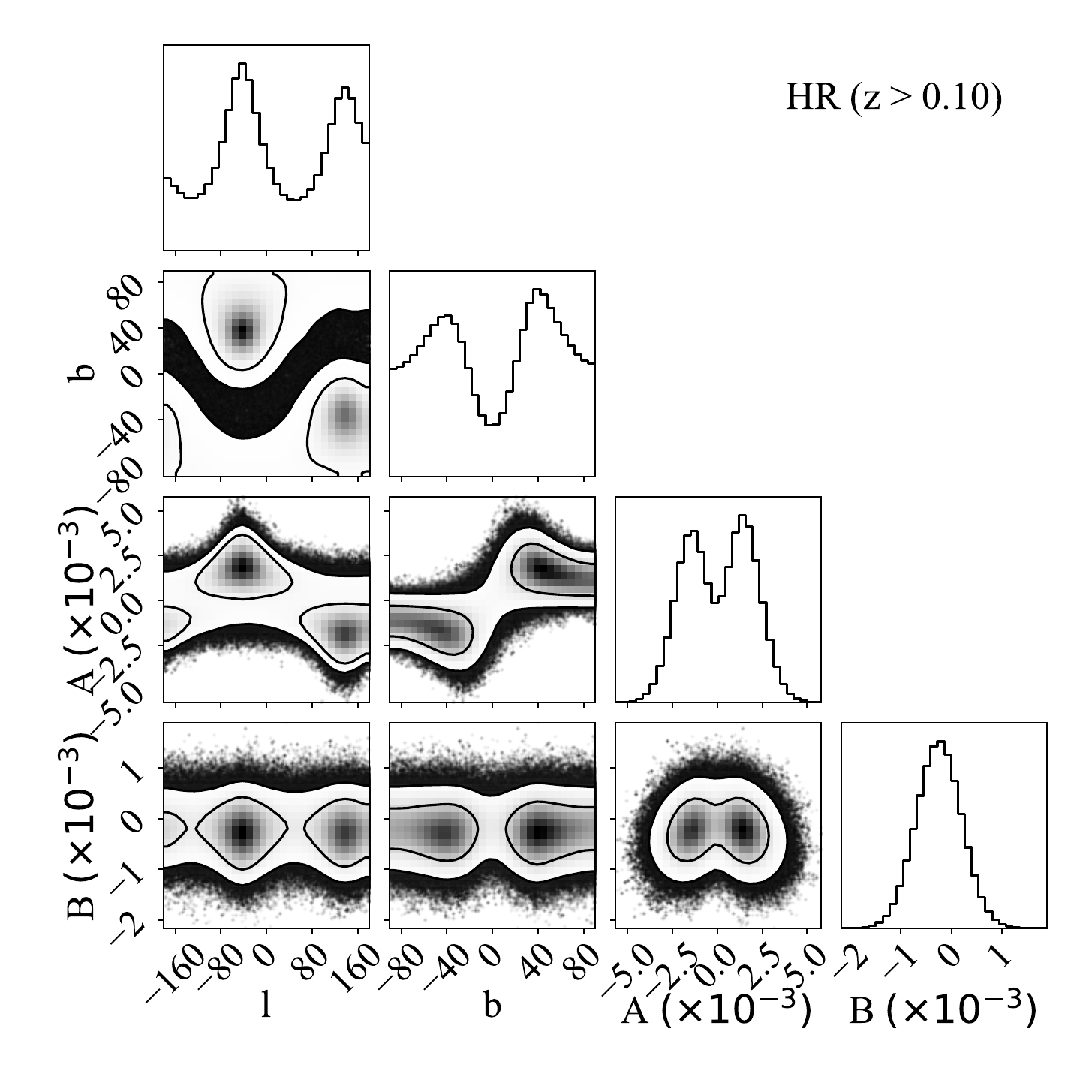}
    \caption{\label{Figdred} Confidence contours (1$\sigma$ and 2$\sigma$) of the dipole parameters (l, b, A, and B) for different redshift ranges (LR and HR). }
\end{figure*}

When the $L_{X}-T$ relation is fixed, the preferred directions ($l$, $b$) are 
\begin{eqnarray*} 
(l, b) = ({257.82^{\circ}}_{-52.88}^{+58.01}, \ \  -31.30{^{\circ}}_{-39.46}^{+35.92}) 
\end{eqnarray*} 
and 
\begin{eqnarray*} 
(l, b) = (80.89{^{\circ}}_{-52.46}^{+60.97}, \ \ 31.75{^{\circ}}_{-40.16}^{+35.19}),
\end{eqnarray*} 
respectively. The corresponding constraints of correction parameters are 
\begin{eqnarray*} 
A = -5.4\pm6.3\times10^{-4}, \ \ B = 0.5\pm3.2\times10^{-4},
\end{eqnarray*} 
and 
\begin{eqnarray*} 
A = 5.2_{-5.8}^{+6.5}\times10^{-4}, \ \ B = 0.4\pm3.2\times10^{-4}.
\end{eqnarray*} 
By comparing the two sets of constraint results, we can find a coupling between the parameter $k$ in the $L_{X}-T$ relation and the corrected parameter $B$; however, fixing the $L_{X}-T$ relation has a weak impact on the parameters ($l$, $b$, and $A$) associated with the anisotropic signal. Therefore, in subsequent research, we fixed the correlation parameters according to the constraints of $L_{X}-T$ relation in Section \ref{LTC} to search for cosmic anisotropic signals.

For the Chandra dataset, two sets of constraints are 
\begin{eqnarray*} 
&l = {328.31^{\circ}}_{-63.56}^{+66.52}, \ \ \ b = {-2.91^{\circ}}_{-52.27}^{+57.78}, \\
\ \ &A = -1.9_{-6.2}^{+6.8}\times10^{-4}, \ \ B = -0.6\pm3.7\times10^{-4},
\end{eqnarray*} 
and 
\begin{eqnarray*} 
&l = {148.32^{\circ}}_{-57.50}^{+60.16}, \ \  b = {4.62^{\circ}}_{-57.03}^{+51.70}, \\
\ \ &A = 2.2_{-6.8}^{+6.1}\times10^{-4}, \ \ B = -0.6\pm3.7\times10^{-4}, 
\end{eqnarray*}
respectively. For the XMM-Newton dataset, the corresponding results are 
\begin{eqnarray*} 
&l = {184.42^{\circ}}_{-24.58}^{+29.01}, \ \  b = {-17.73^{\circ}}_{-21.53}^{+18.55}, \\
\ \ &A = -32.0_{-18.0}^{+15.0}\times10^{-4}, \ \ B = -3.9\pm8.1\times10^{-4},
\end{eqnarray*} 
and 
\begin{eqnarray*}
&l = {5.0^{\circ}}_{-25.33}^{+28.81}, \ \  b = {17.61^{\circ}}_{-18.68}^{+21.56},  \\
\ \ &A = 32.0_{-15.0}^{+18.0}\times10^{-4}, \ \ B = -3.9\pm8.0\times10^{-4}. 
\end{eqnarray*} 
The numerical results are placed in Table \ref{T2}. In addition to considering the effects of datasets on the results, we also investigated the impact of redshift on the anisotropic signal, and the results are described in the next section.

\subsection{Anisotropic signal from different redshift subsamples}
In order to fully discuss the factors that might have influenced the anisotropic signal, we divide the total sample into two subsamples based on the redshift value of 0.10 to investigate the effect of redshift on the results. According to the constraints of the $L_{X}-T$ correlation in Section \ref{LTC}, we fixed the parameters $k$, $s$ and $\sigma_{int}$, and then to find the preferred direction ($l$, $b$) as well as the dipole parameters $A$ and $B$. The final constraints are shown in Fig. \ref{Figdred}. By searching over a half of sphere, we obtain the constraints of each direction and its corresponding correction parameters. The corresponding results are shown in Fig. \ref{fig5}.

For the LR dataset, the corresponding results are as follow: \newline
\begin{eqnarray*} 
&l = {247.66^{\circ}}_{-49.20}^{+58.30}, \ \ b = {-1.93^{\circ}}_{-43.45}^{+45.43}, \\ \nonumber
\ \ &A = -6.5_{-8.0}^{+9.7}\times10^{-4},  \ \ B = -0.9\pm4.2\times10^{-4},
\end{eqnarray*} 
and 
\begin{eqnarray*} 
&l = {73.01^{\circ}}_{-46.50}^{+62.39}, \ \ b = {2.21^{\circ}}_{-44.55}^{+42.77}, \\ \nonumber
\ \ &A = 6.6_{-9.6}^{+8.3}\times10^{-4},  \ \ B = -0.8\pm4.2\times10^{-4}.
\end{eqnarray*} 
For the HR dataset, the two sets of constraints are
\begin{eqnarray*} 
&l = {139.84^{\circ}}_{-35.47}^{+39.62}, \ \ b = {-43.07^{\circ}}_{-25.53}^{+21.52}, \\ \nonumber
\ \ &A = -16.2_{-8.1}^{+9.0}\times10^{-4},  \ \ B = -2.6\pm4.2\times10^{-4},
\end{eqnarray*}
and 
\begin{eqnarray*} 
&l = {320.57^{\circ}}_{-35.15}^{+40.04}, \ \ b = {43.90^{\circ}}_{-21.64}^{+25.52}, \\ \nonumber
\ \ &A = 16.3\pm8.8\times10^{-4},  \ \ B = -2.5\pm4.3\times10^{-4}.
\end{eqnarray*}
All numerical results are summarized in Table \ref{T2}.

\subsection{Statistical isotropic results} \label{stat}
According to two schemes (Bootstrap and Randomized), we got two sets of simulation results $A_{B}$ and $A_{R}$. Utilizing 1000 simulated datasets, we performed statistical isotropic analyses for the different datasets. For any combination of schemes and datasets, the results ($A_{B}$ and $A_{R}$) from 1000 simulated datasets are all distributed in the range of ($-10^{-3}$, $10^{-3}$). 

\begin{figure*}[ht]
    \centering 
    \includegraphics[width=0.26\textwidth]{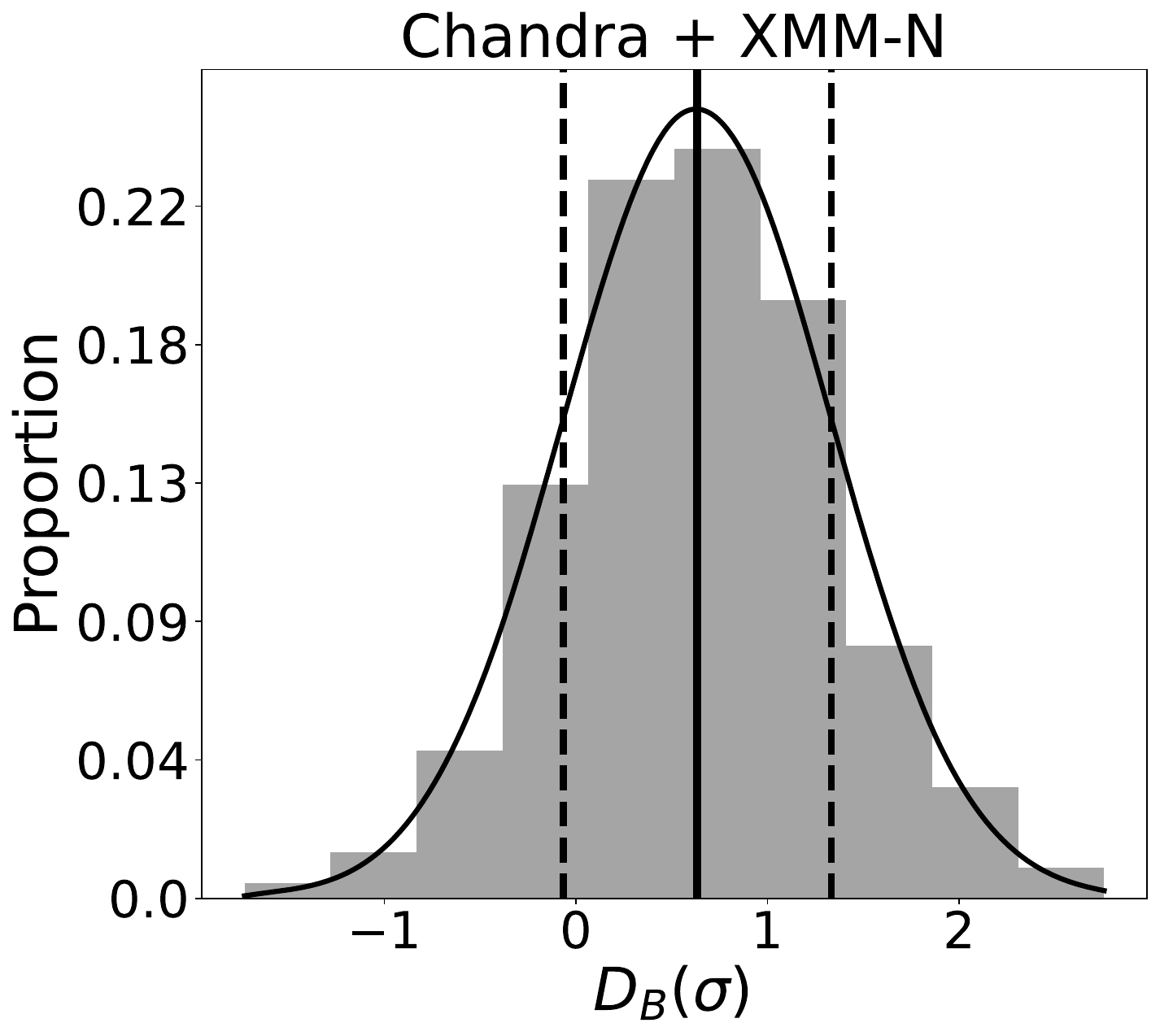} 
    \includegraphics[width=0.26\textwidth]{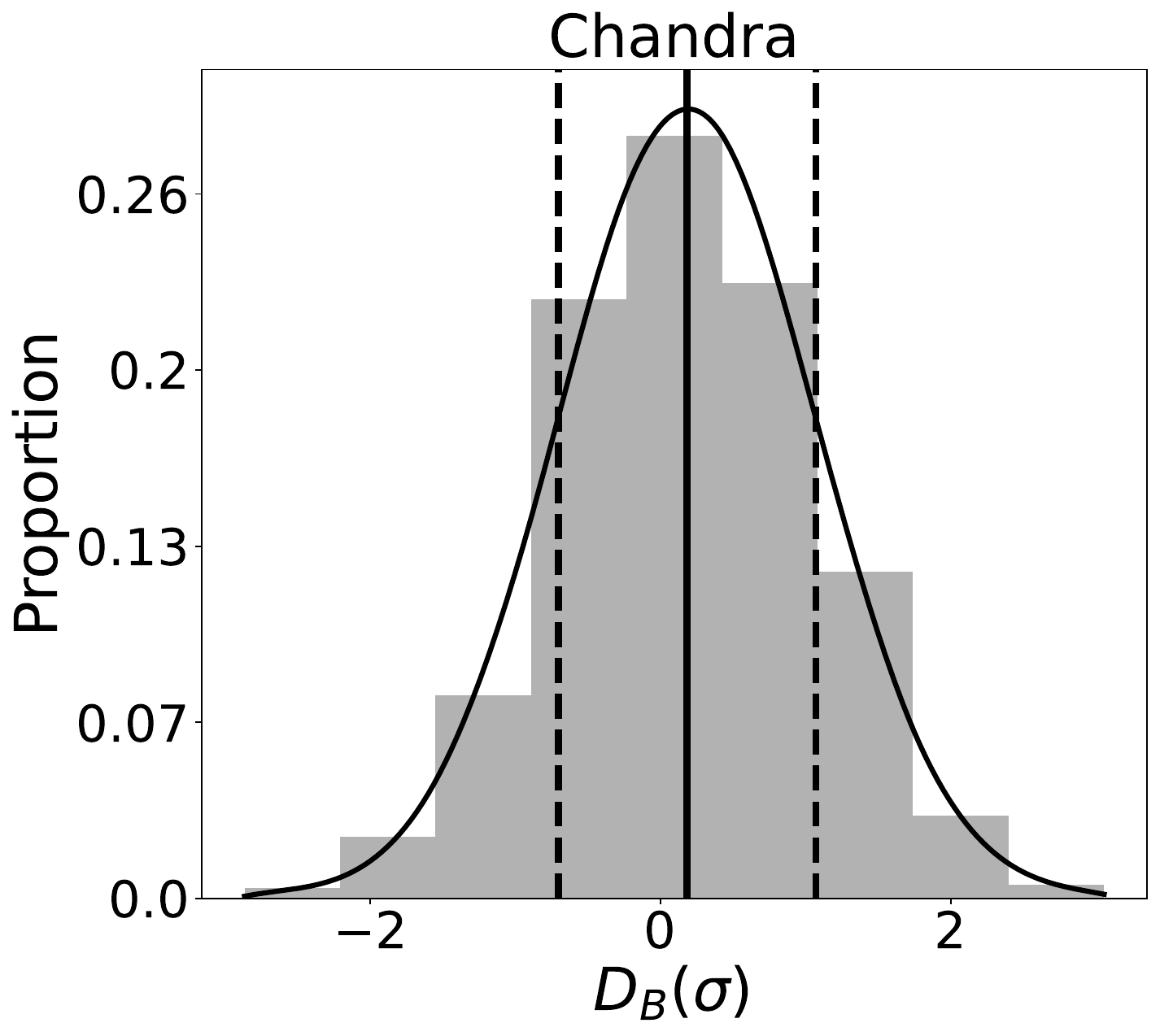} 
    \includegraphics[width=0.26\textwidth]{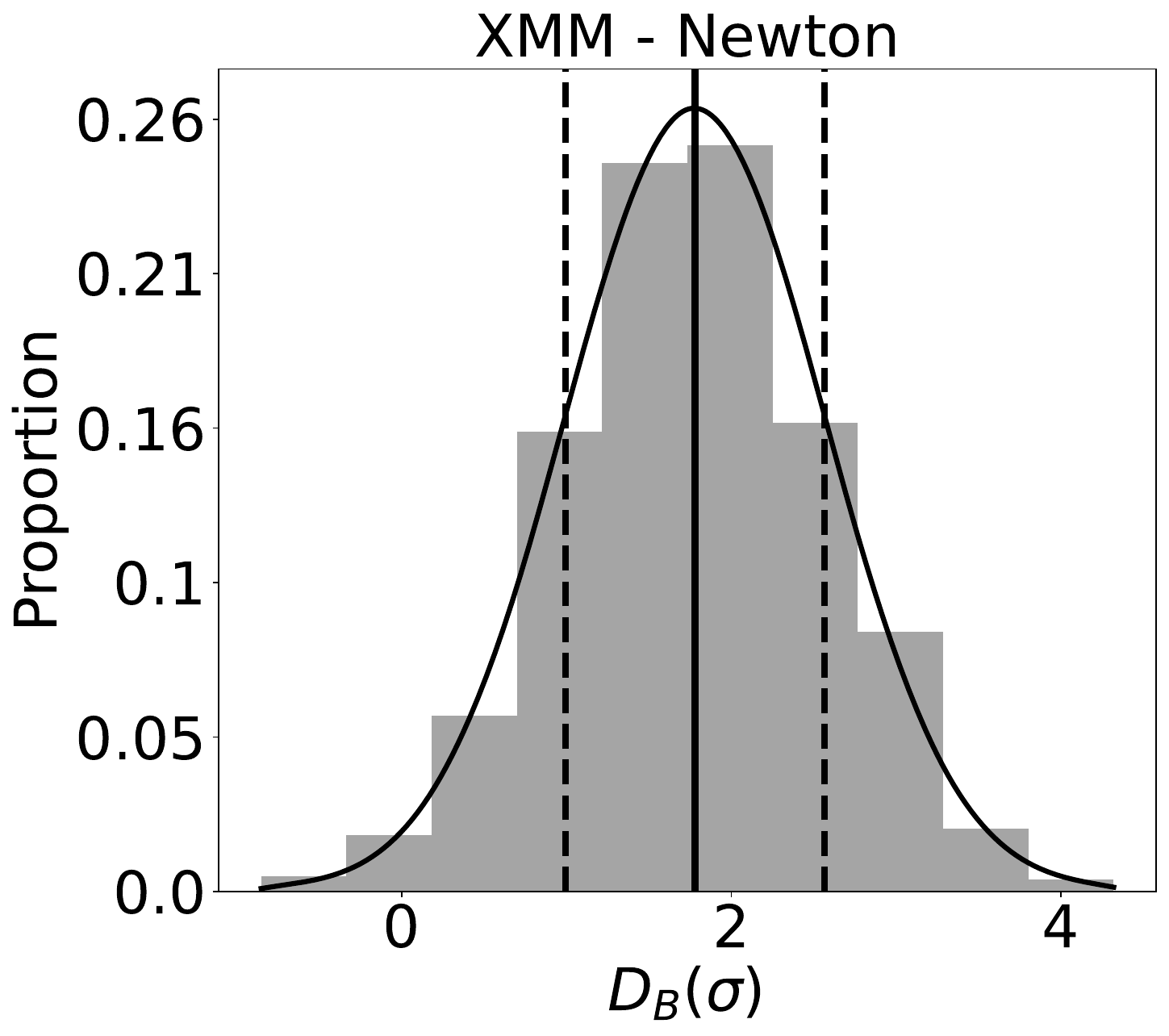}\\
    \includegraphics[width=0.26\textwidth]{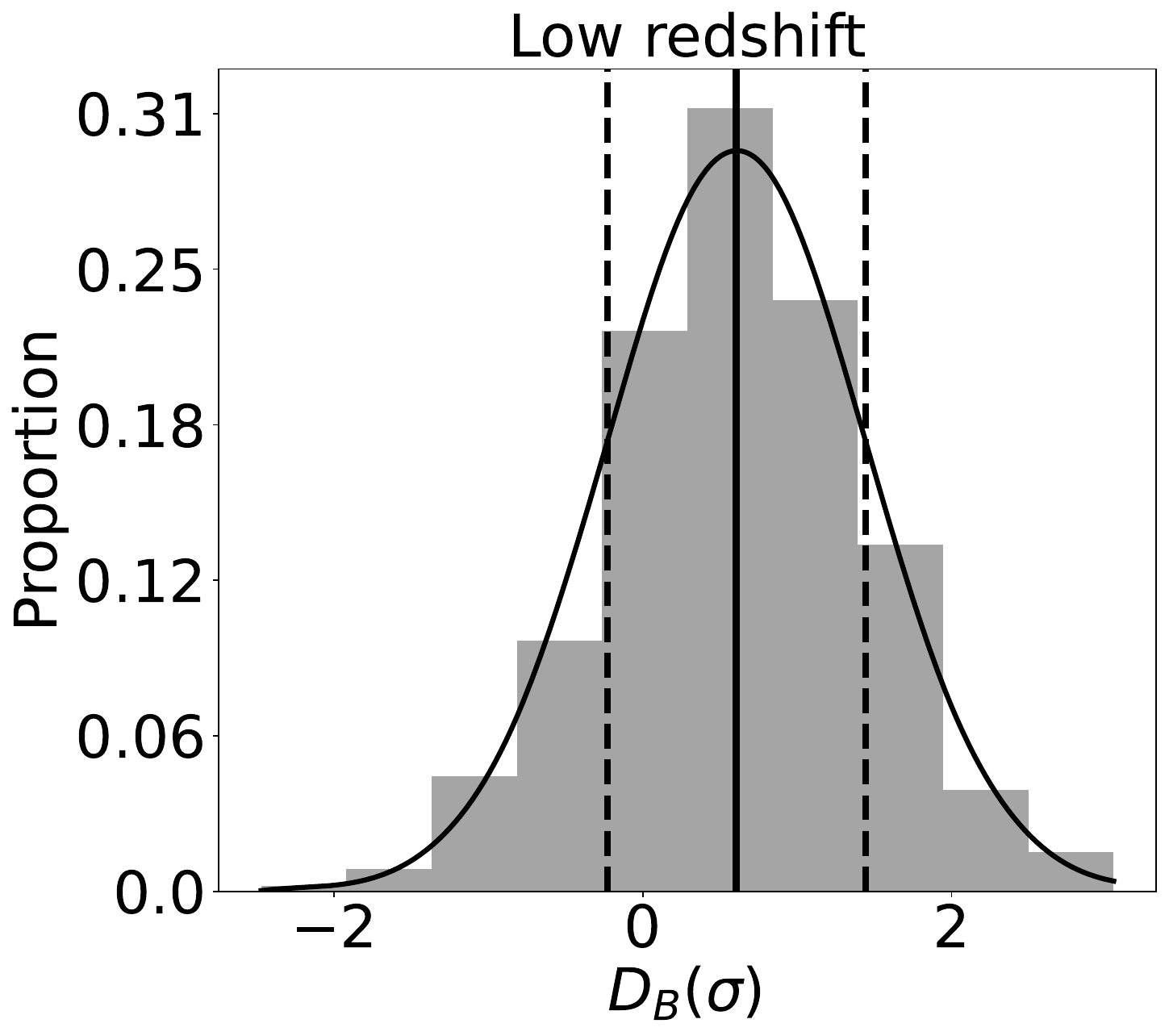}
    \includegraphics[width=0.26\textwidth]{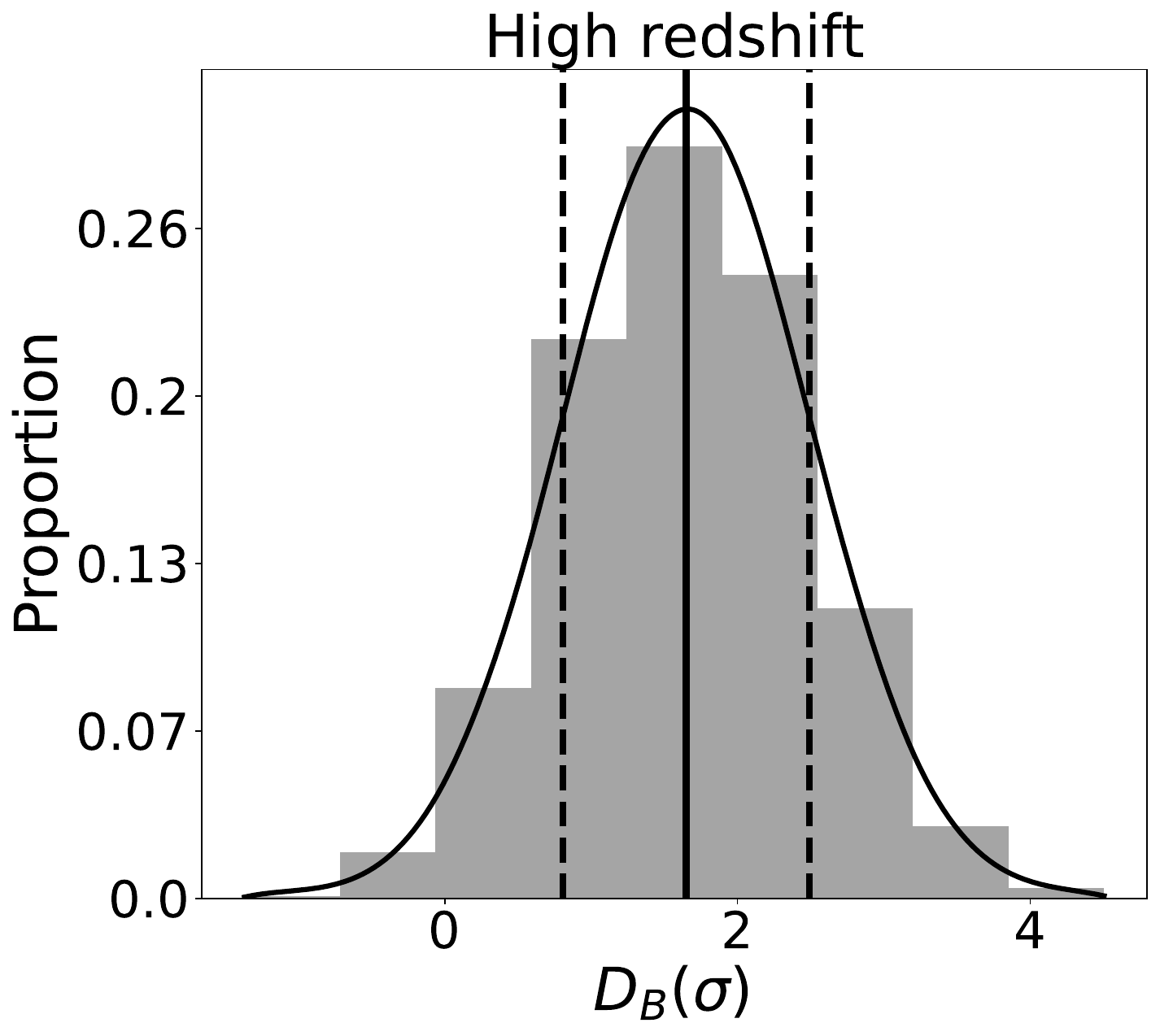}
    \caption{\label{Figmock} Statistical results in Bootstrap scheme. The black curve is the best fit to the Gaussian function. The solid black and vertical dashed lines are commensurate with the mean and the standard deviation, respectively. The statistical significance ($D$) of the Chandra, XMM-Newton, Chandra + XMM-Newton, LR and HR datasets are 0.2$\sigma$, 1.8$\sigma$, 0.6$\sigma$, 0.6$\sigma$ and 1.7$\sigma$, respectively.} 
\end{figure*}

\begin{figure*}[ht]
    \centering 
    \includegraphics[width=0.26\textwidth]{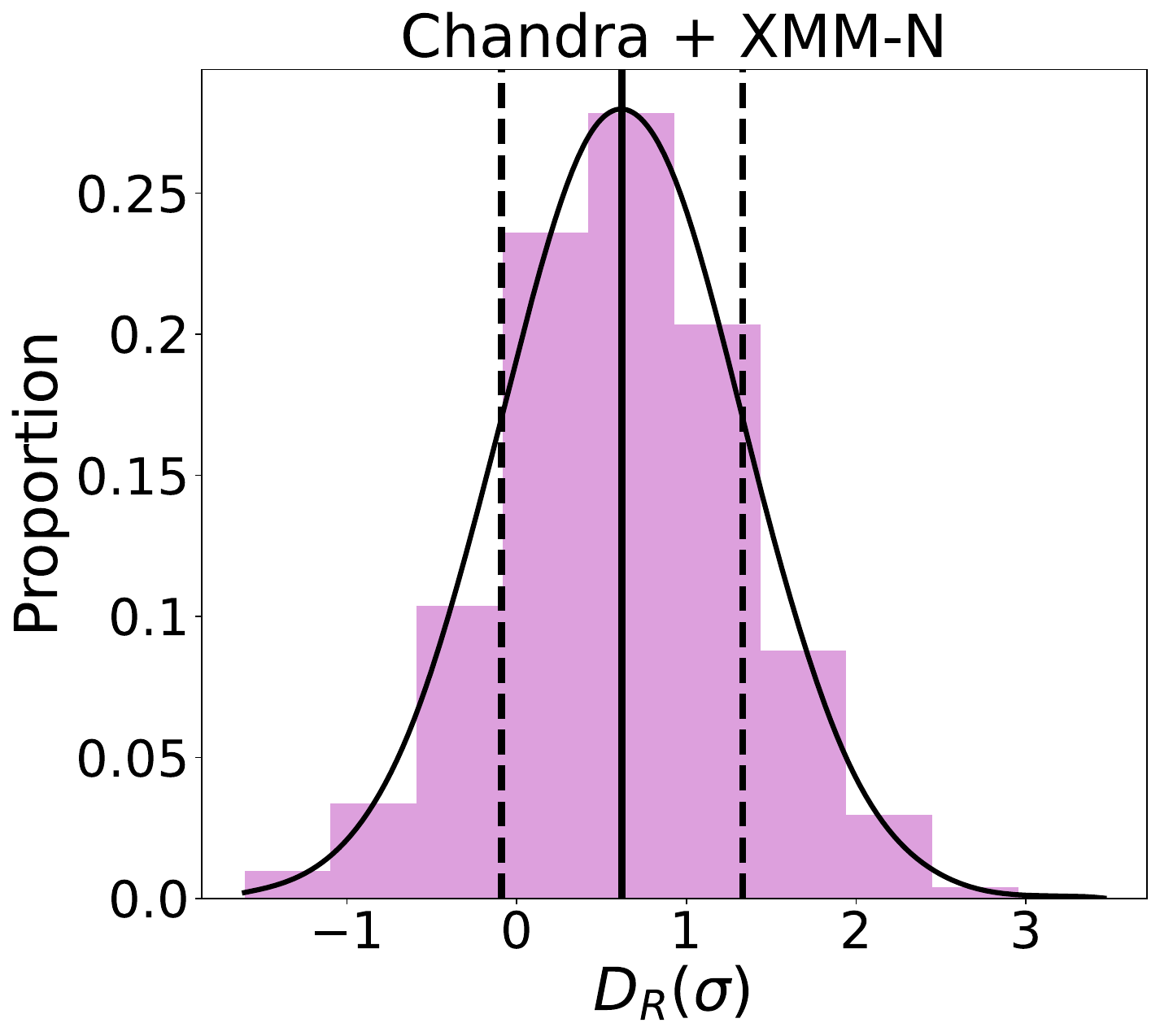} 
    \includegraphics[width=0.26\textwidth]{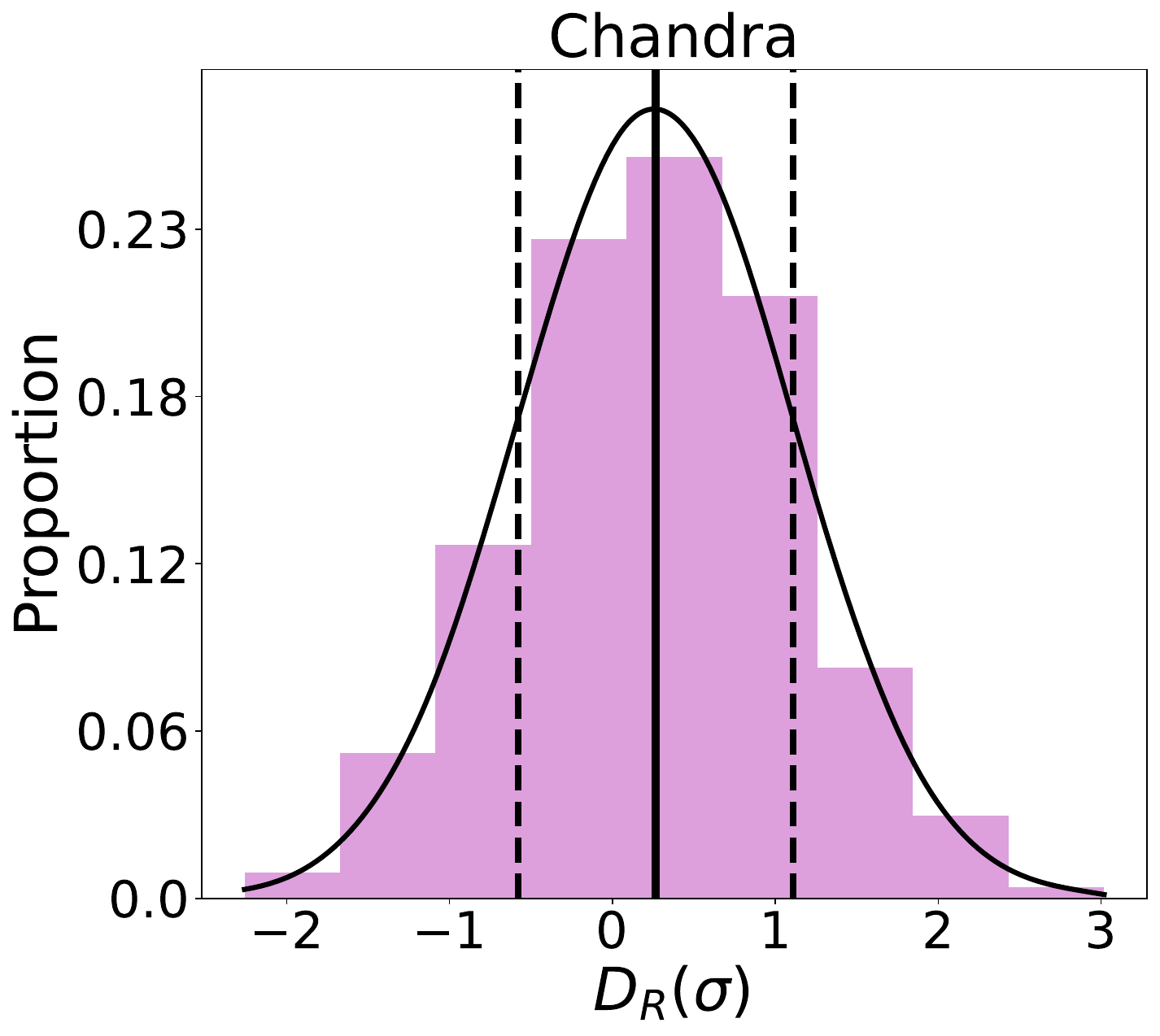} 
    \includegraphics[width=0.26\textwidth]{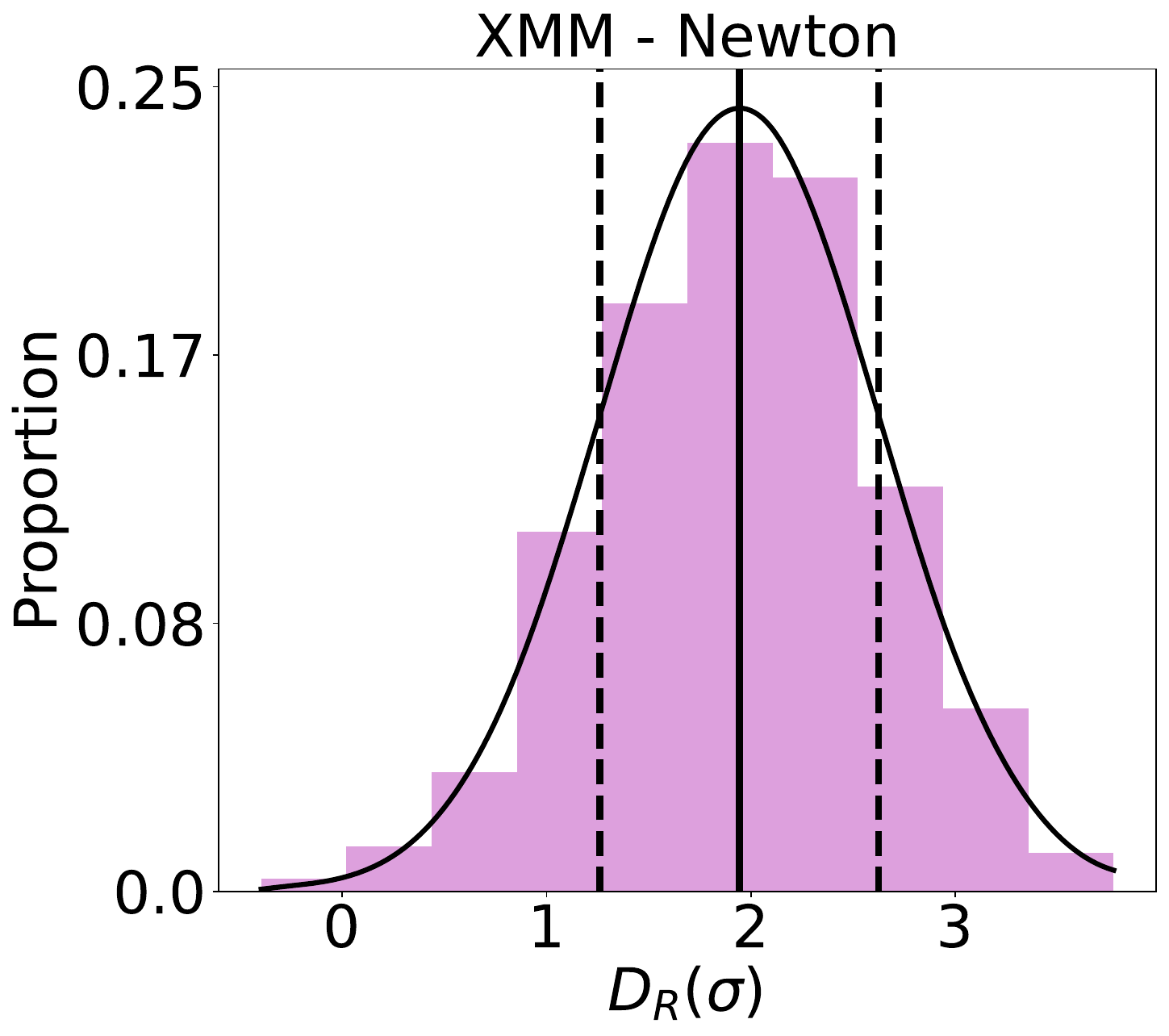} \\
    \includegraphics[width=0.26\textwidth]{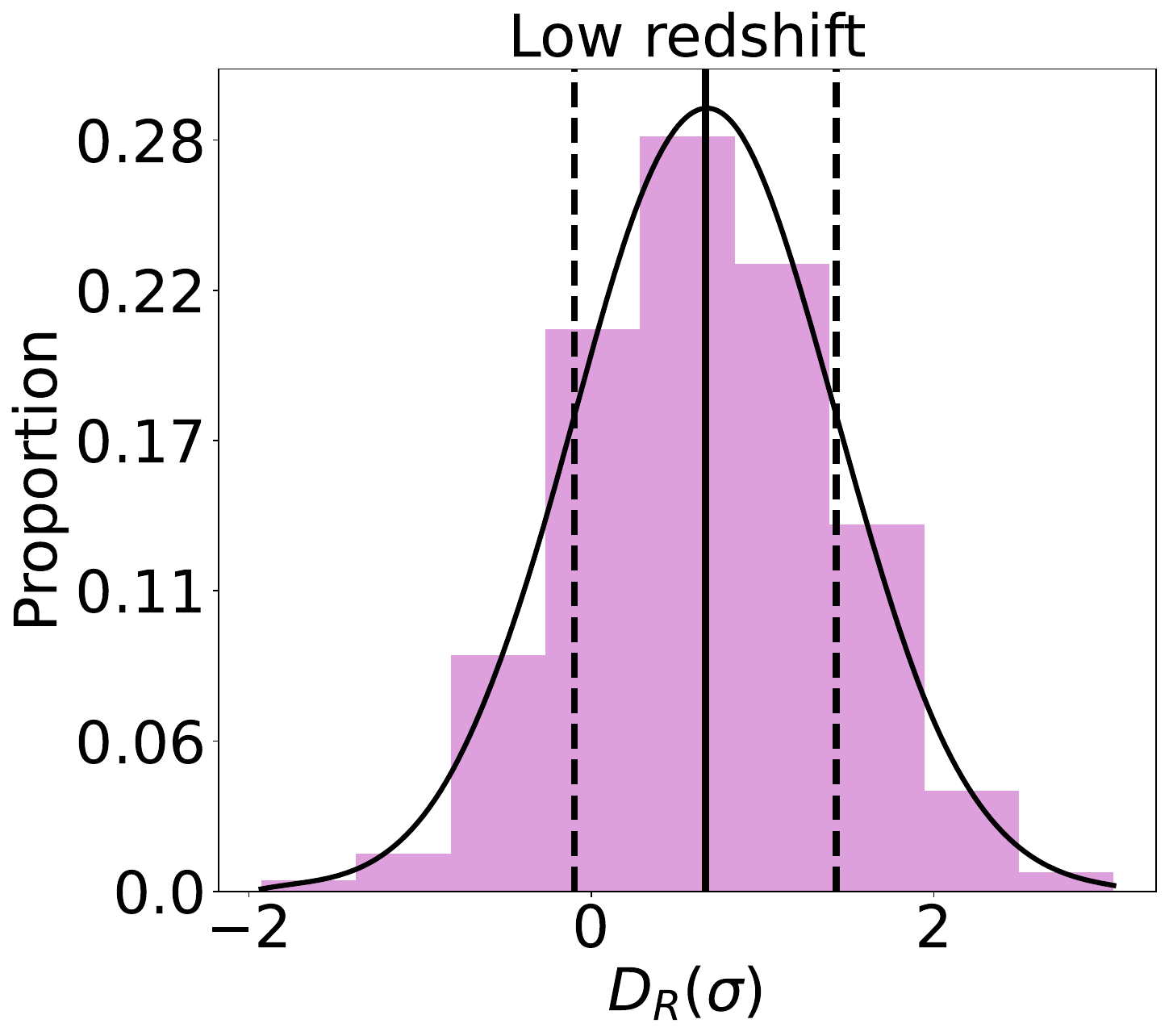}
    \includegraphics[width=0.26\textwidth]{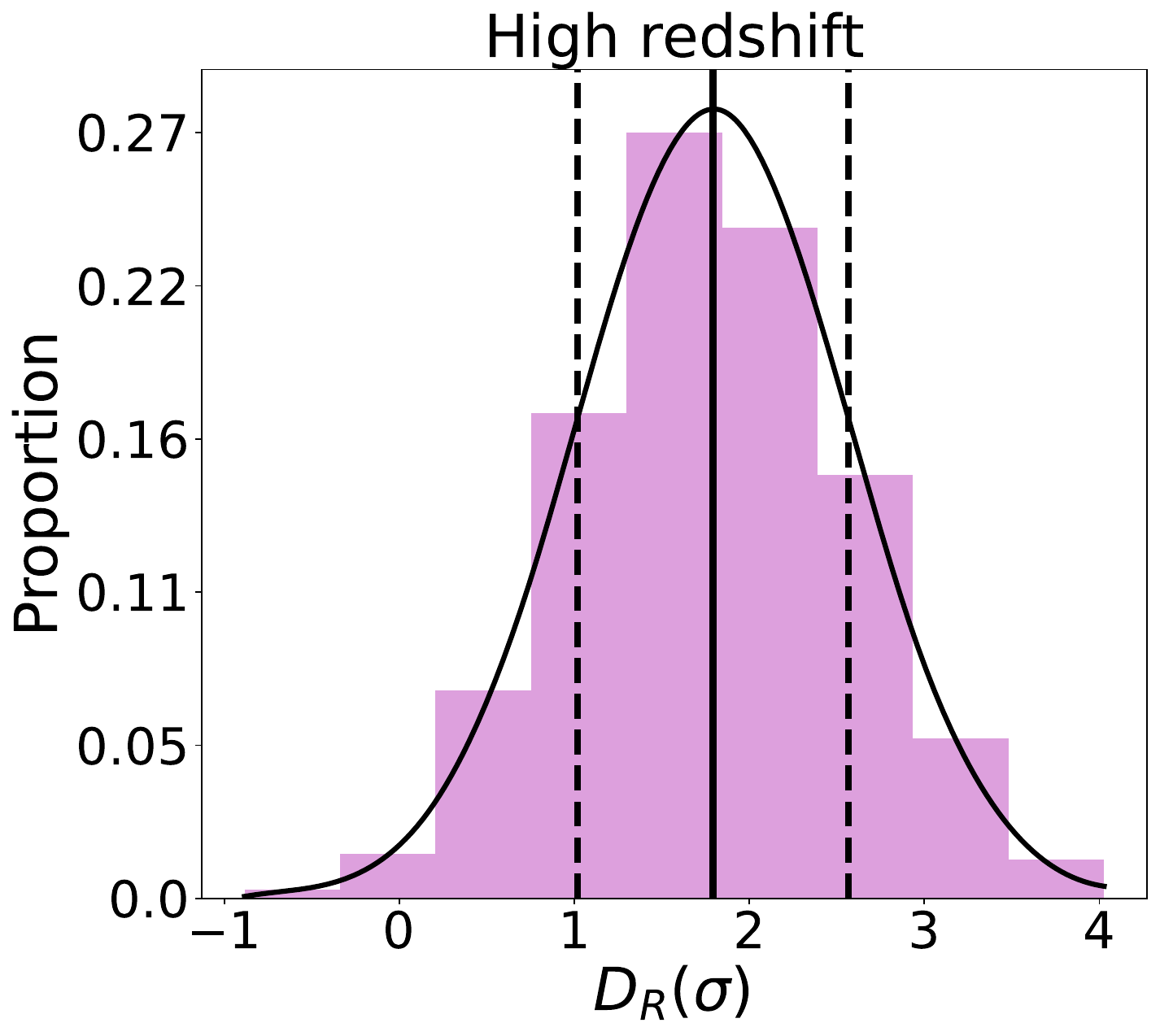}
    \caption{\label{Figiso} Statistical results in the Randomized scheme. The black curve is the best fit to the Gaussian function. The solid black and vertical dashed lines are commensurate with the mean and the standard deviation, respectively. The statistical significance ($D$) of the Chandra, XMM-Newton, Chandra + XMM-Newton, LR and HR datasets are 0.3$\sigma$, 1.9$\sigma$, 0.6$\sigma$, 0.7$\sigma$ and 1.8$\sigma$, respectively.}
\end{figure*}

The simulation results ($A_{B}$ and $A_{R}$) can be described by a Gaussian function with the expectation $\mu_{A}$ and standard deviation $\sigma_{A}$. According to these statistical results, we can derive the statistical significances of the real data. For the Bootstrap scheme, the statistical significance ($D$) of the Chandra, XMM-Newton, and Chandra + XMM-Newton datasets are 0.2$\sigma$, 1.8$\sigma$, and 0.6$\sigma$, respectively. For the LR and HR datasets, the corresponding statistical significances are 0.6$\sigma$ and 1.7$\sigma$, respectively. The statistical results of the Bootstrap scheme are shown in Fig. \ref{Figmock}. For the Randomized scheme, the statistical significance of the Chandra, XMM-Newton, and Chandra + XMM-Newton datasets are 0.3$\sigma$, 1.9$\sigma$, and 0.6$\sigma$, respectively. The statistical significance of the LR and HR datasets are 0.7$\sigma$ and 1.8$\sigma$, respectively. The statistical results of the Randomized scheme are shown in Fig. \ref{Figiso}. All eigenvalues (expectation $\mu_{A}$ and statistical significance $D$) for these two schemes are collected in Table \ref{T3}. 

\begin{figure*}[ht]
    \centering
    \includegraphics[width=0.48\textwidth]{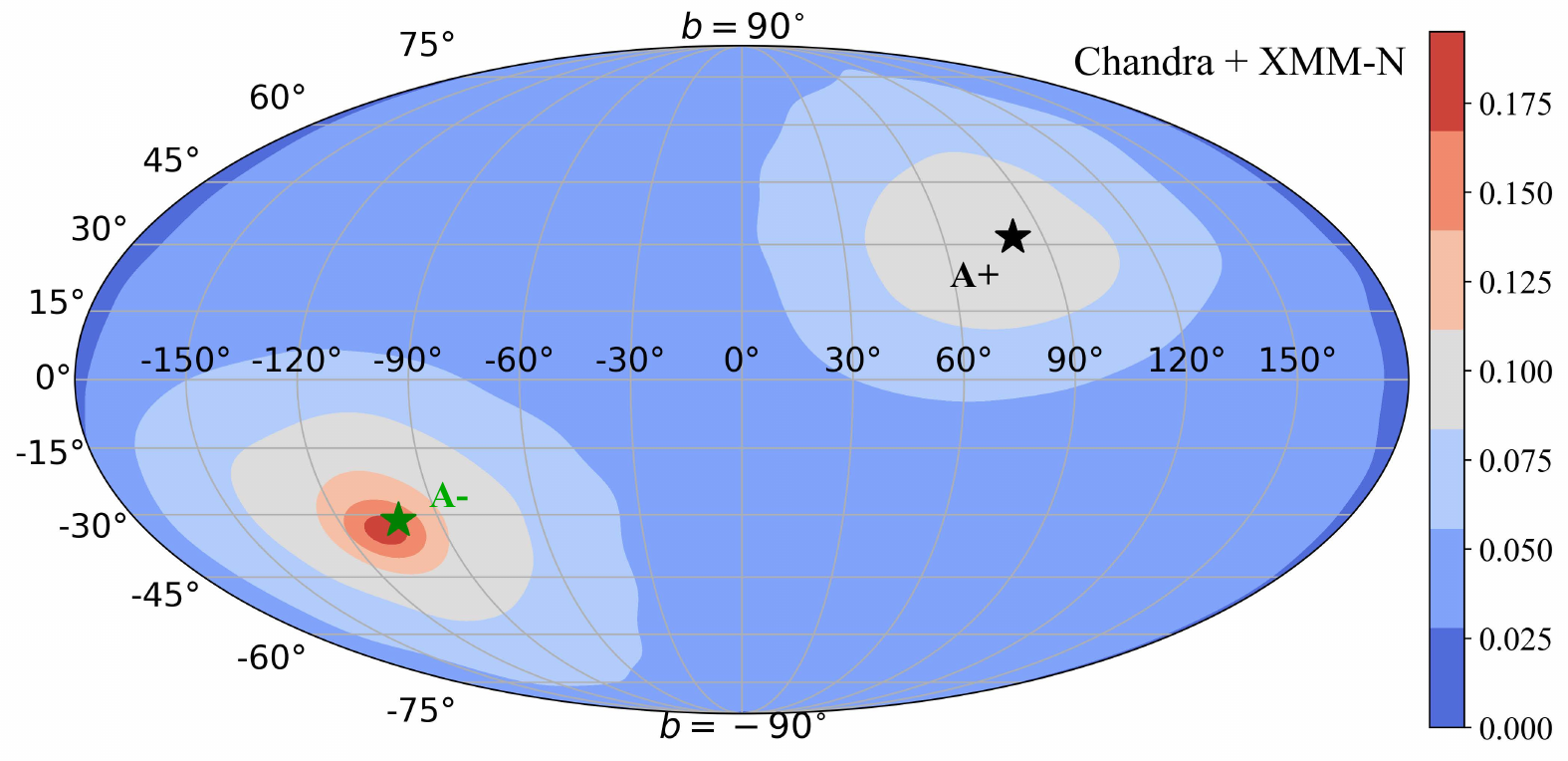} \\
    \includegraphics[width=0.48\textwidth]{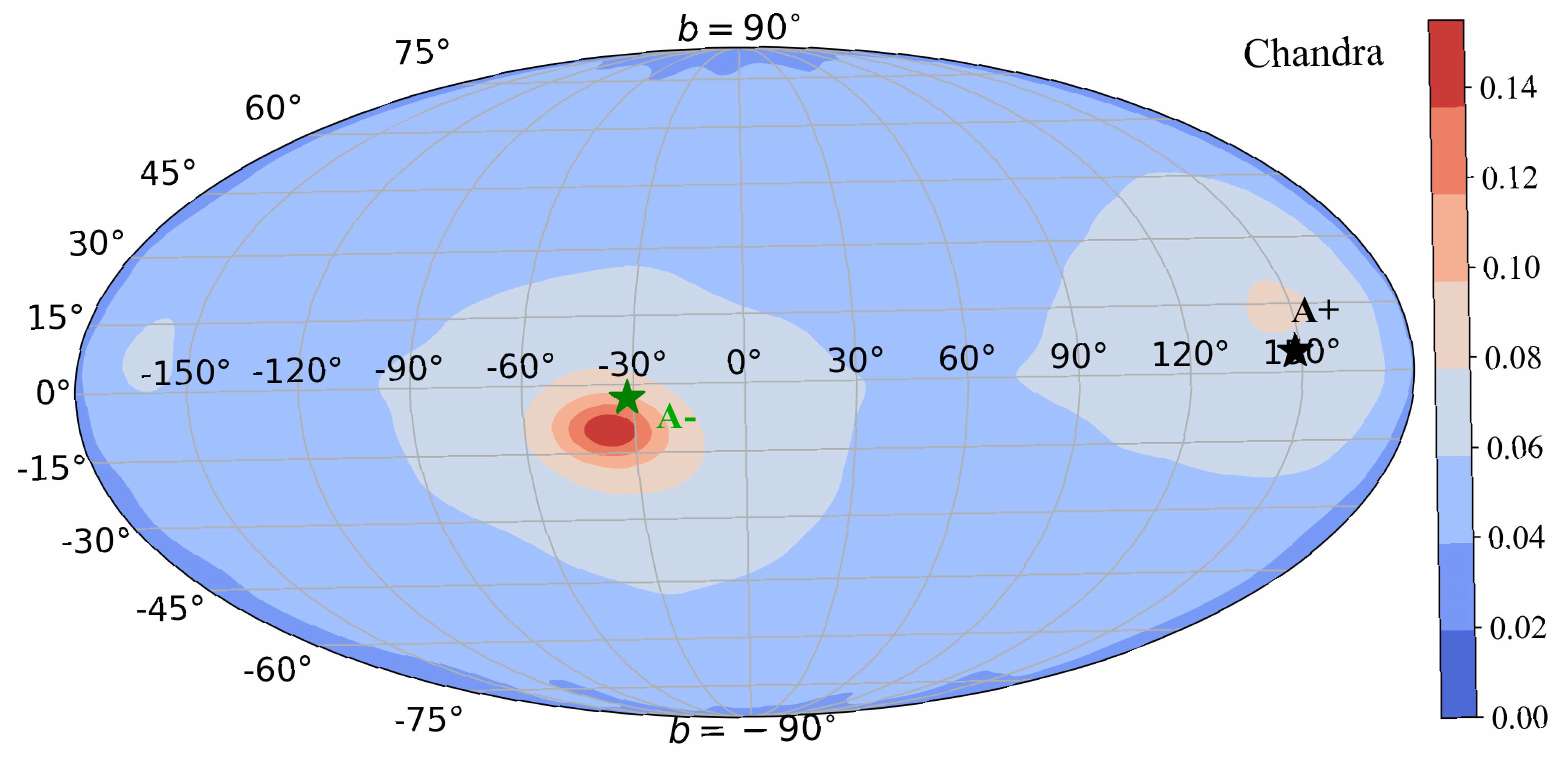}
    \includegraphics[width=0.48\textwidth]{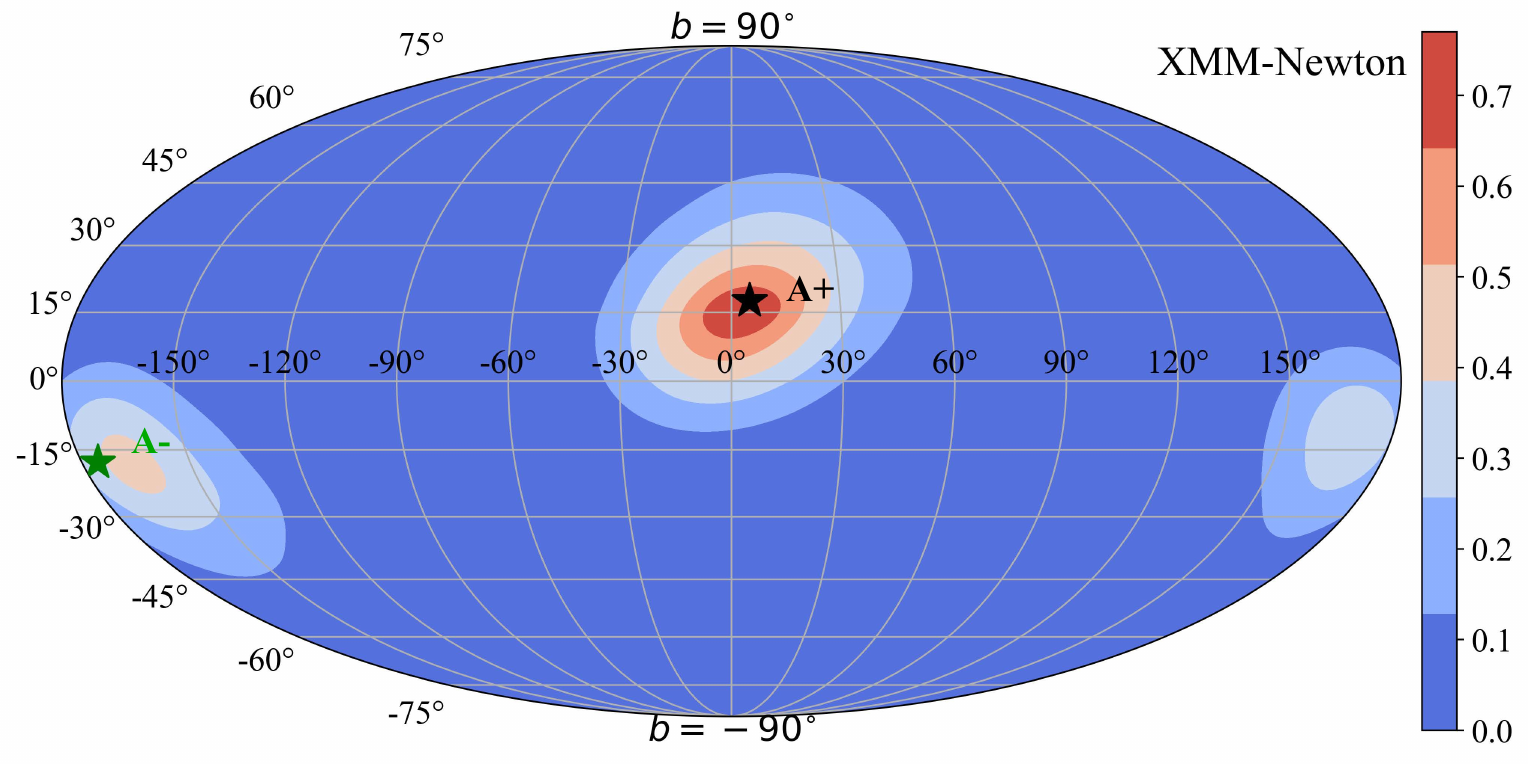} 
    \includegraphics[width=0.48\textwidth]{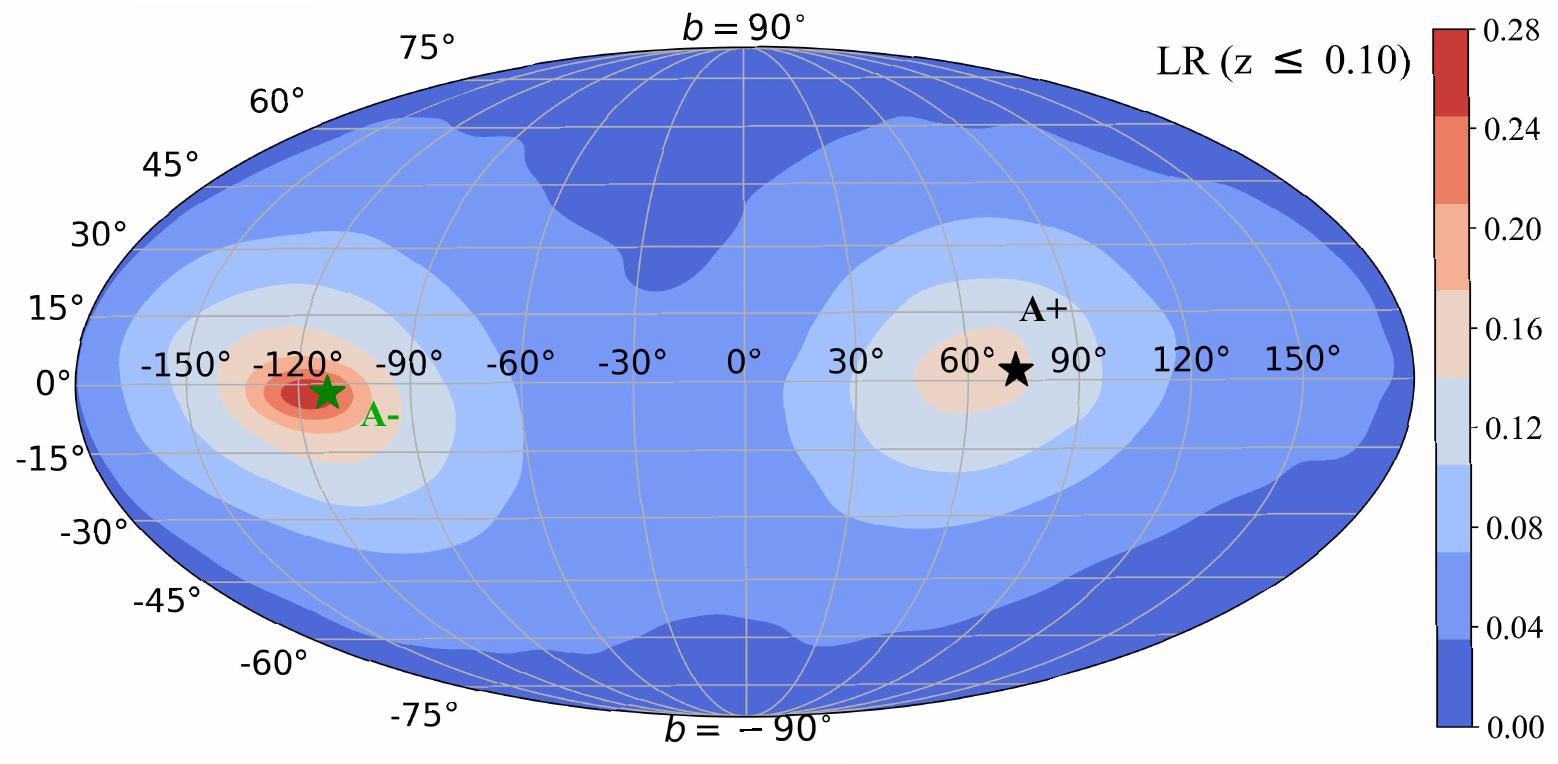}
    \includegraphics[width=0.48\textwidth]{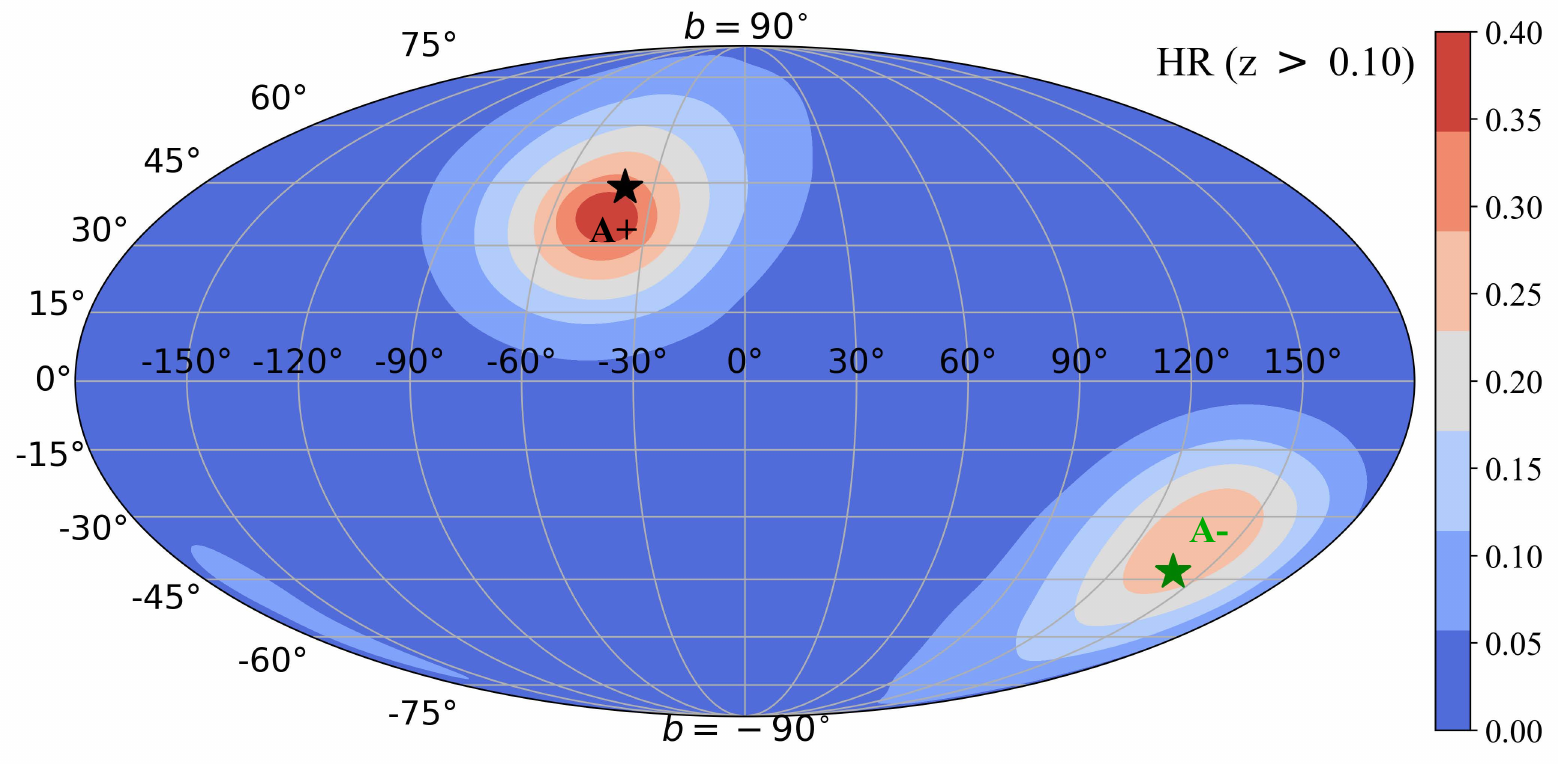}
    \caption{\label{Figdfd1} Direction ($l$, $b$) in the galactic coordinate system. The contours show the MCMC sampling results come from sampling based on the entire space of parameters $l$ ($-180^{\circ} \leq l \leq 180^{\circ}$). The stars represent the directions obtained by the hemispherical search. Green star represent the preferred direction of the cosmic anisotropy. } 
\end{figure*}

\section{Discussion} \label{dis}
Before exploring the preferred direction of the cosmic anisotropy using galaxy clusters, it is necessary to constrain the $L_{X}-T$ correlation of different datasets. Through comparative investigations of different datasets, we discovered some interesting phenomena. There is a significant difference between the  constraints of $L_{X}-T$ correlation from the Chandra and XMM-Newton datasets. The difference is reflected in the parameter $s$, which reaches 4.01$\sigma$. The discrepancies might originate from the observation itself or from systematic errors related to the observation equipment. Analysis of the redshift subsamples (LR $z \leq 0.10$ and HR $z > 0.10$) reveals that the parameter $k$ for the LR dataset aligns more closely with the XMM-Newton result. This is unexpected, as the Chandra dataset accounts for 75\% of the LR dataset; one would therefore anticipate the LR constraints to be more consistent with those from Chandra. In addition, the differences between different redshifted subsamples (LR and HR) are also quite significant. The discrepancy in parameter $k$ is 0.88 (5.18$\sigma$), whereas the difference in $s$ is not statistically significant at 0.35 (2.29$\sigma$). This hints that the $L_{X}-T$ correlation might evolve with redshift.

Based on the previous constraints of the $L_{X}-T$ correlation, we fixed the $L_{X}-T$ relationship parameters ($k$, $s$ and $\sigma_{int}$), and then constrained the dipole parameters ($l$, $b$, $A$ and $B$). By comparing negative values of $A$ and their corresponding preferred directions ($l$, $b$), we find significant differences between the degree of anisotropy and the preferred directions for the Chandra and XMM-Newton datasets. The XMM-Newton datasets gives stronger anisotropy signal (larger value of $|A|$) and better constrains the preferred directions. The main difference in the preferred directions is in longitude, which is 143.89$^{\circ}$. Such stark discrepancies may be due to variations in the redshift range and spatial distribution of the datasets, as we mentioned in Section \ref{data}. Compared to the XMM-Newton dataset, the Chandra dataset covers a wider redshift range and has a more uniform spatial distribution. Meanwhile, the differences in results across different redshift ranges (LR and HR) also support the view that redshift significantly affects the intensity and direction of anisotropy. However, we could not rule out the influence of differences in observation equipment or unknown factors introduced during data processing. At present, it is impossible to determine the source of the discrepancies.

In order to more intuitively understand these preferred directions in different datasets, we mapped the MCMC sampling of preferred direction ($l$, $b$) in the galactic coordinate system utilizing the $plt.contour()$ function\footnote{https://matplotlib.org/stable/gallery/images$_{-}$contours$_{-}$and$_{-}$fields/ir regulardatagrid.html}, as shown in Fig. \ref{Figdfd1}. Color-coded values represent sample fraction per unit area. Larger values indicate tighter results. From Fig. \ref{Figdfd1}, it is easier to spot differences in preferred directions obtained from different datasets. We have also marked the directions obtained from the hemispherical search in Fig. \ref{Figdfd1}. It can be found that the changes in search methods causes a slight shift (within 1$\sigma$ range) in the direction. It should be noted here that the 1$\sigma$ error is not reflected in Fig. \ref{Figdfd1}.

To clarify the effect of inhomogeneity on the anisotropic signal and examine whether dipole correction parameter $A$ are consistent with statistical isotropy, we also performed statistical isotropic analyses. The statistical results ($A$, $\mu_{A}$ and $D$) are described in detail in Section \ref{stat} and collected in Table \ref{T3}. For the Bootstrap scheme, the statistical results ($A_{B}$ and $\mu_{A_{B}}$) indicate that the level of anisotropy contributed by the non-uniform distribution of galaxy clusters is relatively weak. This indicates that the level of anisotropy reflected in the real data is largely contributed by the non-uniformity of the cosmic structure. The Randomized scheme, which places the real data evenly on the celestial sphere, yields statistical results similar to the Bootstrap scheme. This indicates that other factors contribute to anisotropy in galaxy clusters, and that their contribution to anisotropy is comparable to the uniformity of the spatial distribution of the data. By using the real results $A$, the simulation results ($A_{B}$ and $A_{R}$), and the statistical results ($\mu_{A_{B}}$ and $\mu_{A_{R}}$), we can obtain the corresponding statistical significance $D$. The maximum statistical significance obtained by the two schemes are 1.8$\sigma$ (Bootstrap) and 1.9$\sigma$ (Randomized), respectively, both derived from the XMM-Newton dataset. If we could use other studies to precisely restrict the preferred direction of anisotropy, could we improve the significance of anisotropy? To verify this idea, we first fixed the preferred direction as the best fit and then re-constrained the correction parameters $A$ and $B$ independently. We then confirmed that fixing the preferred direction can improve the level of anisotropy in the detection. The corresponding results ($A^{*}$ and $B^{*}$) are collected in Table \ref{T2}. 

\subsection{Effect of sample size and redshift range on the anisotropic signal}
From the analysis and discussion of the above results, it can be found that substantial discrepancies between the Chandra and XMM-Newton results. These discrepancies may stem from factors such as differences in redshift coverage, differences in sky distribution, instrumental calibration effects, and sample-selection effects. Compared to the XMM-Newton sample, the Chandra sample contains a larger number of galaxy clusters and covers a wider redshift range. Consequently, the observed differences may simply reflect variations in sample composition rather than stemming from the instruments themselves or cosmological effects. To investigate this issue, we conducted a controlled matched-sample analysis. Importance sampling is employed to generate multiple Chandra subsamples that match the XMM-Newton sample in terms of both sample size and redshift distribution. Subsequently, the $L_{X}-T$ relation of the Chandra subsamples is constrained to perform the dipole-fitting analysis. This procedure could largely remove the effects of sample size and redshift range, allowing a much cleaner comparison between the two instruments. 

\begin{figure}[htp]
    \centering
    \includegraphics[width=0.45\textwidth]{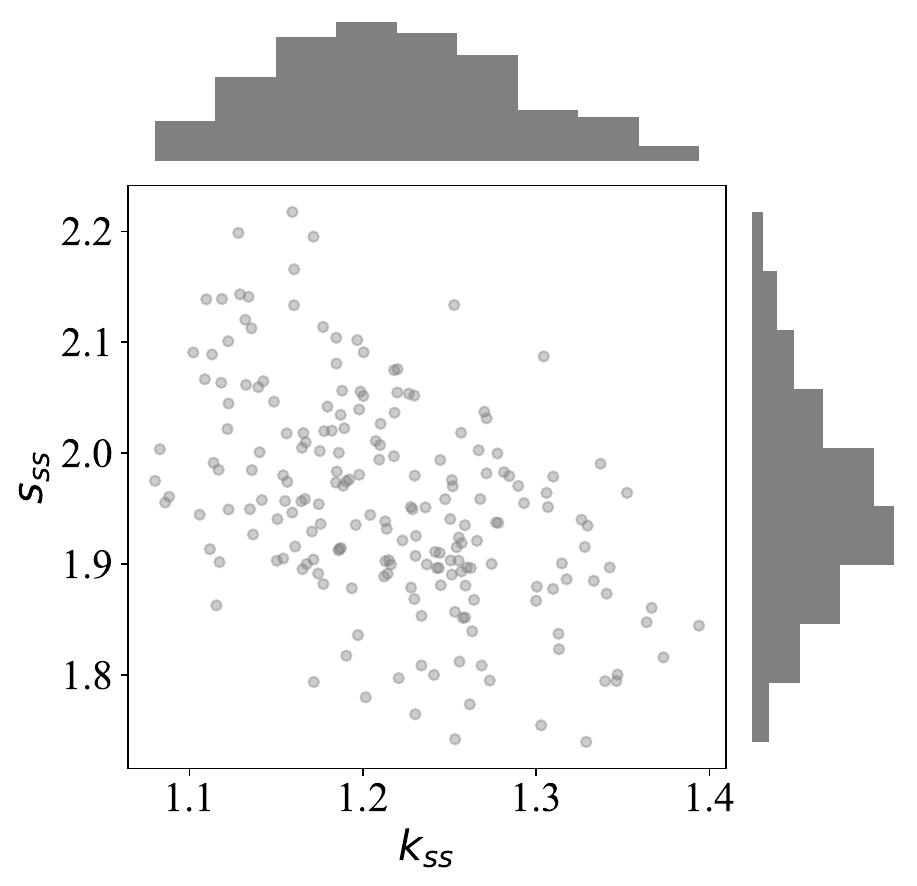}
    \caption{\label{Figks} Distribution of $L_{X}-T$ relation parameters ($k$ and $s$) from 200 Chandra subsamples. The distributions of both $k_{ss}$ and $s_{ss}$ exhibit a Gaussian profile, yielding best-fit values of $k_{ss} = 1.22 \pm 0.07$ and $s_{ss} = 1.95 \pm 0.09$. }
\end{figure}

\begin{figure*}[htp]
    \centering
    \includegraphics[width=0.29\textwidth]{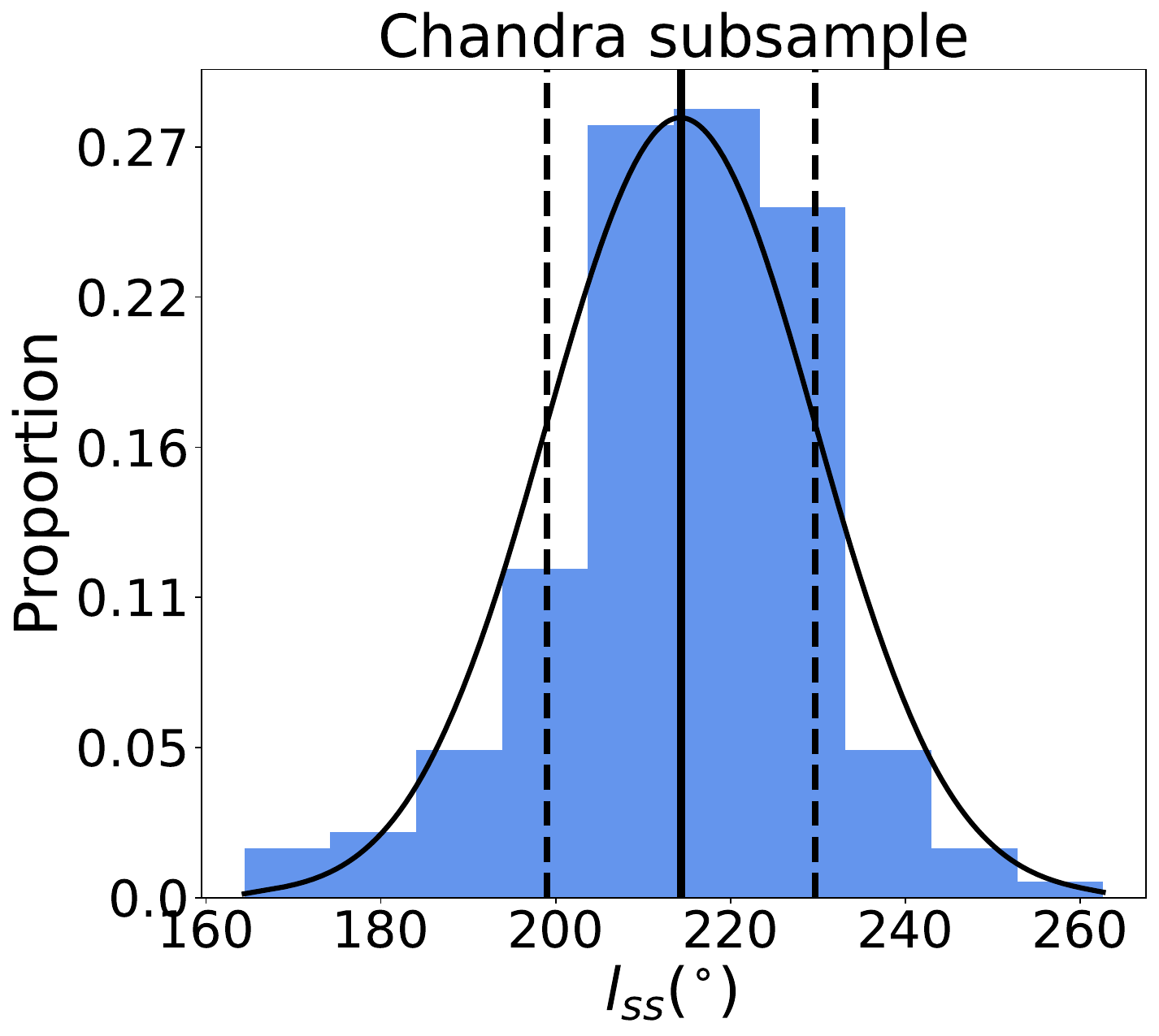}
    \includegraphics[width=0.29\textwidth]{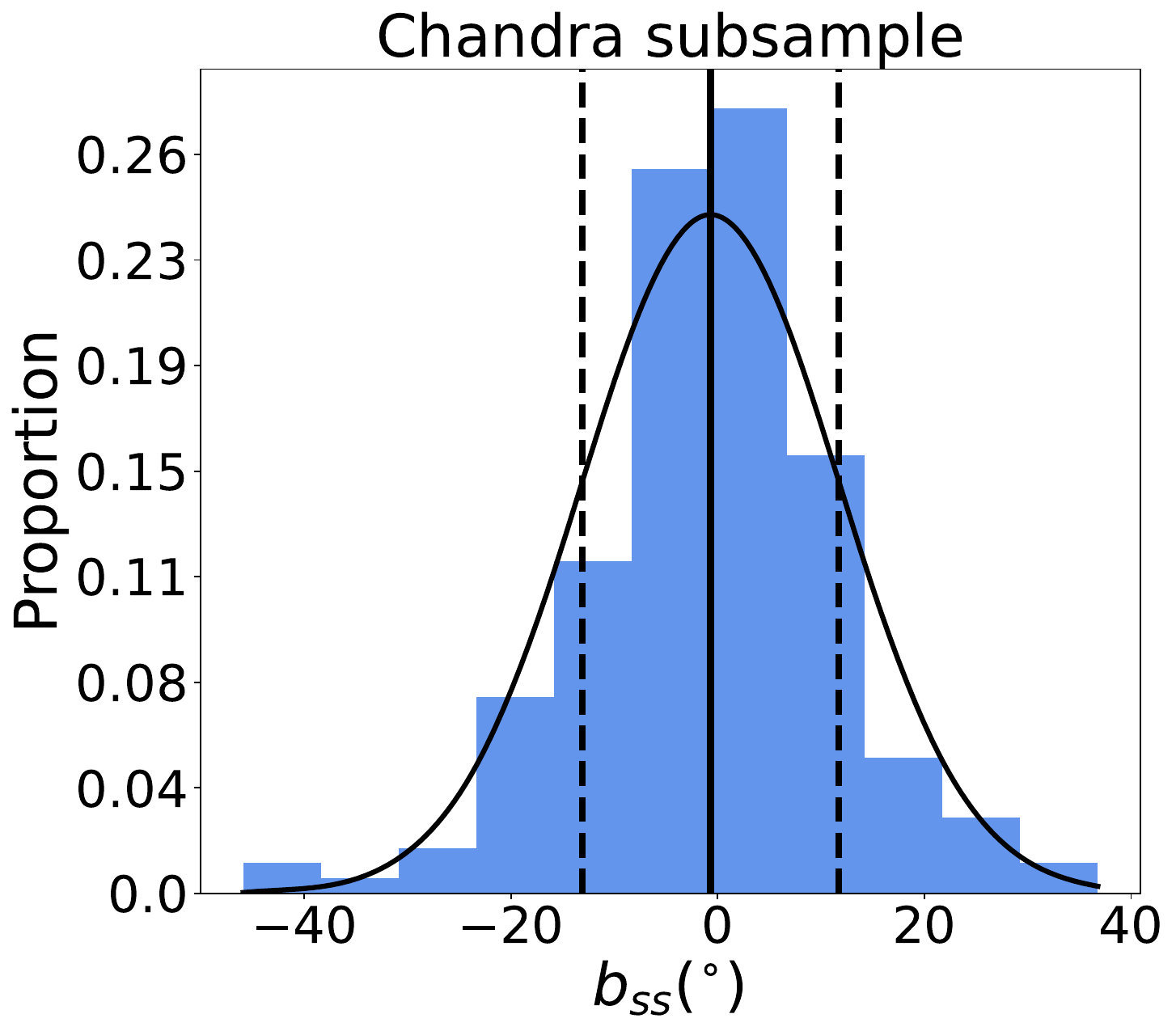}
    \includegraphics[width=0.29\textwidth]{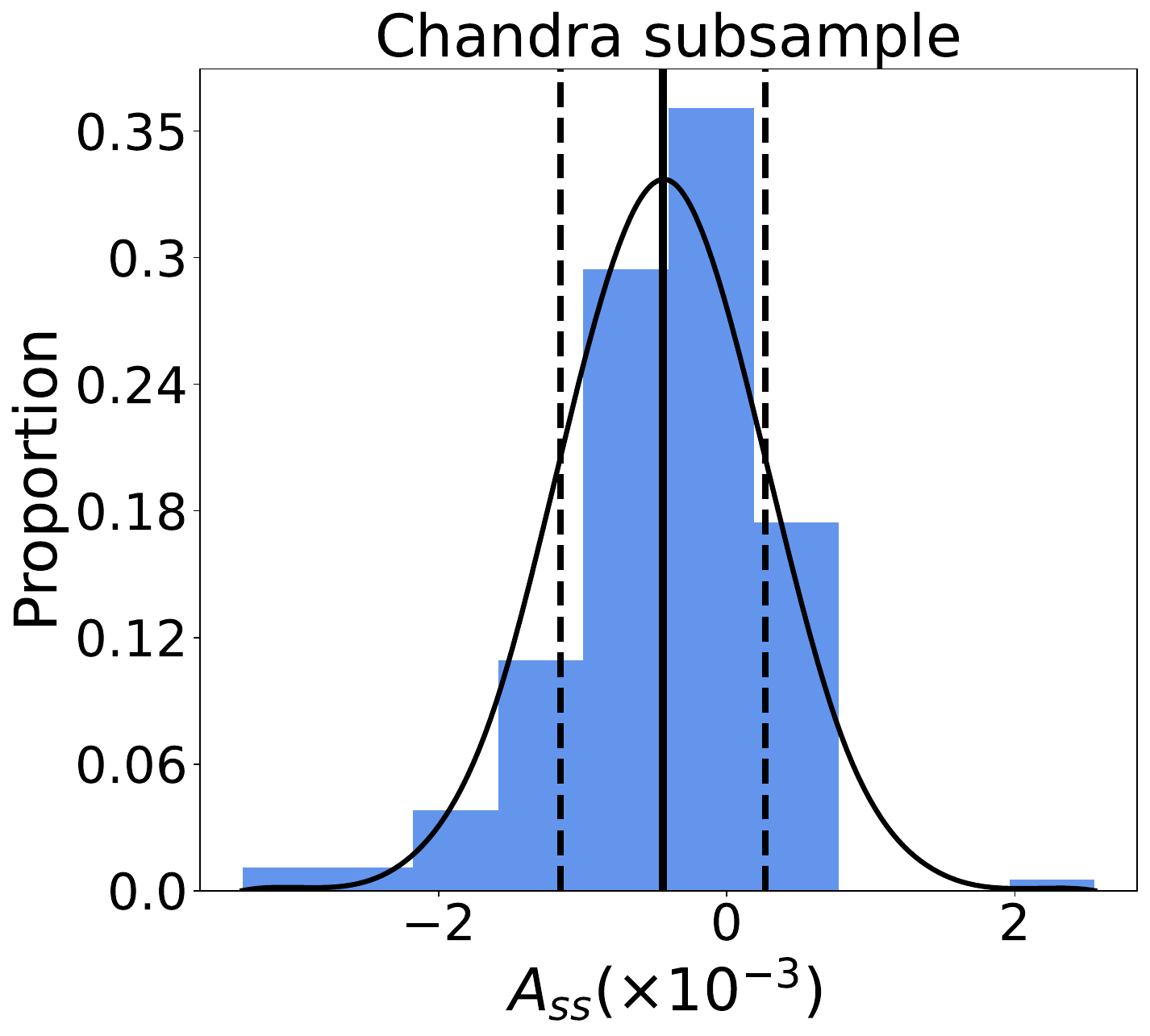}
    \caption{\label{Figss} Distribution of dipole correction parameters ($l$, $b$, $A$) from 200 Chandra subsamples. The distributions of correction parameters exhibit a Gaussian profile, yielding best-fit values of $l_{ss}$ = 214.34$^{\circ}$$\pm$15.36$^{\circ}$, $b_{ss}$ = $-$0.68$^{\circ}$$\pm$12.4$^{\circ}$ and $A_{ss}$ = $-$4.4$\pm$0.71 $\times$10$^{-4}$.} 
\end{figure*}

We generated 200 Chandra subsamples to perform a controlled matched-sample analysis. The distribution of the best fit of $L_{X}-T$ relation parameters (k, s, $\sigma_{int}$) can be described by a Gaussian distribution, where $k_{ss}$ = 1.22$\pm$0.07, $s_{ss}$ = 1.95$\pm$0.09, and $\sigma_{int, ss}$ = 0.23$\pm$0.02. We present the results of the $L_{X}-T$ relation using scatter plots with marginal histograms, as shown in Fig \ref{Figks}. The corresponding dipole fitting results are given for $l_{ss}$ = 214.34$^{\circ}$$\pm$15.36$^{\circ}$, $b_{ss}$ = -0.68$^{\circ}$$\pm$12.4$^{\circ}$, $A_{ss}$ = -4.4$\pm$0.71 $\times$10$^{-4}$, and $B_{ss}$ = -0.71$\pm$0.68 $\times$10$^{-4}$. Figure \ref{Figss} displays the results related to the anisotropy signal. Comparing the Chandra subsample results with the dipole fitting results of Chandra and XMM-Newton reveals that reducing the sample size and redshift range makes the anisotropy directions detected by both instruments tend to be consistent, but the $A$-value, reflecting the anisotropy intensity, does not change significantly. More pronounced anisotropic signal intensity was found in the XMM-Newton dataset with similar sample size and redshift range.

\subsection{Comparison with previous work}
\begin{figure*}[htp]
    \centering
    \includegraphics[width=0.7\textwidth]{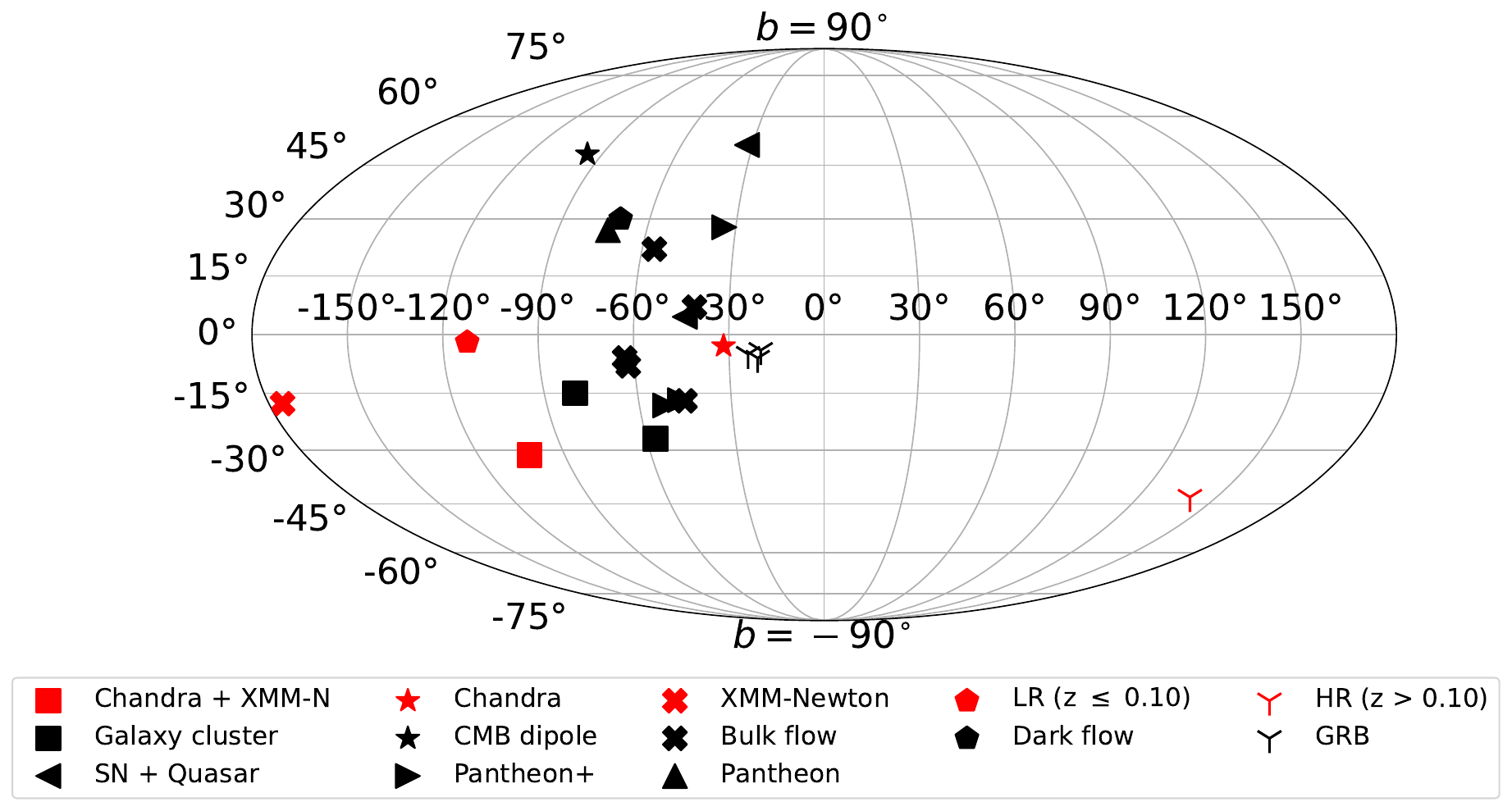}
    \caption{\label{Figpw} Distribution of preferred directions (l, b) from the previous research. Red represents the preferred direction of this work. Black represents the preferred direction given by other independent observations including the galaxy cluster \cite{2021A&A...649A.151M}, CMB dipole \cite{2016A&A...594A...1P,2020A&A...641A...1P}, bulk flow \cite{2012MNRAS.420..447T,2017MNRAS.468.1420F,2023MNRAS.524.1885W}, dark flow \cite{2022JHEAp..34...49A}, GRB \cite{2022MNRAS.511.5661Z}, SN + Quasar \cite{2020A&A...643A..93H}, Pantheon+\cite{2024A&A...681A..88H,2024A&A...689A.215H,2024ApJ...967...47L} and Pantheon \cite{2020PhRvD.102b3520K}.}
\end{figure*}

In the section, we compared our preferred directions with those of the previous research. First, we compared our preferred directions with \cite{2020A&A...636A..15M} that utilize the same galaxy cluster sample. Utilizing the sky scanning method, Migkas et al. \cite{2020A&A...636A..15M} found strong anisotropies ($\leq 4\sigma$ confidence level) toward Galactic coordinates (280$^{\circ}$, $-20$$^{\circ}$) from the Chandra + XMM-Newton dataset. Our preferred direction (${257.82^{\circ}}_{-52.88}^{+58.01}$, $-31.30{^{\circ}}_{-39.46}^{+35.92}$), inferred from the same dataset via the novel DF method, is consistent with it within $1.49\sigma$. Note that this result only considers one-sided error; if both-sided error is considered, the value will decrease. Our analysis yields a negligible anisotropy signal ($0.63\sigma$ and $0.62\sigma$), in sharp contrast to the stronger significance found by Migkas et al. \cite{2020A&A...636A..15M}.

Compared to the Chandra, LR and total datasets, we found more pronounced anisotropic signals in the XMM-Newton and HR datasets. The preferred directions obtained from the XMM-Newton and HR datasets are quite different from previous studies, including CMB dipole \cite{2016A&A...594A...1P,2020A&A...641A...1P}, bulk flow \cite{2012MNRAS.420..447T,2017MNRAS.468.1420F,2023MNRAS.524.1885W}, dark flow \cite{2022JHEAp..34...49A}, Pantheon \cite{2020PhRvD.102b3520K}, Pantheon+ \cite{2024A&A...681A..88H,2024A&A...689A.215H,2024ApJ...967...47L}, GRB \cite{2022MNRAS.511.5661Z}, SN + Quasar \cite{2020A&A...643A..93H}. To facilitate understanding, we aggregated these results with the ones we obtained, marking them on the galactic coordinate system, as shown in Fig. \ref{Figpw}. Comparisons that cover different redshift ranges and exhibit drastically different statistical significance results based solely on the anisotropic directions shown in Fig. \ref{Figpw} have significant limitations. However, the significant discrepancies between the research results also reflect that the actual structure of the Universe may be far more complex than current models predict. This indicates that rigorous testing of the cosmological principle remains a long and arduous task.

\begin{table*}[htp]
\centering
	\caption{Best fitting results of the dipole parameter ($l$, $b$, $A$ and $B$) from different datasets. The symbol $*$ represents $A$ and $B$ when the preferred direction ($l$, $b$) fixed as the best fitting results. \label{T2}}
    \renewcommand{\arraystretch}{1.7}
	\begin{tabular}{c|cccc|cc}     
			\hline\hline
			 Sample & $l$ & $b$  & $A$ & $B$  & $A^{*}$ & $B^{*}$ \\
               & [$^{\circ}$] & [$^{\circ}$] & $\times$10$^{-4}$ & $\times$10$^{-4}$ & $\times$10$^{-4}$ & $\times$10$^{-4}$ \\ \hline	
            Chandra  & 328.31$_{-63.56}^{+66.52}$ &  $-$2.91$_{-52.27}^{+57.78}$  & $-$1.9$_{-6.2}^{+6.8}$ & $-$0.6$\pm$3.7 & $-$5.9$\pm$7.5 & $-$0.3$\pm$3.6 \\
            & 148.32$_{-57.50}^{+60.16}$ &  4.62$_{-57.03}^{+51.70}$  & 2.2$_{-6.8}^{+6.1}$ & $-$0.6$\pm$3.7  & -- & --  \\
            \hline
            XMM-Newton & 184.42$_{-24.58}^{+29.01}$  &  $-$17.73$_{-21.53}^{+18.55}$  & $-$32.0$_{-18.0}^{+15.0}$  & $-$3.9$\pm$8.1  & $-$41$\pm$14 & $-$5.8$\pm$6.6 \\
            & 5.0$_{-25.33}^{+28.81}$  & 17.61$_{-18.68}^{+21.56}$  & 32.0$_{-15.0}^{+18.0}$ & $-$3.9$\pm$8.0  & -- & -- \\
            \hline
            Chandra + XMM-N &  257.82$_{-52.88}^{+58.01}$  &  $-$31.30$_{-39.46}^{+35.92}$ & $-$5.4$\pm$6.3  & 0.5$\pm$3.2  & $-$8.8$\pm$5.7 & 0.5$\pm$3.2  \\ 
            & 80.89$_{-52.46}^{+60.97}$  & 31.75$_{-40.16}^{+35.19}$  & 5.2$_{-5.8}^{+6.5}$ & 0.4$\pm$3.2  & -- & -- \\
            \hline
            LR (z $\leq$ 0.10)  & 247.66$_{-49.20}^{+58.30}$  &  $-$1.93$_{-43.45}^{+45.43}$ & $-$6.5$_{-8.0}^{+9.7}$  & $-$0.9$\pm$4.2  & $-$14.1$\pm$8.3 & $-$0.9$\pm$4.1  \\ 
             &  73.01$_{-46.50}^{+62.39}$  & 2.21$_{-44.55}^{+42.77}$  & 6.6$_{-9.6}^{+8.3}$ & $-$0.8$\pm$4.2 & -- & -- \\
            \hline
            HR (z $>$ 0.10)  &  139.84$_{-35.47}^{+39.62}$  & -43.07$_{-25.53}^{+21.52}$  & $-$16.2$_{-8.1}^{+9.0}$ & $-$2.6$\pm$4.2 & $-$20.5$\pm$7.8 & $-$3.3$\pm$4.2 \\
             & 320.57$_{-35.15}^{+40.04}$  &  43.90$_{-21.64}^{+25.52}$ & 16.3$\pm$8.8  & $-$2.5$\pm$4.3 & -- & --  \\ 
		\hline\hline
		\end{tabular}
\end{table*}

\begin{table*}[htp]
\centering
	\caption{Analysis results of the statistical isotropy. $A$ represents the dipole magnitude obtained from the real data. $\mu_{A}$ represents the expected value and standard deviation of the statistical isotropic results. $D$ represent the statistical significance of the real data compared to the statistical results $\mu_{A}$, and its value is rounded to one decimal place}. \label{T3} 
    \renewcommand{\arraystretch}{1.7}
		\begin{tabular}{c|c|cc|cc}
			\hline\hline
               & Real data & \multicolumn{2}{c|}{Bootstrap scheme} & \multicolumn{2}{c}{Randomized scheme} \\ 
		Sample & $A$   & $\mu_{A_{B}}$ & $D$  & $\mu_{A_{R}}$ & $D$  \\ 
        & $\times$10$^{-4}$ & $\times$10$^{-5}$ & ($\sigma$)  & $\times$10$^{-5}$ & ($\sigma$)  \\ \hline
		Chandra & $-$1.9$_{-6.2}^{+6.8}$  & -3.1$\pm$77.1 & 0.2$\pm$0.9  &  2.2$\pm$67.6  & 0.3$\pm$0.8  \\
        XMM-Newton & $-$32.0$_{-18.0}^{+15.0}$ & 5.5$\pm$143.9 & 1.8$\pm$0.8  & 4.0$\pm$113.8  & 1.9$\pm$0.7 \\
        Chandra + XMM-N & $-$5.4$\pm$6.3 & -0.1$\pm$59.5 & 0.6$\pm$0.7 & -2.0$\pm$59.6  & 0.6$\pm$0.7  \\ 
        LR (z $\leq$ 0.10) & $-$6.5$_{-8.0}^{+9.7}$ & -2.0$\pm$86.8 & 0.6$\pm$0.8  & -0.3$\pm$74.3  & 0.7$\pm$0.8  \\ 
        HR (z $>$ 0.10)& $-$16.2$_{-8.1}^{+9.0}$ & -1.5$\pm$82.0 & 1.7$\pm$0.8 & 4.8$\pm$71.9  & 1.8$\pm$0.8   \\ 
			\hline\hline
		\end{tabular} 
\end{table*}

\section{Summary} \label{sum}
In this paper, we utilized the DF method to the galaxy cluster for the first time to search for the preferred direction of the cosmic anisotropy. By using the $L_{X}-T$ correlation, we were able to apply the DF method to the galaxy cluster. The Chandra + XMM-Newton dataset (313 galaxy clusters) gives the preferred direction as $(l, b)$ = (${257.82^{\circ}}_{-52.88}^{+58.01}$, $-31.30{^{\circ}}_{-39.46}^{+35.92}$), with corresponding anisotropy level ($A$) of $-$5.4 $\pm$ 6.3 $\times$ 10$^{-4}$. The Chandra and XMM-Newton datasets yielded very distinct results, with the latter exhibiting a stronger anisotropic signal. This significant difference might stem from the real physical differences or from the selection effects caused by device differences. Meanwhile, the reanalyses with different redshift subsamples (LR and HR) shows that the $L_{X}-T$ correlation and the cosmic anisotropy might evolve with redshift. This hints that when exploring anisotropy using the spatial distribution differences of the $L_{X}-T$ relation, the difference in redshift distribution on the sky can now masquerade as anisotropies, thereby affecting the detected anisotropy signal. Based on the results of statistical isotropic analysis (Bootstrap and Randomized), we find that the real magnitude ($A$) is much larger than the simulated magnitude ($\mu_{A_{B}}$ and $\mu_{A_{R}}$). This suggests that the anisotropic signals are more likely to originate from the non-uniformity of the cosmic structure. Furthermore, these two schemes yield anisotropy of the same magnitude, with $\mu_{A_{R}}$ still deviates from zero. This hints that other factors might exist within galaxy clusters that contribute to anisotropy, and that these factors contribute to anisotropy in a manner comparable to the non-uniformity of the spatial distribution. Currently, the maximum statistical significance obtained by the two schemes is 1.8$\sigma$ (Bootstrap) and 1.9$\sigma$ (Randomized), both derived from the XMM-Newton dataset.

Overall, the DF method can be well integrated with the galaxy clusters and applied to the detection of cosmic anisotropy. Final results indicate the presence of anisotropic signals within the galaxy clusters. In the future, we can further confirm and understand this signal using high-quality galaxy cluster data from the e-ROSITA \cite{2012arXiv1209.3114M,2024A&A...688A.107M,2025arXiv251114356R,2025A&A...704A.346Z}.

\section*{DATA AVAILABILITY}
The data used in the paper are publicly available in Tables C.1. in literature \citep{2020A&A...636A..15M}.

\section*{Conflict of Interest}
No conflict of interest exists in this contribution.

\appendix
\section{Additional materials}
\setcounter{figure}{0}          
\renewcommand{\thefigure}{A\arabic{figure}}
Figures \ref{fig4} and \ref{fig5} show the constraints of the dipole parameter ($l$, $b$, $A$ and $B$) obtained using the hemispherical search. In other words, the constraints in Figs. \ref{Figdy} and \ref{Figdred} are split into two separate constraints. Figure \ref{figfree} displays the constraints of the anisotropy signal for the Chandra + XMM-Newton sample, with the $L_{X}-T$ relation parameters treated as free parameters. In this case, the free parameters include the $L_{X}-T$ relation parameters ($k$, $s$, $\sigma_{int}$) and the correction parameters ($l$, $b$, $A$, $B$). As can be seen from the figure, parameters $k$ and $B$ exhibit a distinct coupling characterized by a negative correlation.

\begin{figure*}[htp]
    \centering 
    \includegraphics[width=0.4\textwidth]{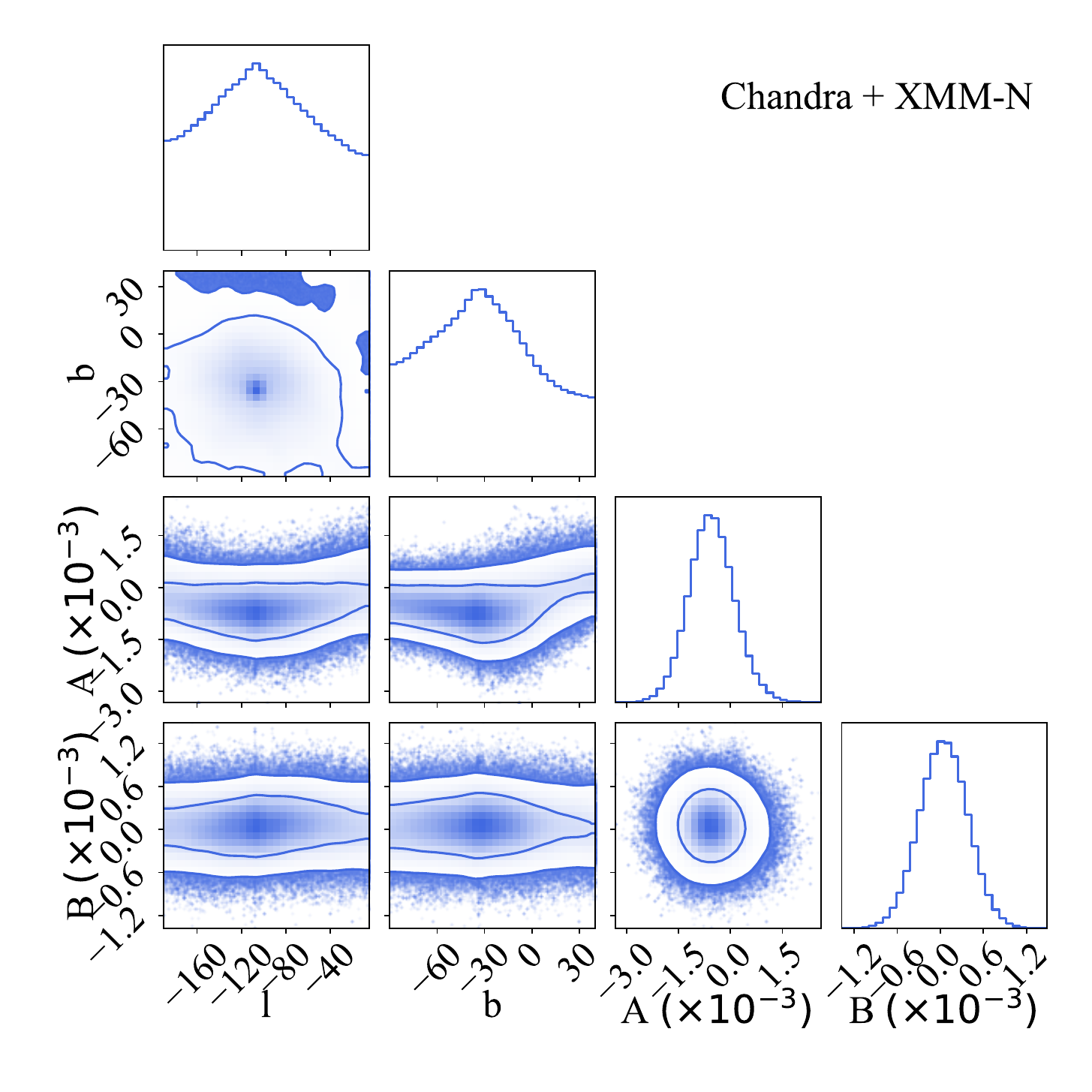}
    \includegraphics[width=0.4\textwidth]{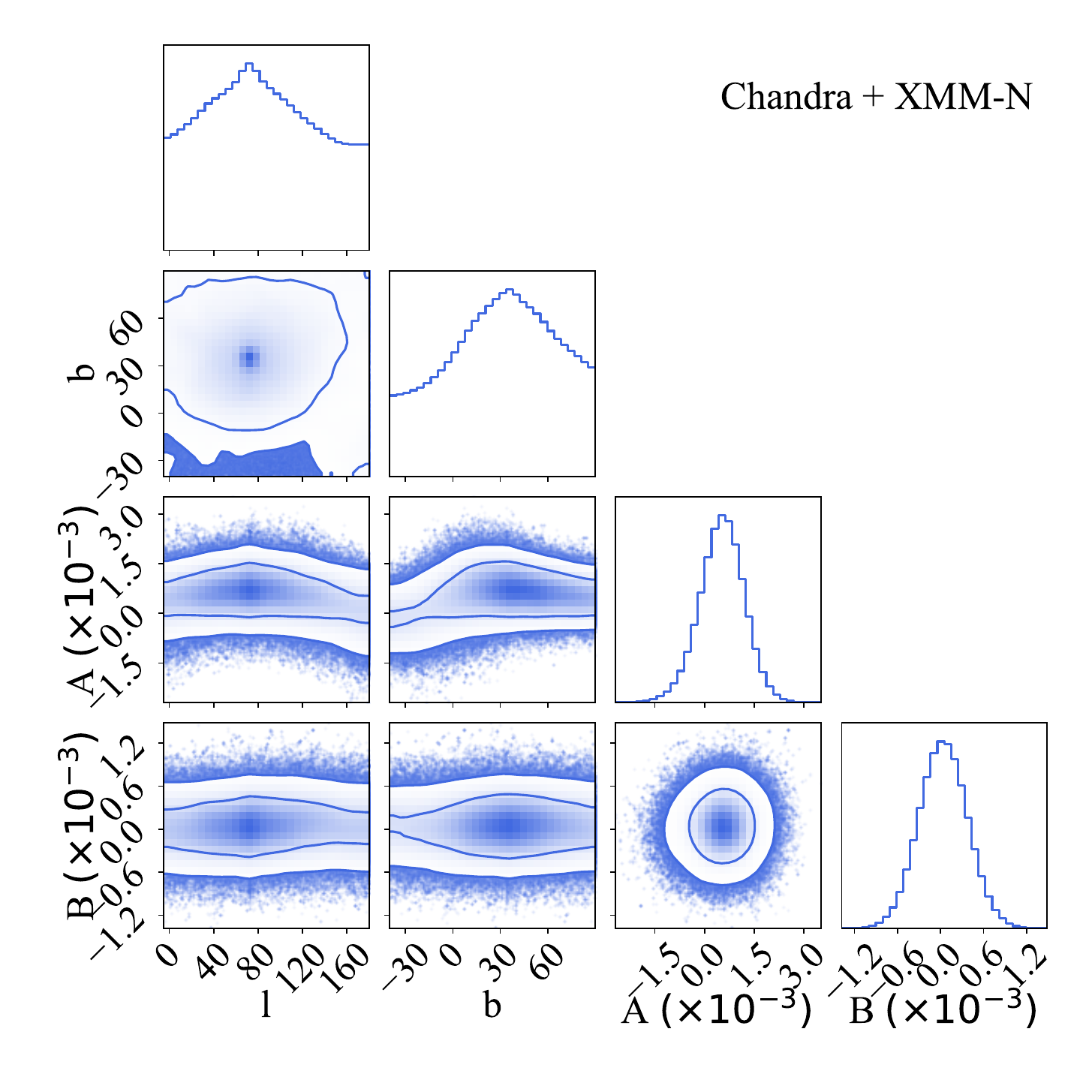} 
    \includegraphics[width=0.4\textwidth]{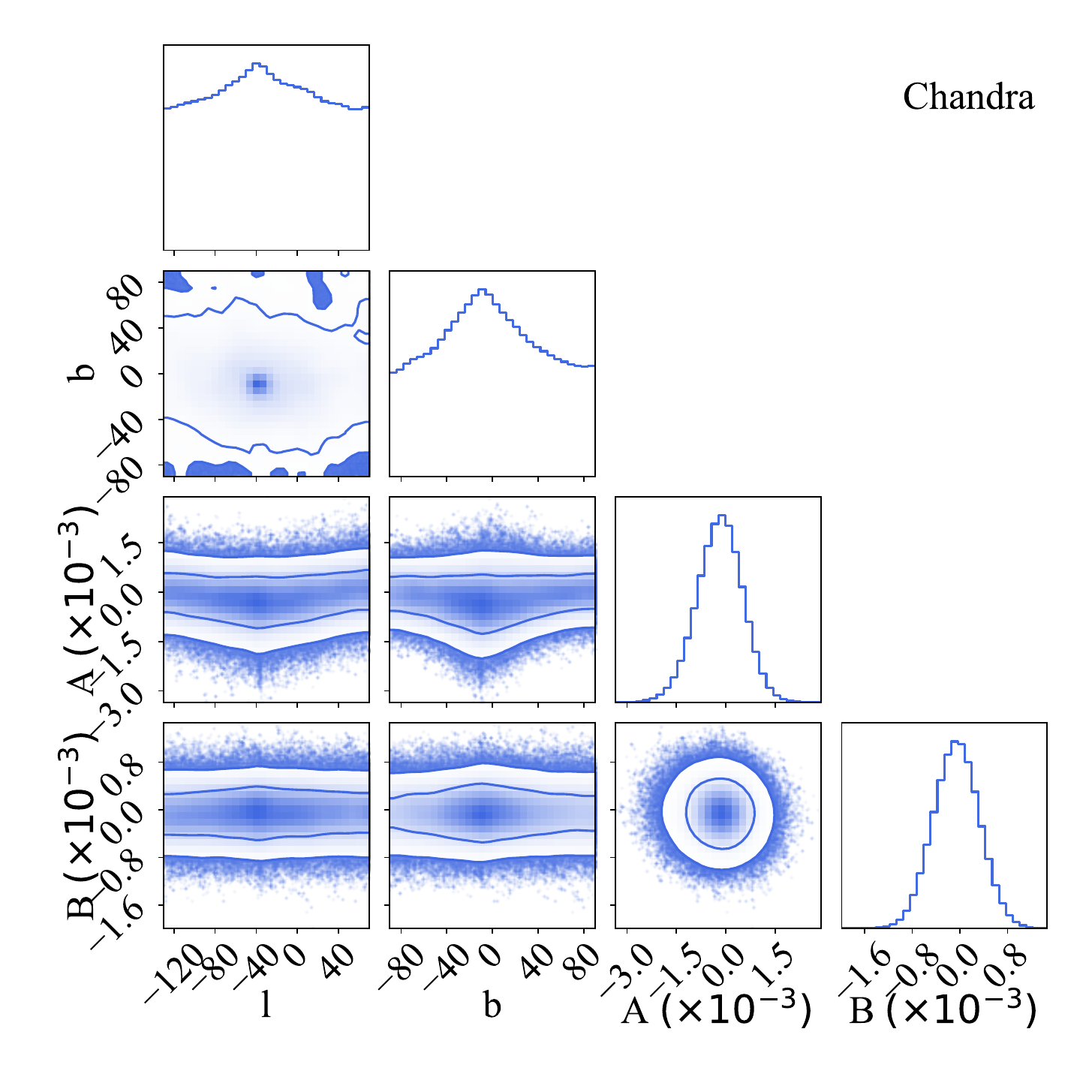}
    \includegraphics[width=0.4\textwidth]{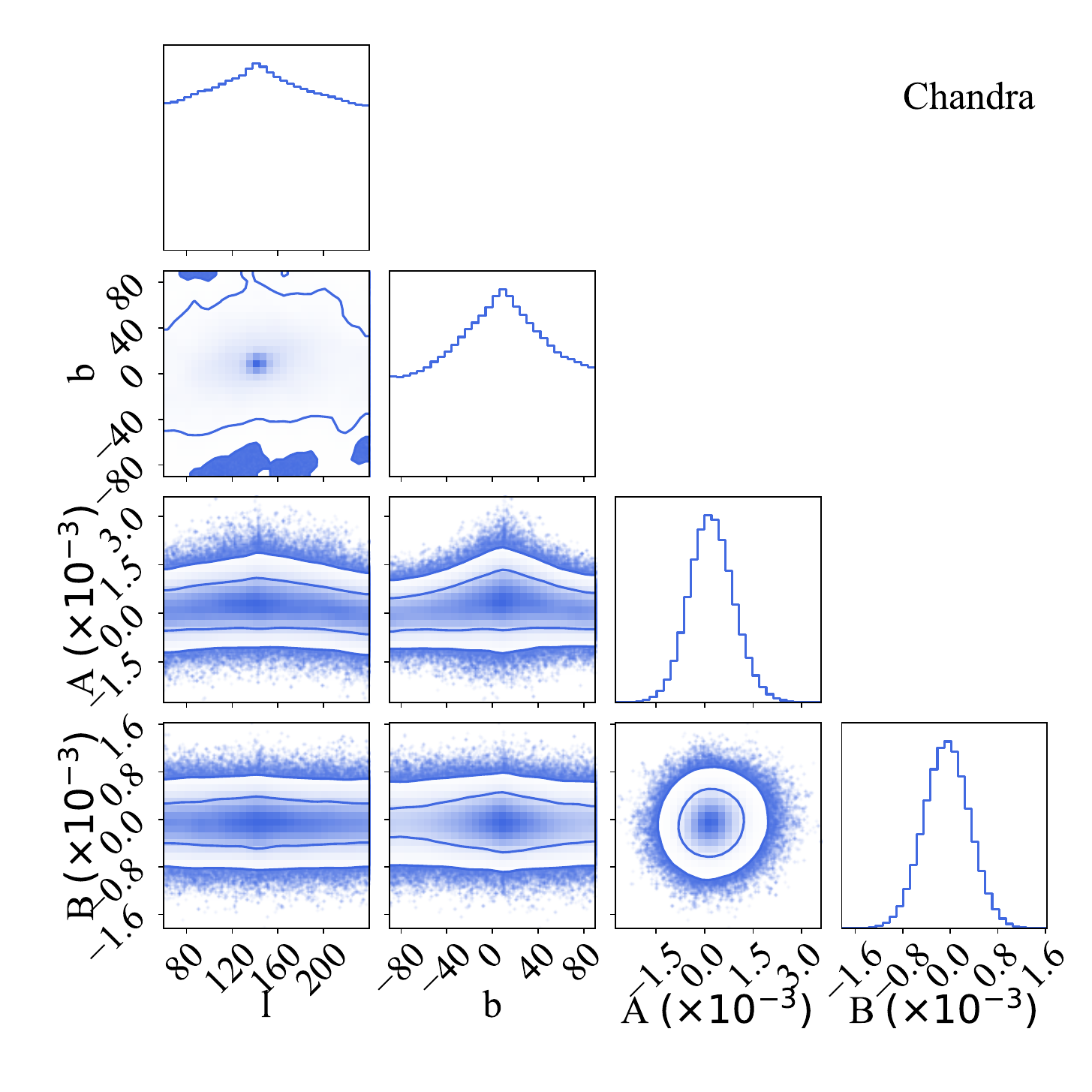}
    \includegraphics[width=0.4\textwidth]{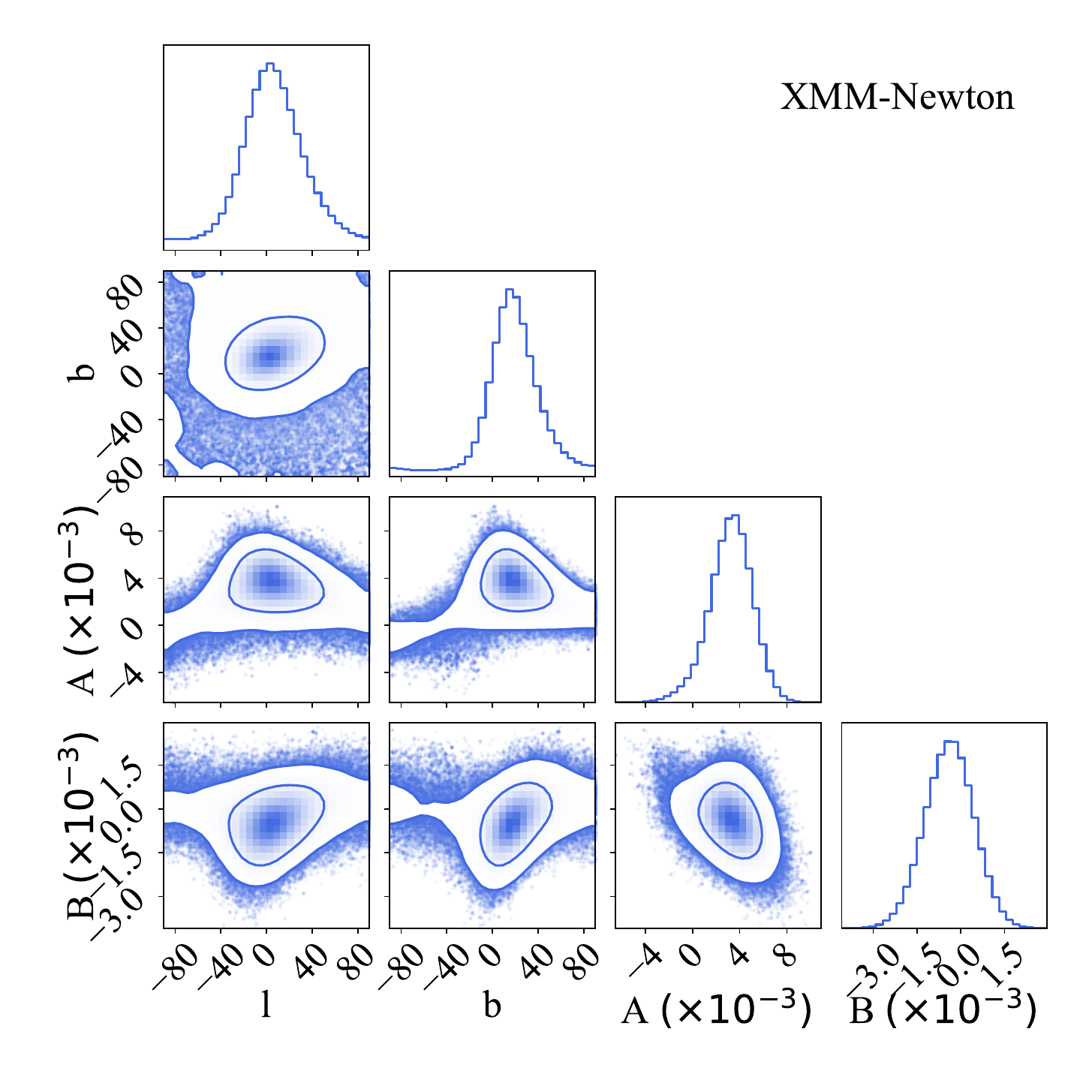}
    \includegraphics[width=0.4\textwidth]{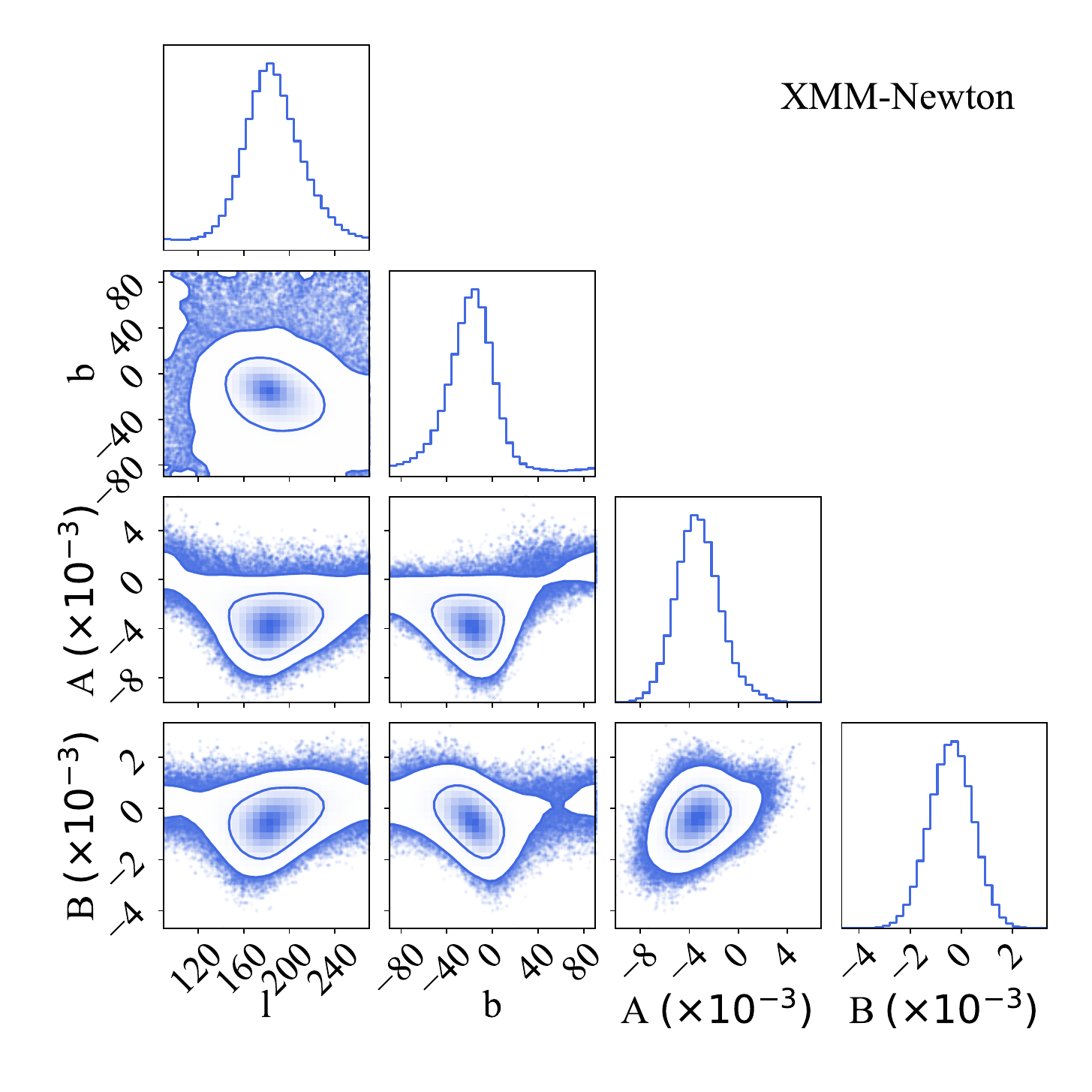}
    \caption{\label{fig4} Confidence contours (1$\sigma$ and 2$\sigma$) of the dipole parameters (l, b, A, and B) in hemispherical searches for different types datasets, including Chandra + XMM-Newton, Chandra and XMM-Newton.}
\end{figure*}

\begin{figure*}[htp]
    \centering 
    \includegraphics[width=0.8\textwidth]{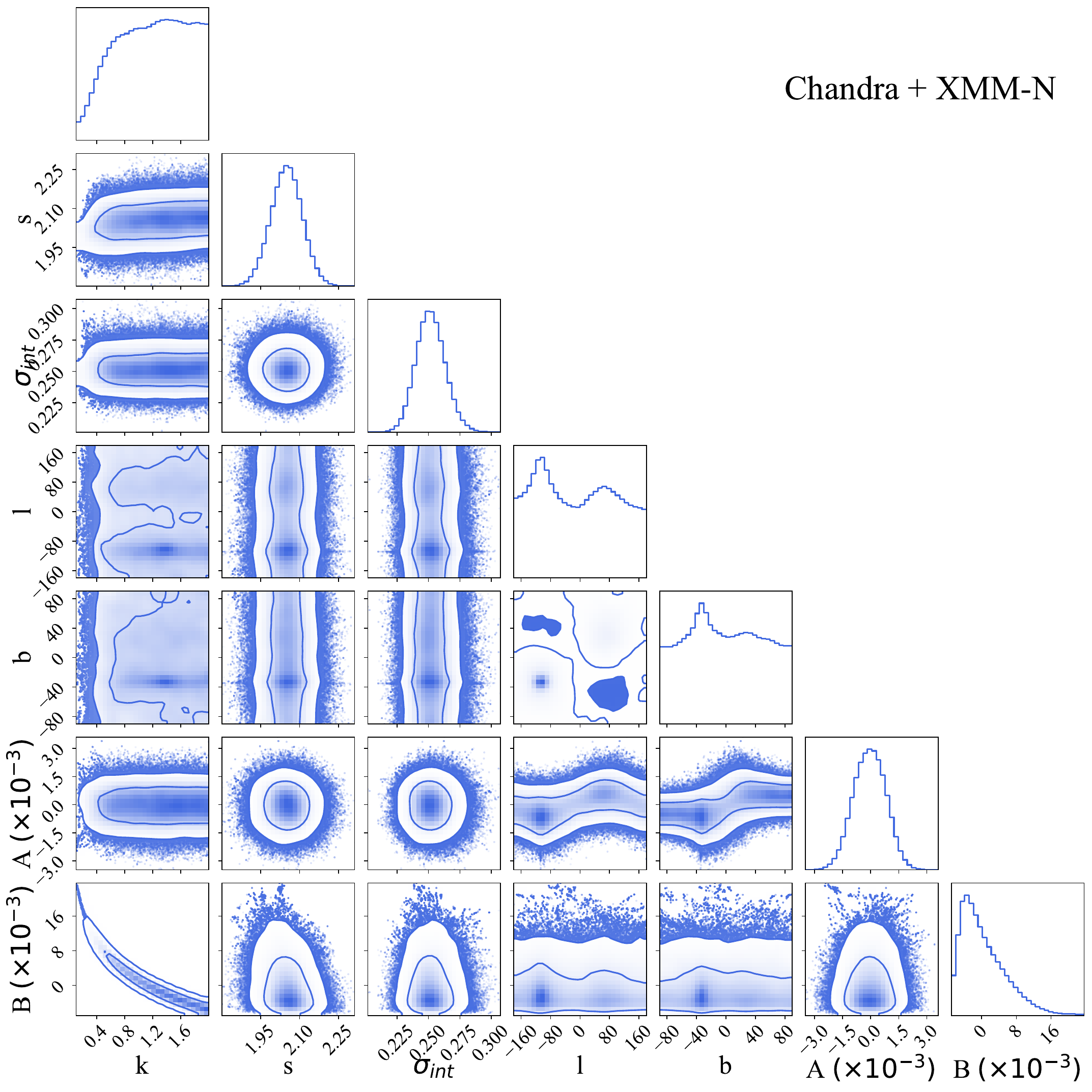}
    \caption{\label{figfree} Confidence contours (1$\sigma$ and 2$\sigma$) of the $L_{X}-T$ relation parameter (k, s, and $\sigma_{int}$) and the dipole parameters (l, b, A, and B) for the Chandra + XMM-Newton dataset.}
\end{figure*}

\begin{figure*}[htp]
    \centering 
    \includegraphics[width=0.38\textwidth]{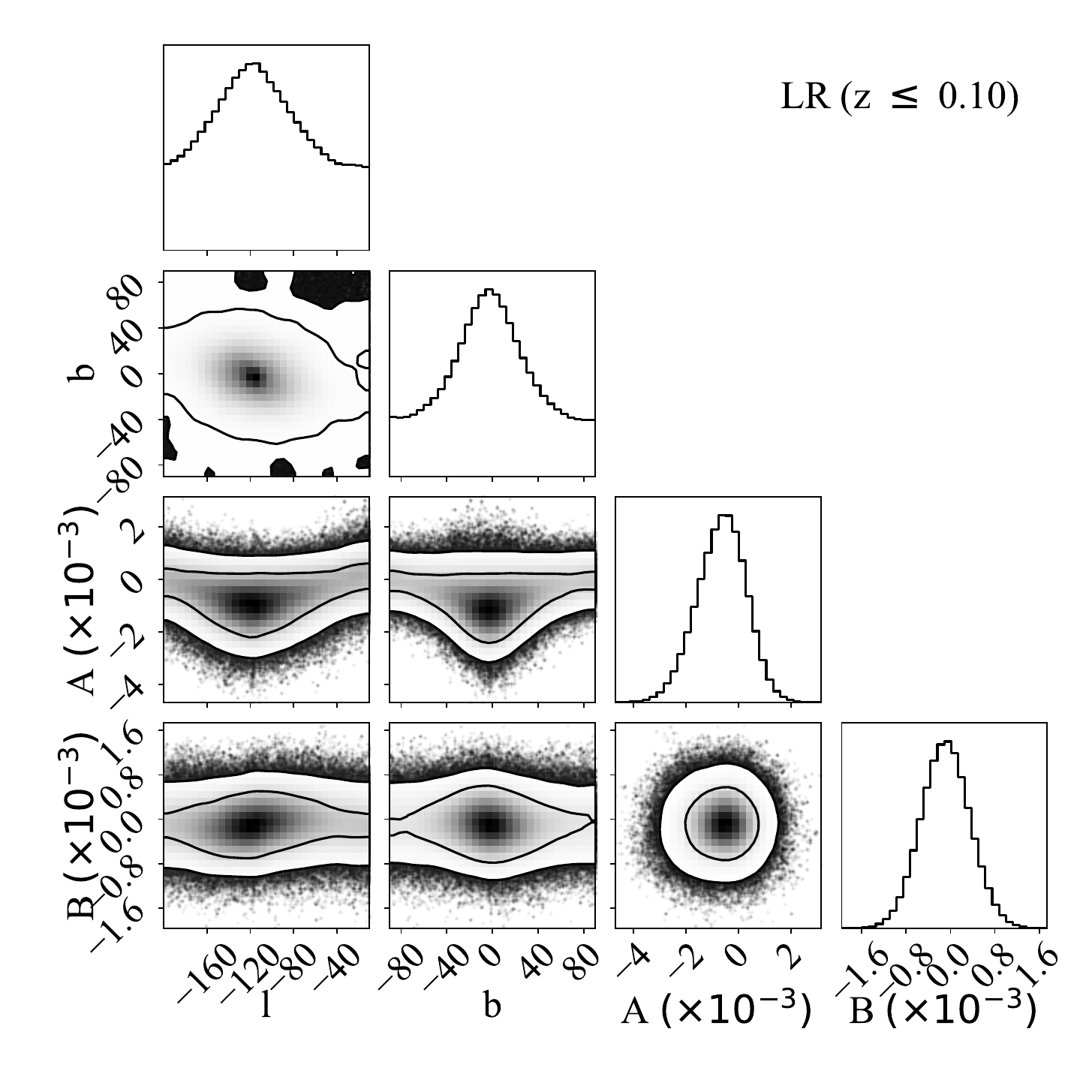}
    \includegraphics[width=0.38\textwidth]{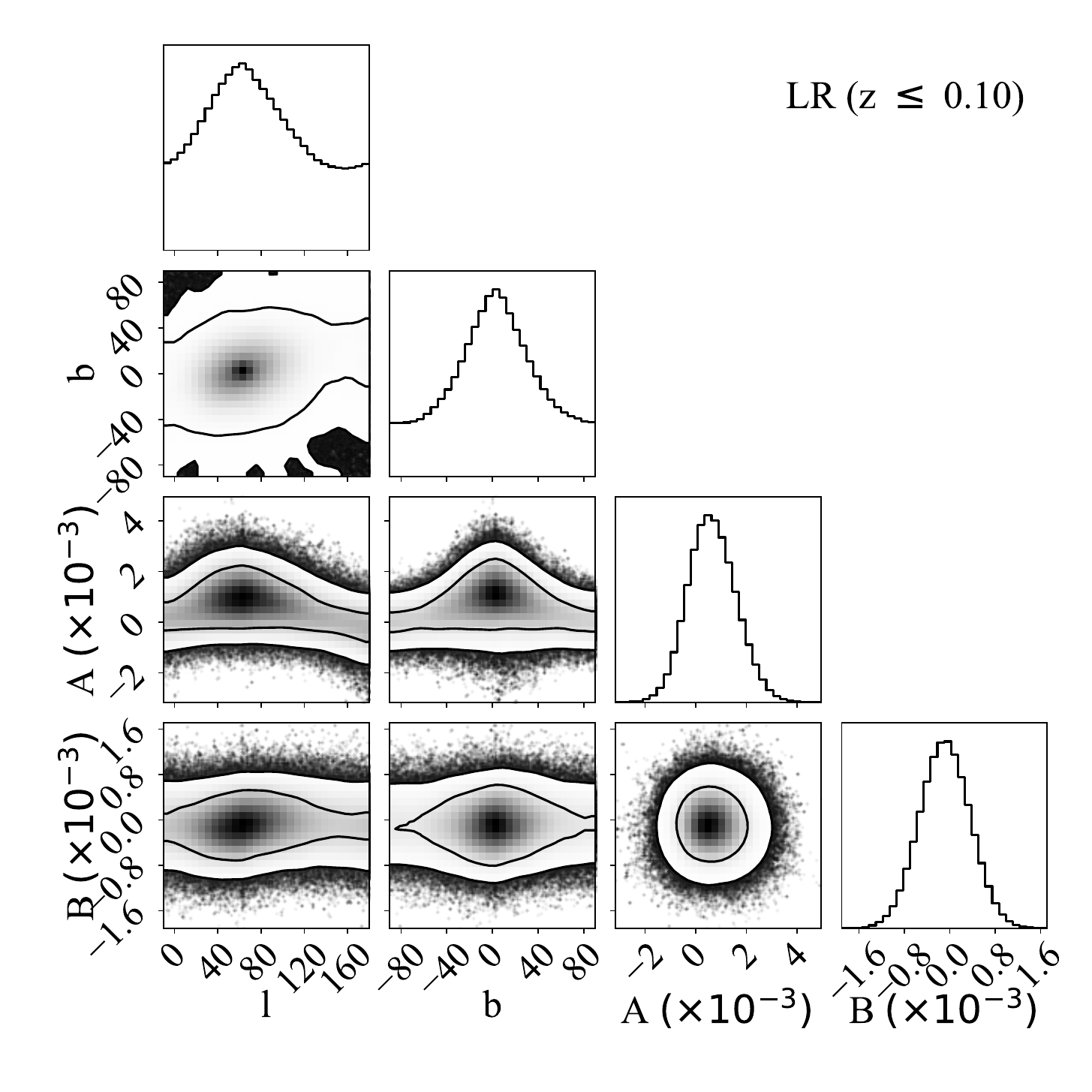} 
    \includegraphics[width=0.38\textwidth]{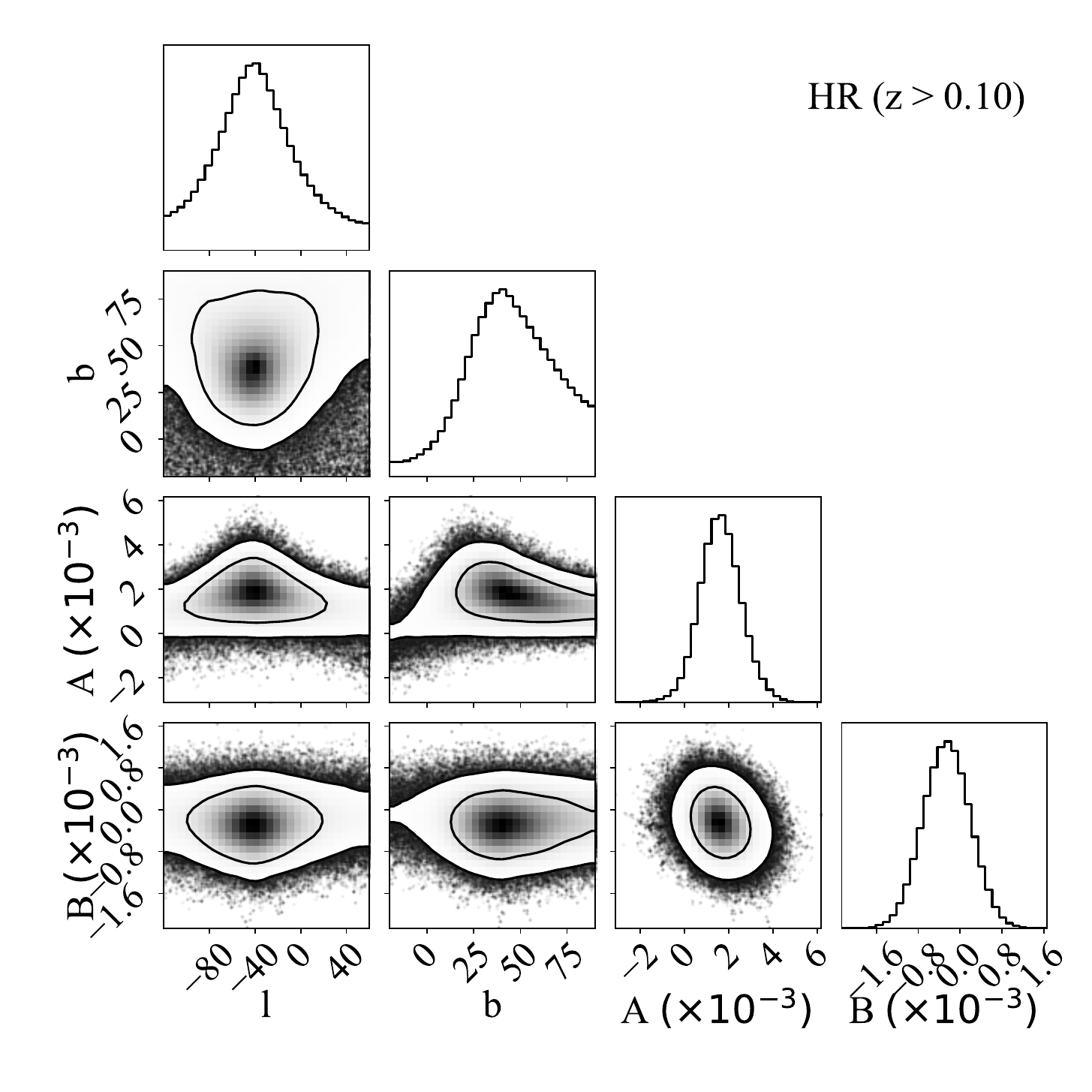}
    \includegraphics[width=0.38\textwidth]{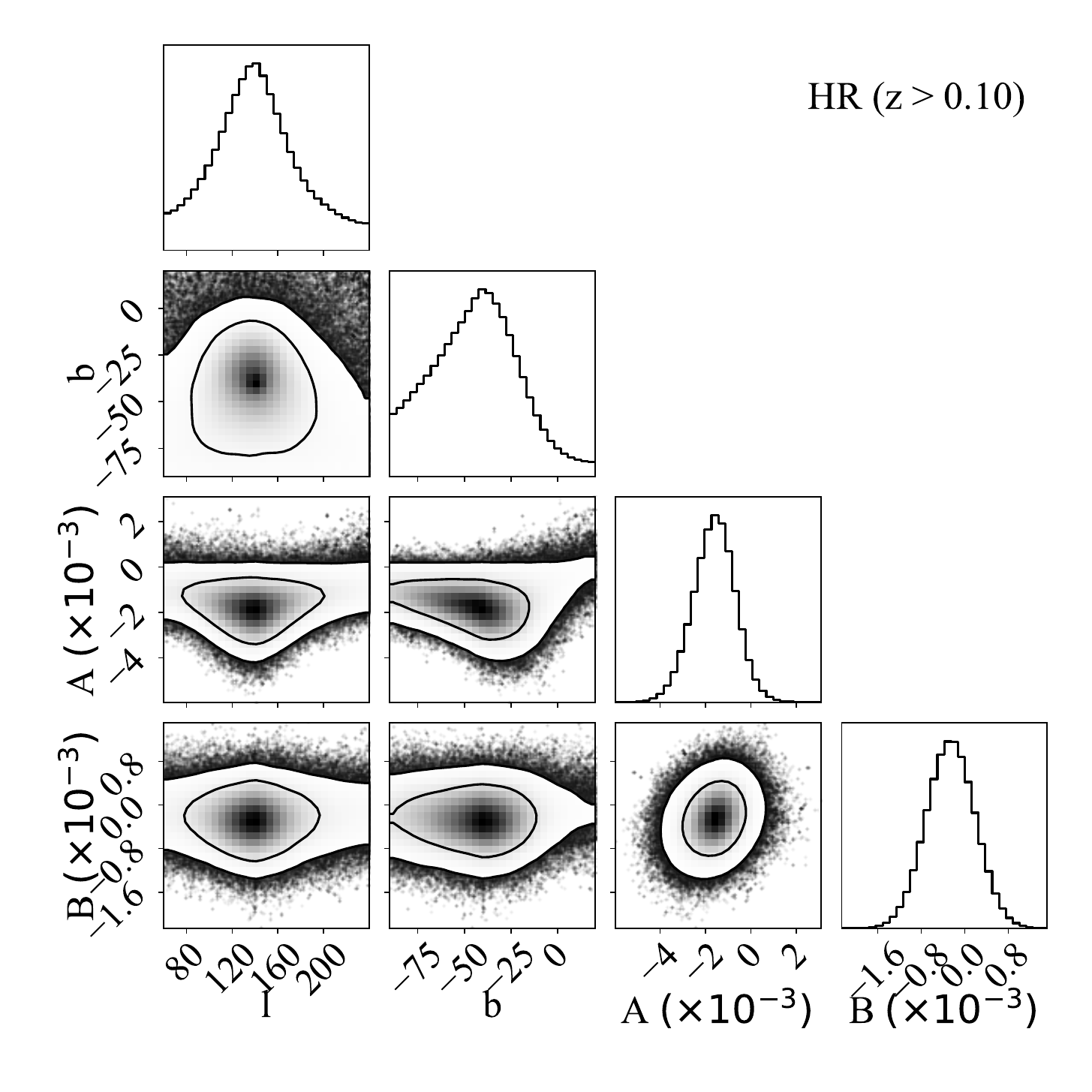}
    \caption{\label{fig5} Confidence contours (1$\sigma$ and 2$\sigma$) of the dipole parameters (l, b, A, and B) in hemispherical searches for different redshift ranges (LR and HR).}
\end{figure*}

\section*{Acknowledgements}
Anonymous reviewer provided many constructive comments, for which we are deeply grateful. This work was supported by the National Natural Science Foundation of China (grant No. 12494575, 12273009, 12373026, 12003020), Shaanxi Fundamental Science Research Project for Mathematics and Physics (grant No. 23JSY015), the Fundamental Research Funds for the Central Universities (No. xzy012026061), and Project funded by China Postdoctoral Science Foundation (No. 2026M793512).

\end{document}